\documentclass[useAMS,usenatbib,usegraphicx]{mn2e}
\pdfoutput=1
\usepackage{color}

\title[IFU spectroscopy of 10 ETG nuclei - IV]
  {IFU spectroscopy of 10 early-type galactic nuclei - IV. Properties of the circumnuclear stellar kinematics}
\author[Ricci et al.]
  {T.V.~Ricci\thanks{tiago.ricci@uffs.edu.br}$^{1,2}$,
  J.E.~Steiner$^2$, R.B.~Menezes$^2$ \\
  $^1$Universidade Federal da Fronteira Sul, Campus Cerro Largo, RS 97900-000, Brasil \\
  $^2$Instituto de Astronomia, Geof\'isica e Ci\^encias Atmosf\'ericas, Universidade de S\~ao Paulo, 05508-900, S\~ao Paulo, Brasil}
\date{Released 2002 Xxxxx XX}

\pagerange{\pageref{firstpage}--\pageref{lastpage}} \pubyear{2002}

\def\LaTeX{L\kern-.36em\raise.3ex\hbox{a}\kern-.15em
    T\kern-.1667em\lower.7ex\hbox{E}\kern-.125emX}

\begin{document}

\label{firstpage}

\maketitle

\begin{abstract}
The study of stellar kinematic properties may provide hints on the formation and evolution of elliptical and lenticular galaxies. Although most previous studies have focused on the large scale of these galaxies, their central regions (scales of $\sim$ 100 pc) may contain important clues about their structure, such as kinematically decoupled cores. This is the fourth paper on a sample of 10 massive ($\sigma$ $>$ 200 km s$^{-1}$) and nearby ($d$ $<$ 31 Mpc) early-type galaxies, observed with the integral field unit of the Gemini South Multi Object Spectrograph. Here, we analyse the properties of the stellar kinematics in the circumnuclear region. We fitted the line-of-sight velocity distribution with a Gauss-Hermite function. In seven galaxies of the sample, we detected a rotation pattern in their radial velocity maps that are anti-correlated with $h_3$. We interpret this as stellar structures in rotation embedded in the bulges of the objects. Comparing the stellar kinematic results with the PCA Tomography results and also with the gas kinematic results of IC 5181, it seems that this object may have a non-axisymmetric potential at its centre. The velocity dispersion maps of four objects have a nuclear peak, which must correspond, in part, to unresolved stellar rotation. In NGC 1404, we detected a kinematic decoupled core with an extension of $\sim$ 200 pc. This galaxy also has a $\sigma$-drop in the centre, which may be related to both stellar components in counterrotation or with a kinematically cold star-forming region. 
\end{abstract}


\section{Introduction} \label{sec:intro}

The study of stellar kinematics in galaxies has been expanding our knowledge about the formation and structure of these objects. In spite of a separation between E and S0s based on photometry, \citet{2011MNRAS.416.1680C} proposed that early-type galaxies (ETGs) have a dichotomy regarding stellar kinematics: fast rotators and slow rotators \citep{2007MNRAS.379..401E,2011MNRAS.414..888E}. Both types are related to the amount of angular momentum measured for the stellar structure of the galaxies. The presence of slow and fast rotators in the local Universe is a consequence of the formation history of ETGs. Different types of mergers result in different configurations of stellar kinematics. \citet{2014MNRAS.444.3357N}, using hydrodynamic cosmologic simulations, proposed several scenarios for the stellar kinematics of the newly formed galaxy as a function of the merger characteristics and the properties of its progenitors (i.e. if the progenitors are gas-rich or gas-poor, if it is a major or a minor merger). One example is that the most massive E galaxies are slow rotators and are mainly round. This is a consequence of several gas-poor minor mergers in an already formed E galaxy. This scenario was also proposed by \citet{2009ApJS..182..216K}.
So far, the study of stellar kinematics using 3D spectroscopy has been mostly focused in large field of views (FOVs). One example is the SAURON project \citep{2001MNRAS.326...23B,2002MNRAS.329..513D}, which analysed 327 galaxies. Using the same instrument as the SAURON project, the ATLAS\textsuperscript{3D} project \citep{2011MNRAS.413..813C} covers the effective radius of a volume-limited sample of 260 ETGs. Also the SAMI Galaxy Survey \citep{2015MNRAS.447.2857B} will analyse 3400 galaxies, covering a broad stellar mass range and environment with $z$ $<$ 0.095. Indeed, \citet{2015MNRAS.454.2050F} presented stellar kinematics data for 106 galaxies from three clusters using the SAMI Pilot Survey. Another survey is MaNGA \citep{2015ApJ...798....7B}, which will analyse $\sim$ 10000 galaxies using an integral field unit (IFU) that is part of the fourth generation of the Sloan Digital Sky Survey programme. The MaNGA instrument has 17 IFUs with diameters that range from 12 to 32 arcsec. 

Although the sizes of the FOVs of such projects are large enough to study the structure of galaxies as a whole, their spatial resolution is $\sim$ 2 arcsec, which is insufficient to resolve the circumnuclear region (scales of $\sim$ 100 pc) of the galaxies. Some other analysis were performed on seeing-limited IFUs, but with a smaller FOV. \citet{2006MNRAS.373..906M} analysed 28 ETGs that were included in the SAURON sample with the OASIS spectrograph operating at the Canada-France-Hawaii Telescope. They showed that ETGs also have decoupled stellar structures with an extension of a few hundred pc. \citet{2006MNRAS.371..170B} observed the inner 5 arcsec region of six nearby galaxies with active galactic nuclei (AGNs) using the Gemini Multi Object Spectrograph (GMOS) installed on the Gemini North telescope. They were able to identify regions of low velocity dispersion within the few hundred pc of three galaxies and they interpreted this as kinematically cold recently formed stars. Although it is not possible to obtain global information of the stellar kinematics from the sample galaxies, the seeing-limited resolution allowed the researchers of both papers to obtain important information within the circumnuclear region of the galaxies.

The aim of this paper is to present new stellar kinematics data of a sample of 10 nearby (d $<$ 31 Mpc) and massive ($\sigma$ $>$ 200 km s$^{-1}$) ETGs observed with the GMOS-IFU at the Gemini South Telescope. This sample has a FOV of 3.5 x 5.0 arcsec$^2$ with seeing-limited spatial resolution of 0.6 - 1.0 arcsec. The excellent data quality of the sample, with a superb seeing-limited resolution, may reveal new information about the stellar kinematics in the circumnuclear region of these objects. In \citet[hereinafter paper I]{2014MNRAS.440.2419R}, this sample was analysed with PCA Tomography \citep{2009MNRAS.395...64S}, which is a technique to extract information from data cubes by means of principal component analysis. We showed that at least seven galaxies of the sample have stellar rotation. We confirm the result in this paper. In \citet{2014MNRAS.440.2442R} and \citet{2015MNRAS.451.3728R} (hereinafter paper II and paper III, respectively), we studied the ionized gas emission in the nuclear (paper II) and circumnuclear (paper III) regions. 

This paper is structured as follows. Section \ref{data_description} briefly describes the data of the galaxy sample; Section \ref{stellar_kinematical_maps} presents the results related to the stellar kinematics of the sample galaxies and information about the fitting procedure of the line-of-sight velocity distribution (LOSVD); Section \ref{ngc1404_case} discusses the case of the galaxy NGC 1404; Section \ref{conclusions_section} presents a summary and the conclusions of this paper. 

\section{Brief description of the data} \label{data_description}

We discussed the sample galaxies in more details in paper I. In short, the sample is composed of 10 nearby ($d$ $<$ 31 Mpc) and massive ($\sigma$ $>$ 200 km s$^{-1}$) ETGs. These objects were observed with GMOS-IFU \citep{2002PASP..114..892A,2004PASP..116..425H}, located at the Gemini-South Telescope, in one-slit mode (programmes GS-2008A-Q51 and GS-2008B-Q21). This setup resulted in a FOV of 3.5 x 5 arcsec$^2$ and a spectral range that covered from the H$\beta$ to the [S II] emission lines (see paper I for more information). All raw data were reduced with the standard Gemini {\scriptsize IRAF}\footnote{{\scriptsize IRAF} is distributed by the National Optical Astronomy Observatories, which are operated by the Association of Universities for Research in Astronomy, Inc., under cooperative agreement with the National Science Foundation.} package. Bias and flat-field corrections, in addition to wavelength and flux calibrations were properly applied to the data. The data cubes were built with spaxels of 0.2 arcsec. Special noise removal treatments developed for data cubes were performed: Butterworth filtering to remove high-frequency noise from the spatial dimension and PCA Tomography to remove low-frequency defects that are present in data cubes. We refer the reader to paper I and also to \citet{2014MNRAS.438.2597M,2015MNRAS.450..369M} for more details on these techniques. All sample galaxies were corrected for galactic reddening assuming $R$ = 3.1 and using the extinction curves of \citet{1989ApJ...345..245C} with the Av measurements of \citet{1998ApJ...500..525S}. The values of the full width at half-maximum (FWHM) of the point spread functions (PSFs) of the data cubes analysed in the present work are shown in paper I\footnote{The data cubes analysed in paper I were deconvolved using the Richardson-Lucy technique. For the present paper, the data cubes were not deconvolved, so one must assume that the FWHM's values of the PSFs of the data cubes are equal to the seeing of the observations, shown in table 2 of paper I.}.

\section{Stellar kinematics of the sample galaxies} \label{stellar_kinematical_maps}

\subsection{Extracting the LOSVD from the sample galaxies} \label{ppxf_meth}

Absorption line profiles may be described by a Gauss Hermite function \citep{1993ApJ...407..525V,1993MNRAS.265..213G}. The first three moments of this function describe the radial velocity (V$_r$), the velocity dispersion ($\sigma$) and the moment $h_3$, which is related to asymmetric deviations from a Gaussian (skewness). In all spectra from the data cubes of the sample galaxies, we used the penalized pixel fitting procedure ({\sc ppxf} - \citealt{2004PASP..116..138C}) to fit the profiles of the absorption lines and extract the LOSVD from each spaxel. This software fits a template spectrum convolved with a Gauss Hermite function to an observed spectrum in the pixel space. We fitted the LOSVD in the spectral range between 4800 and 6800\AA\ and we used the stellar population base of \citet{2009MNRAS.398L..44W}. This base contains 120 population model spectra with a resolution of 2.51\AA\ \citep{2011A&A...531A.109B,2011A&A...532A..95F}, ages between 3 and 12 Gyr and abundances [Fe/H] = $-$0.5, $-$0.25, 0.0 and 0.2 and [$\alpha$/Fe] = 0.0, 0.2 and 0.4. For the fitting procedure, we used only the populations that account for at least for 0.1\% of the total light flux of the circumnuclear stellar component in each galaxy. The previous determination of the stellar populations in the circumnuclear region of the sample galaxies had been previously done in papers II and III, using spectral synthesis with the {\sc starlight} code \citep{2005MNRAS.358..363C}. We also used Legendre polynomials of order 4 in {\sc ppxf} to correct low-frequency mismatches between the models and the observed spectra. We masked all emission lines, i.e., these regions were not considered for the fit. Some fit examples for the sample galaxies are shown in Appendix \ref{starlight_results}. For each spectrum of the data cubes of the sample galaxies, we obtained values for radial velocity, velocity dispersion and the Gauss-Hermite moment $h_3$. The real velocity dispersion of each spectrum was calculated as $\sigma^2$ = $\sigma_m^2$ - $\sigma_i^2$ + $\sigma_b^2$, where $\sigma_m$ is the dispersion measured with {\sc ppxf}, $\sigma_i$ is the instrumental resolution, which are $\sigma_i$ = 0.55\AA\ (30 km s$^{-1}$  in $\lambda$ = 5577\AA) for the data cubes from the programme GS-2008A-Q51 and $\sigma_i$ = 0.76\AA\ (41 km s$^{-1}$  in $\lambda$ = 5577\AA) for the data cubes from the programme GS-2008B-Q21 (note that the instrumental resolution is constant in wavelength, but it varies in velocity along the spectra of the data cubes), and $\sigma_b$ is the template resolution. 

We calculated the statistical errors associated with the kinematic parameters using a Monte Carlo (MC) simulation. First, we added Gaussian noise to the optimal template, obtained with {\sc ppxf} for a given spectrum in a data cube. We used the error spectrum of each spaxel of each data cube to calculate the Gaussian noise. Then, we applied {\sc ppxf} to this optimal template with noise added. The procedure was repeated 25 times in each spaxel. The errors related to the values of the radial velocity, velocity dispersion and $h_3$ are just the standard deviation of the values obtained with the MC simulation. By doing this, we obtained error maps for each kinematic parameter.

We built maps of the radial velocity, velocity dispersion and the Gauss Hermite moment $h_3$ of the stellar components from the circumnuclear region of the sample galaxies. They are shown in Fig. \ref{cinematica_estelar_1}. These maps are described in the following subsections.

\begin{figure*}
\hspace{-1.2cm}
\vspace{1.0cm}
\includegraphics[scale=0.7]{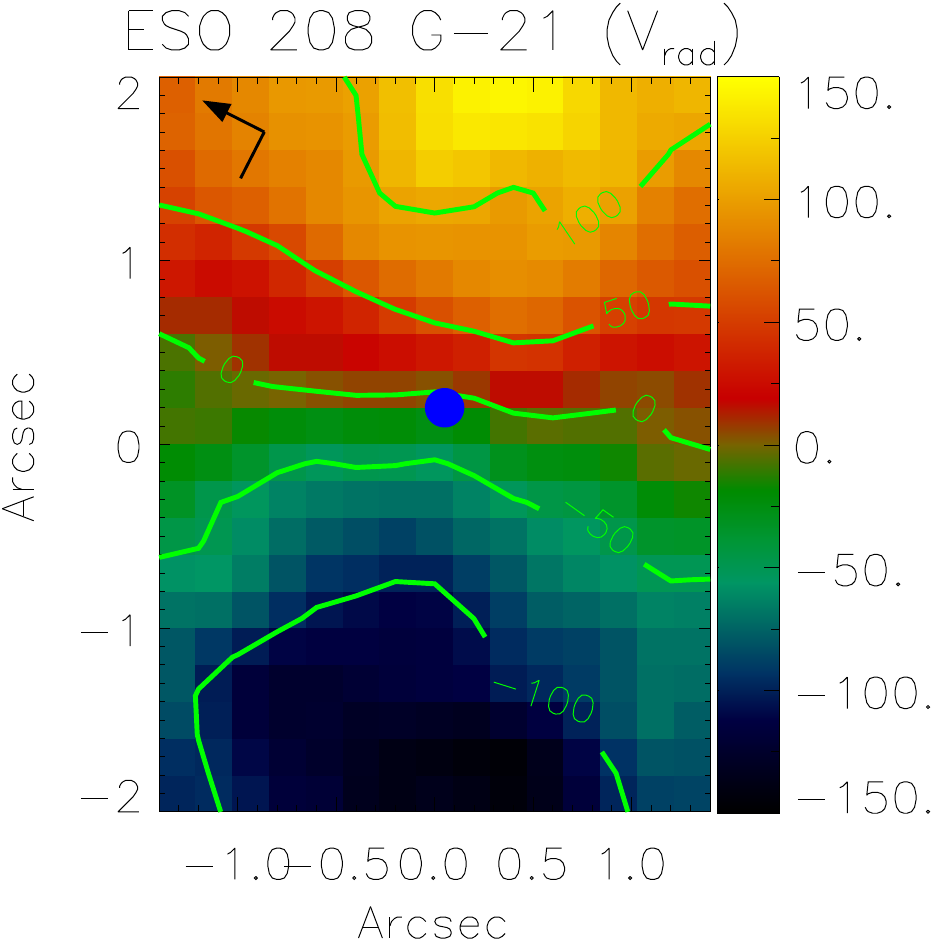}
\hspace{0.0cm}
\vspace{-0.5cm}
\includegraphics[scale=0.7]{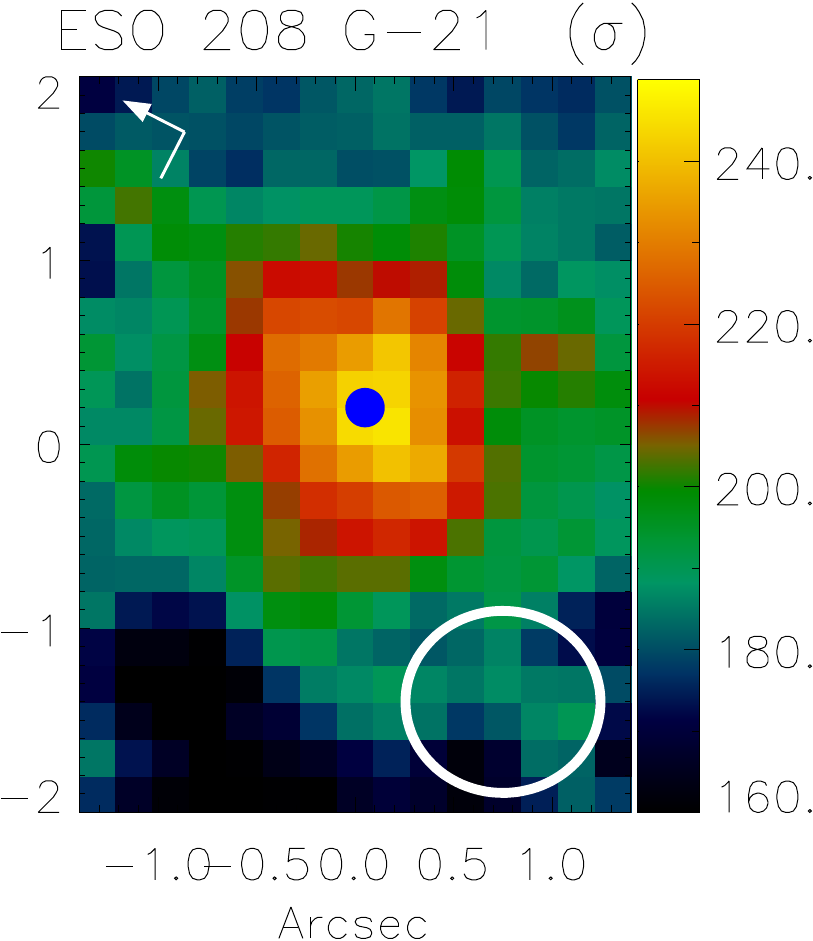}
\hspace{0.0cm}
\includegraphics[scale=0.7]{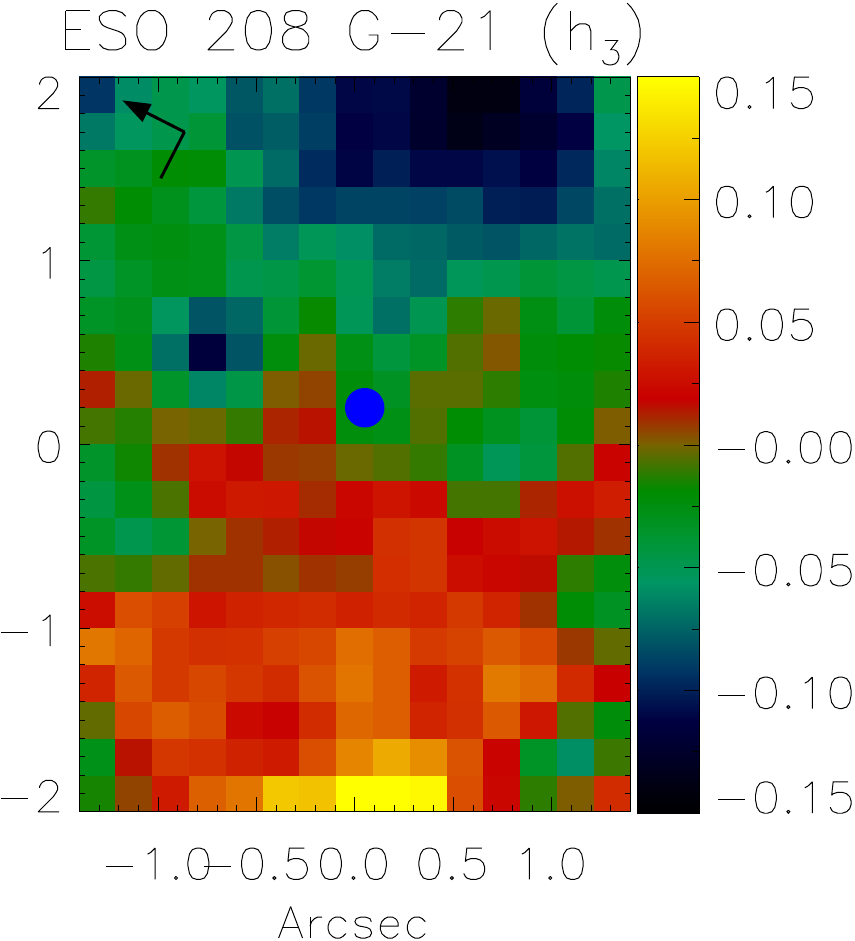} \\
\hspace{-1.4cm}
\includegraphics[scale=0.7]{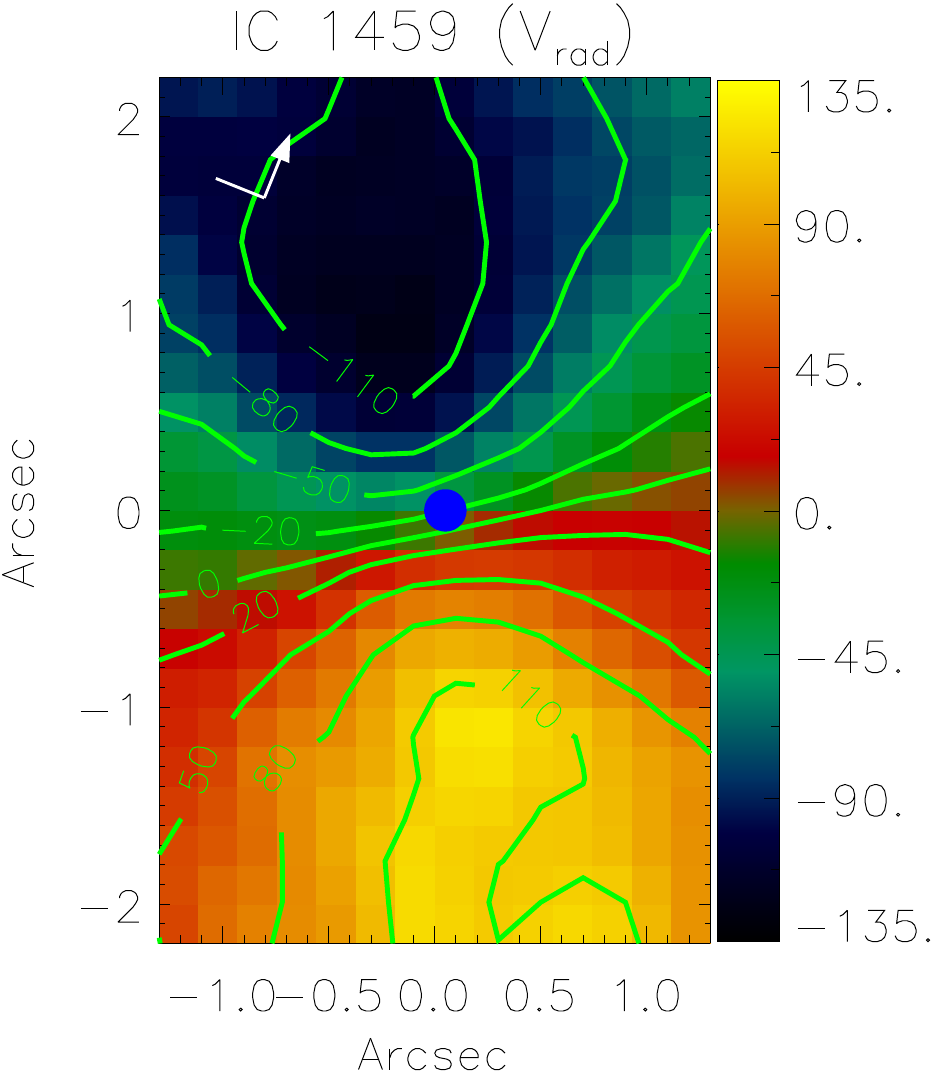}
\hspace{0.0cm}
\vspace{0.2cm}
\includegraphics[scale=0.7]{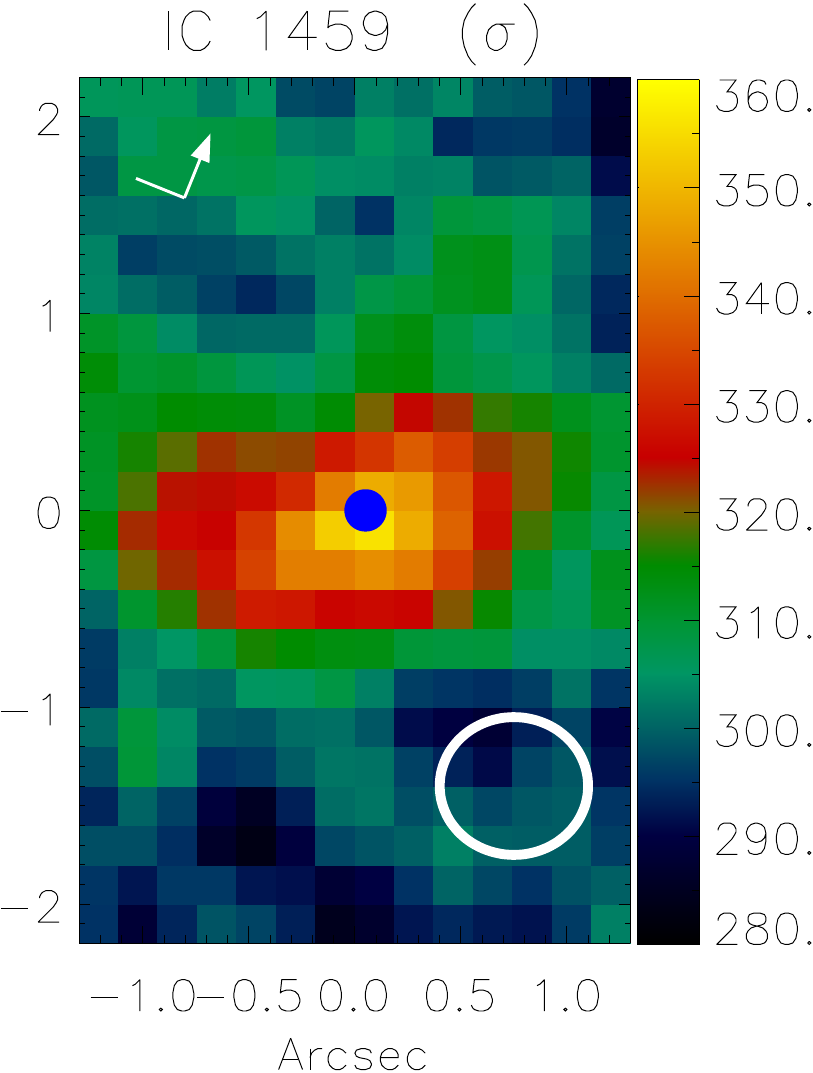}
\hspace{0.0cm}
\includegraphics[scale=0.7]{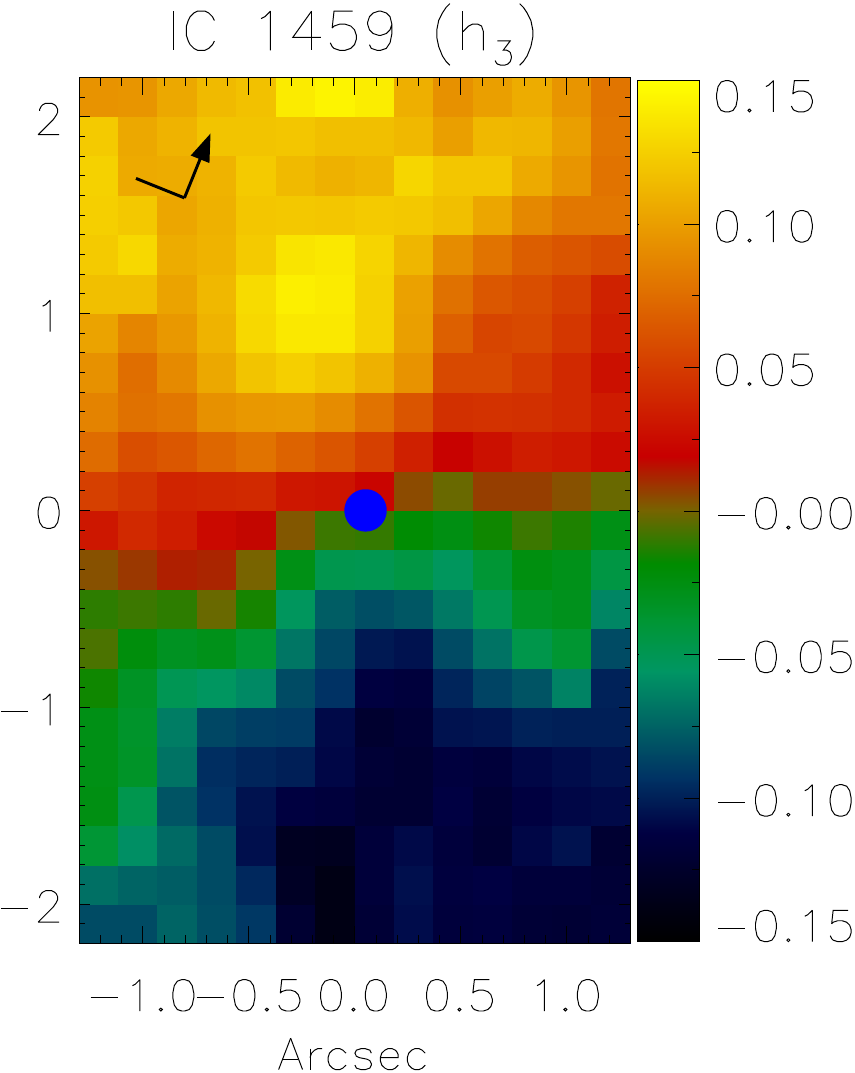} \\
\hspace{-1.4cm}
\includegraphics[scale=0.7]{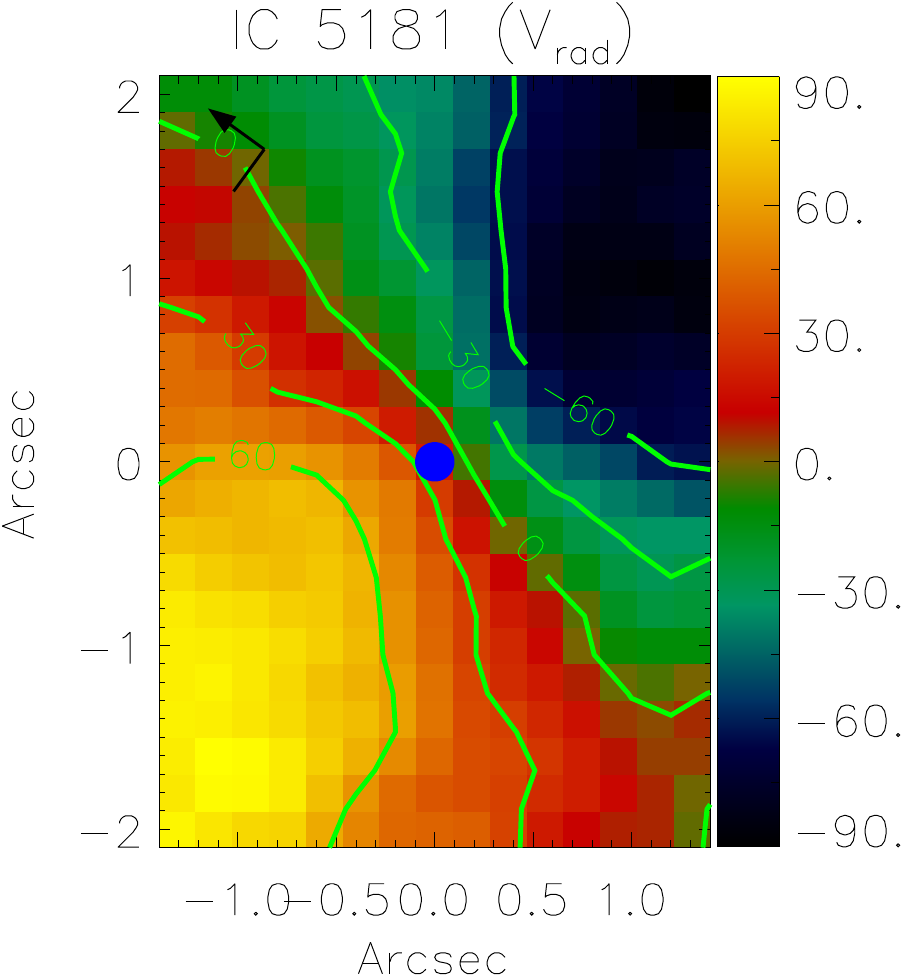}
\hspace{0.0cm}
\includegraphics[scale=0.7]{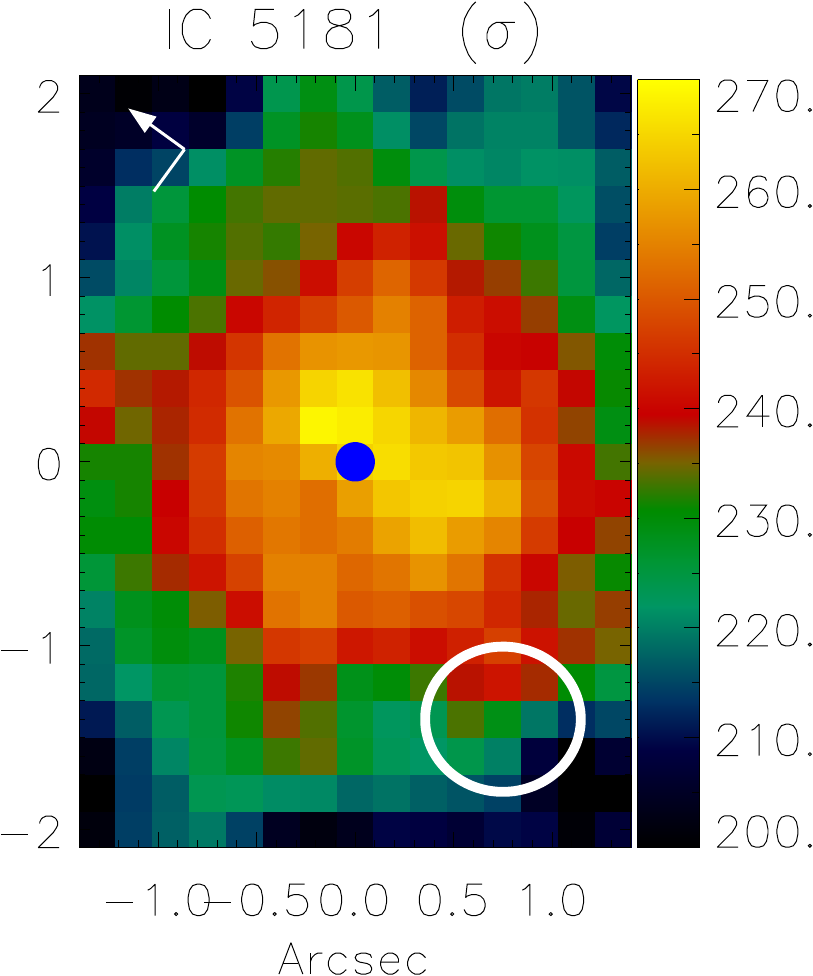}
\hspace{0.0cm}
\includegraphics[scale=0.7]{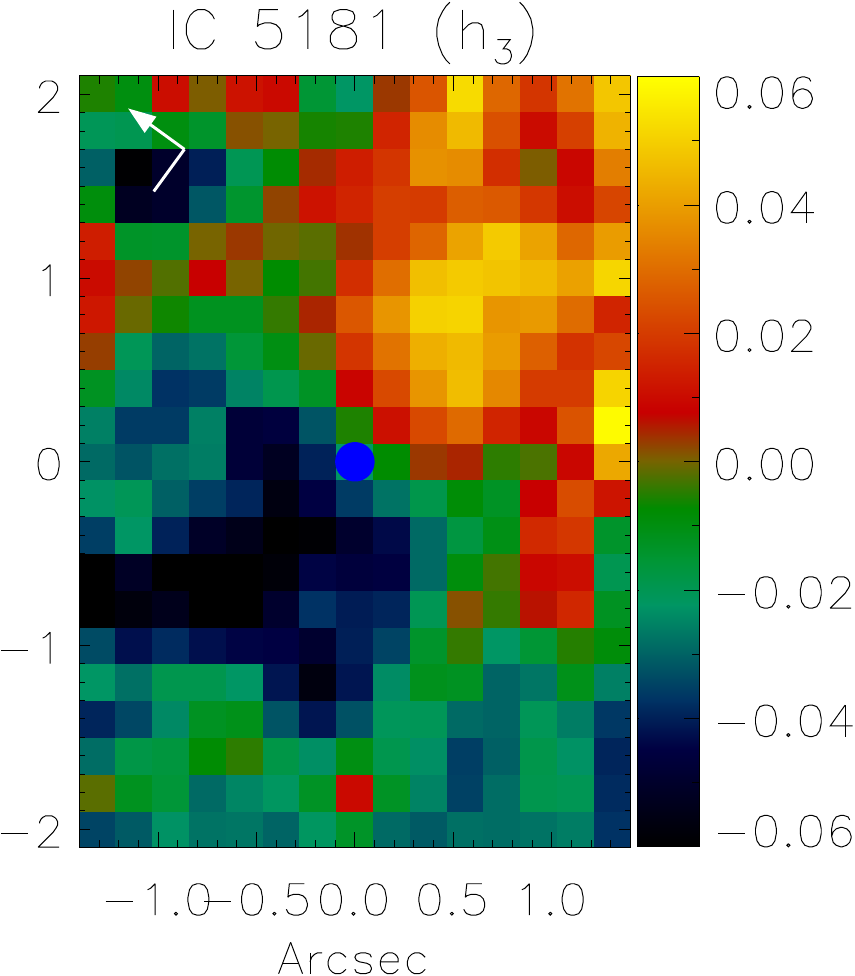}

\caption{Stellar kinematic maps of the circumnuclear region of the sample galaxies. From left to right: radial velocity, velocity dispersion, and Gauss Hermite moment $h_3$. The blue circles set the position of the AGNs of the galaxies, while the green circles represent the photometric centre of the stellar structure of NGC 1399 and NGC 1404. The white circles represent the FWHM of the seeing of the observations, whose values are shown in paper I. \label{cinematica_estelar_1} 
 } 
\end{figure*}

\addtocounter{figure}{-1}

\begin{figure*}
\hspace{-1.2cm}
\vspace{1.0cm}
\includegraphics[scale=0.7]{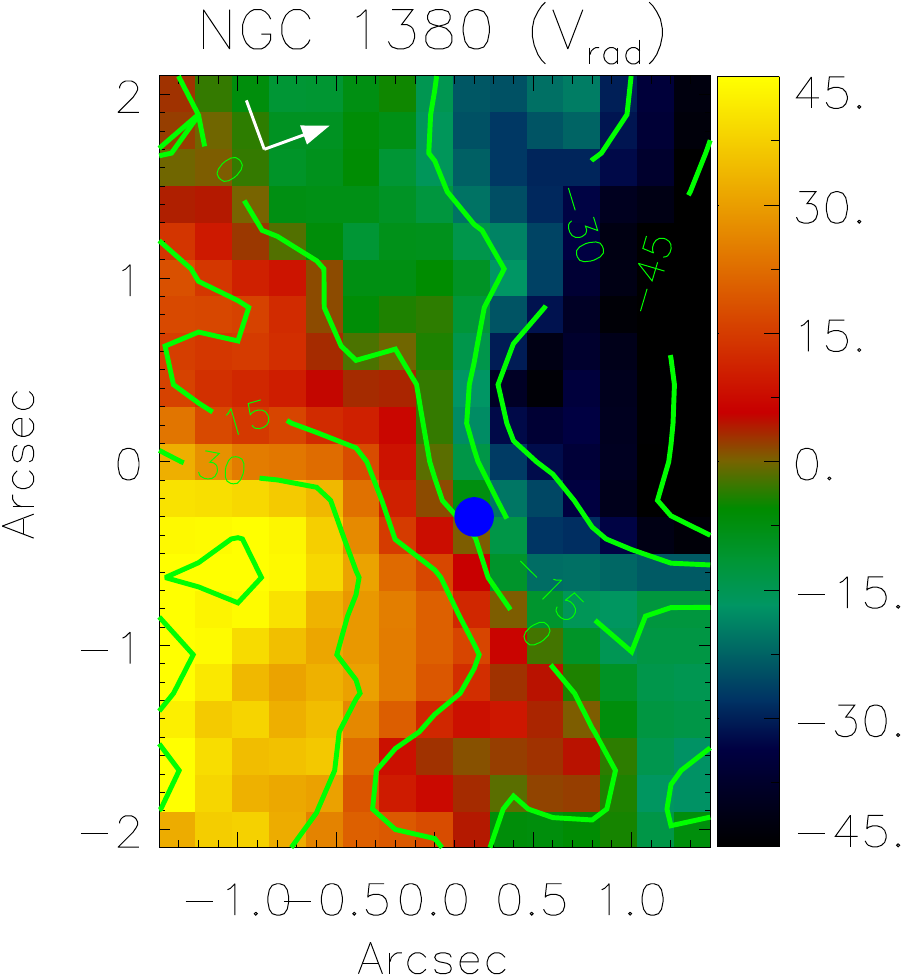}
\hspace{0.0cm}
\vspace{-0.3cm}
\includegraphics[scale=0.7]{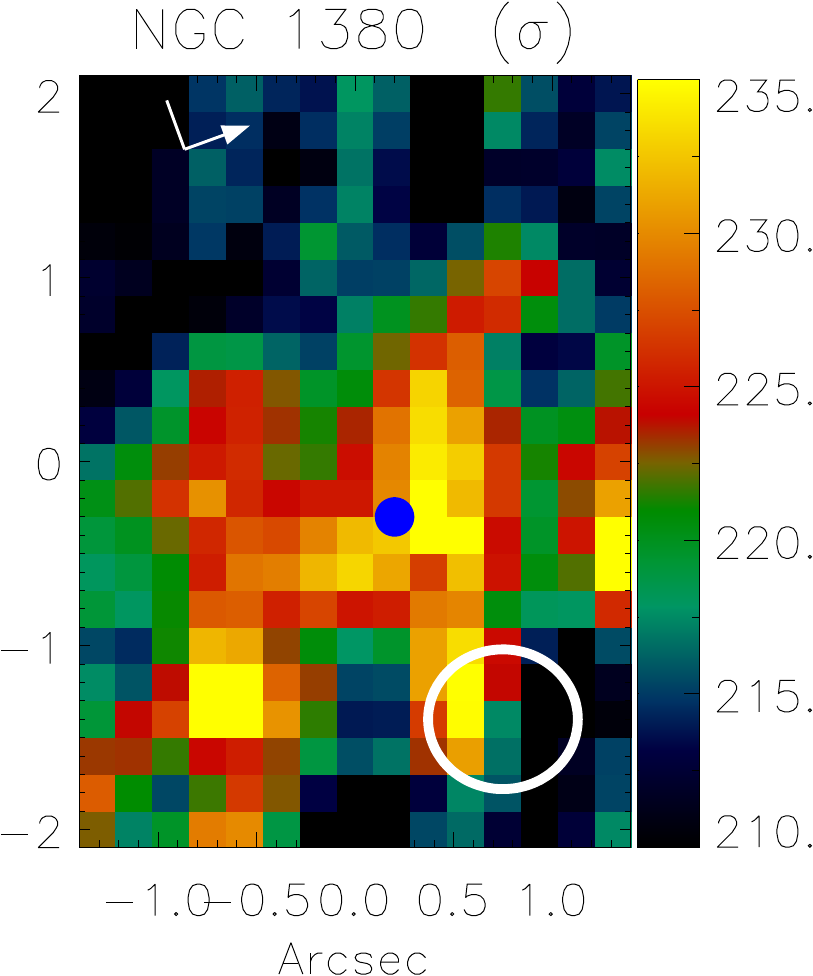}
\hspace{0.0cm}
\includegraphics[scale=0.7]{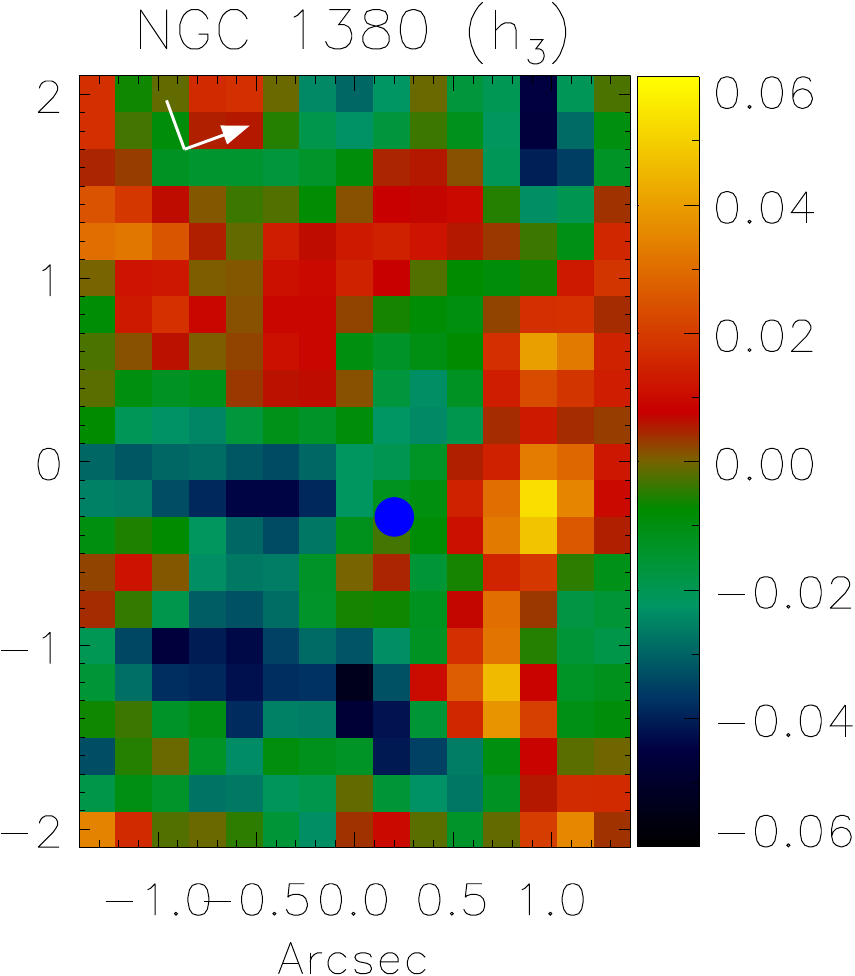} \\
\hspace{-1.4cm}
\includegraphics[scale=0.7]{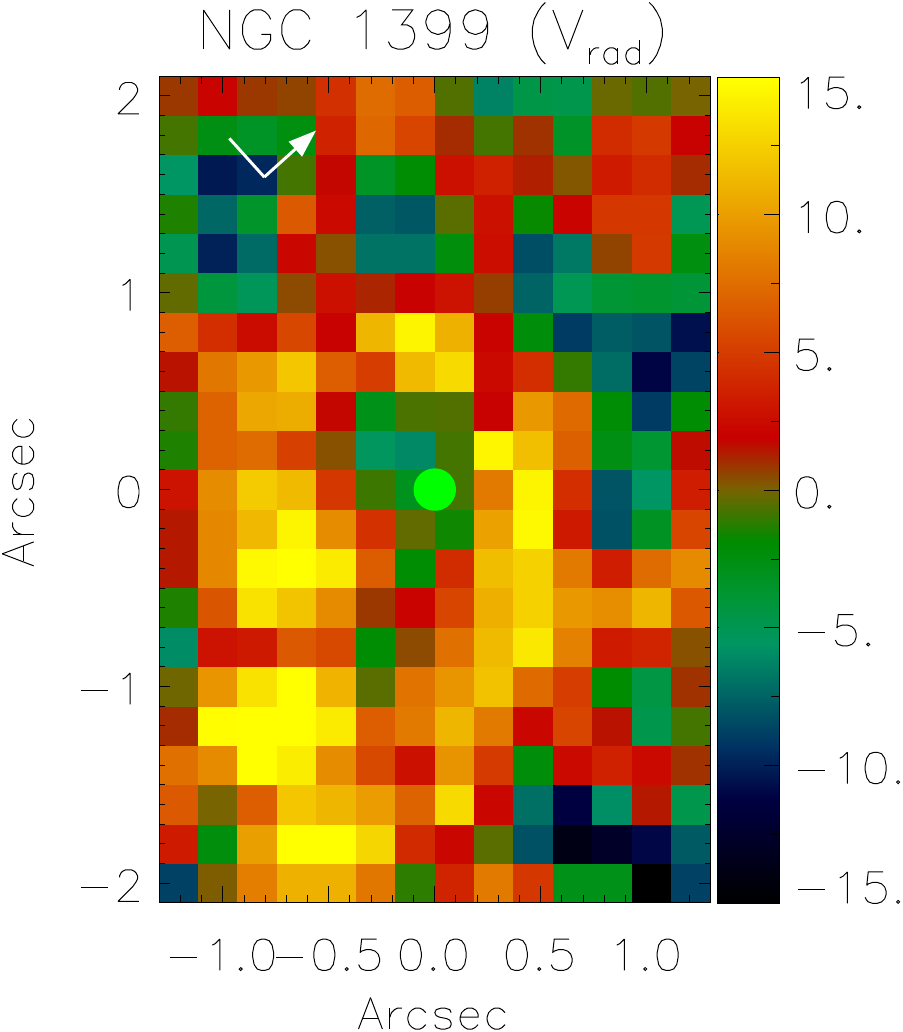}
\hspace{0.0cm}
\vspace{0.2cm}
\includegraphics[scale=0.7]{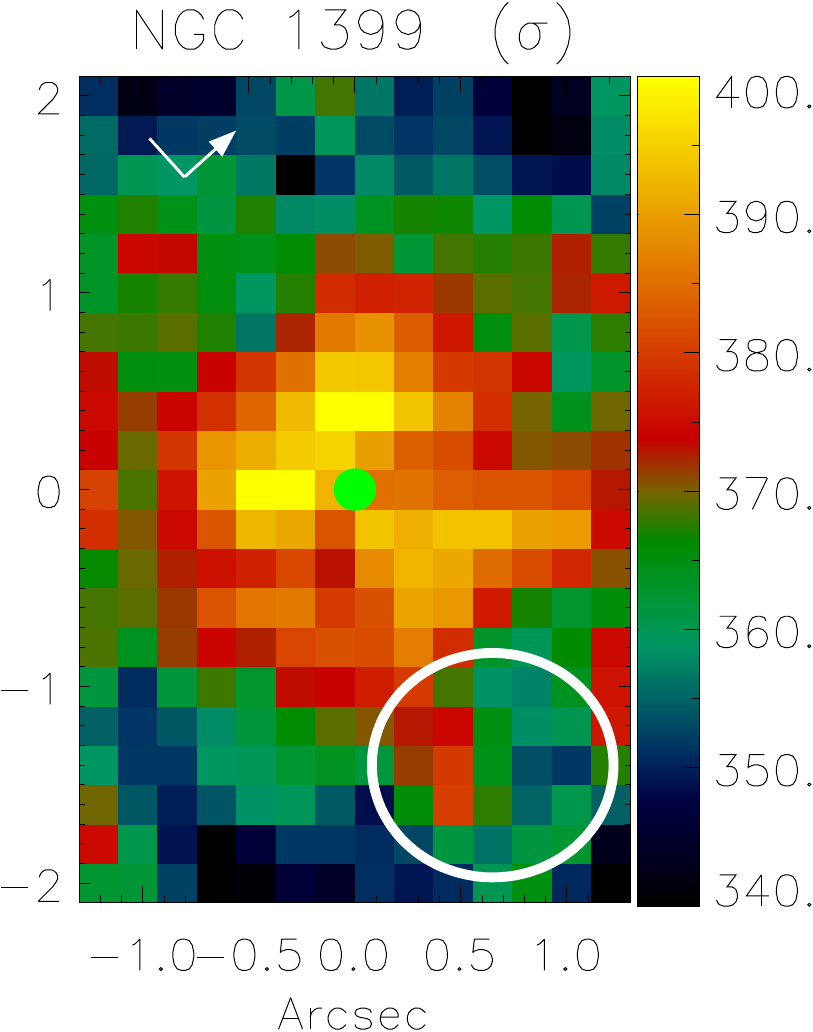}
\hspace{0.0cm}
\includegraphics[scale=0.7]{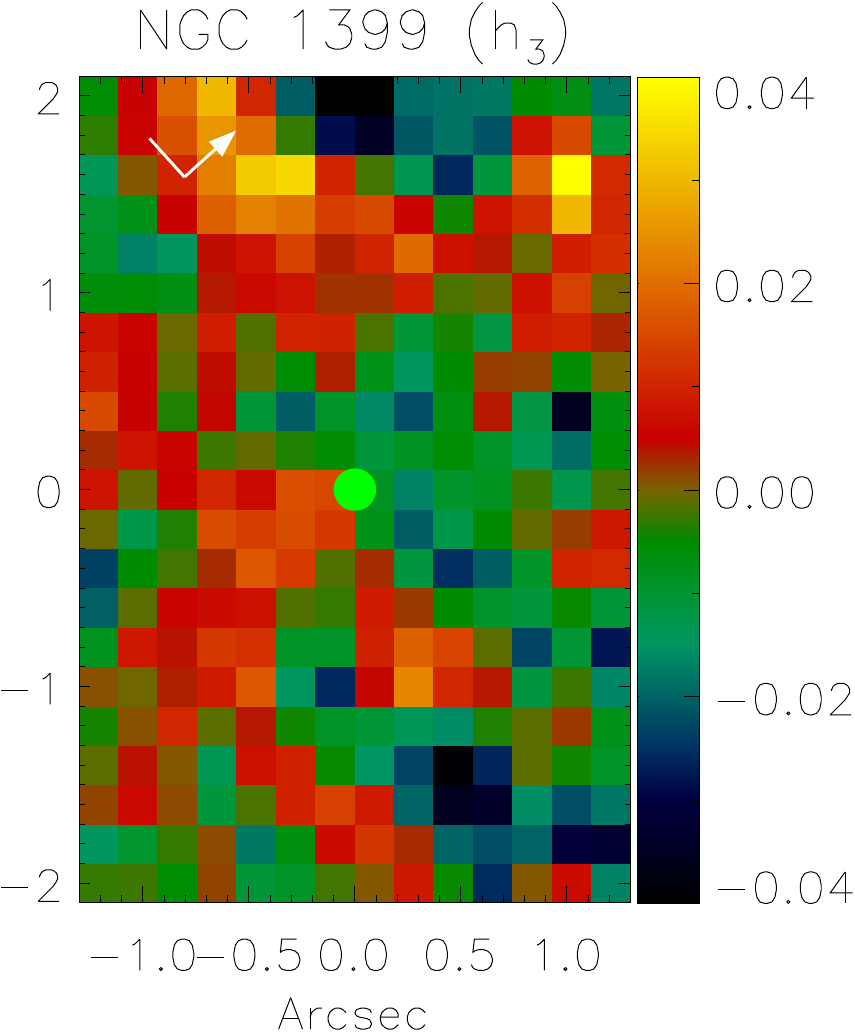} \\
\hspace{-1.4cm}
\includegraphics[scale=0.7]{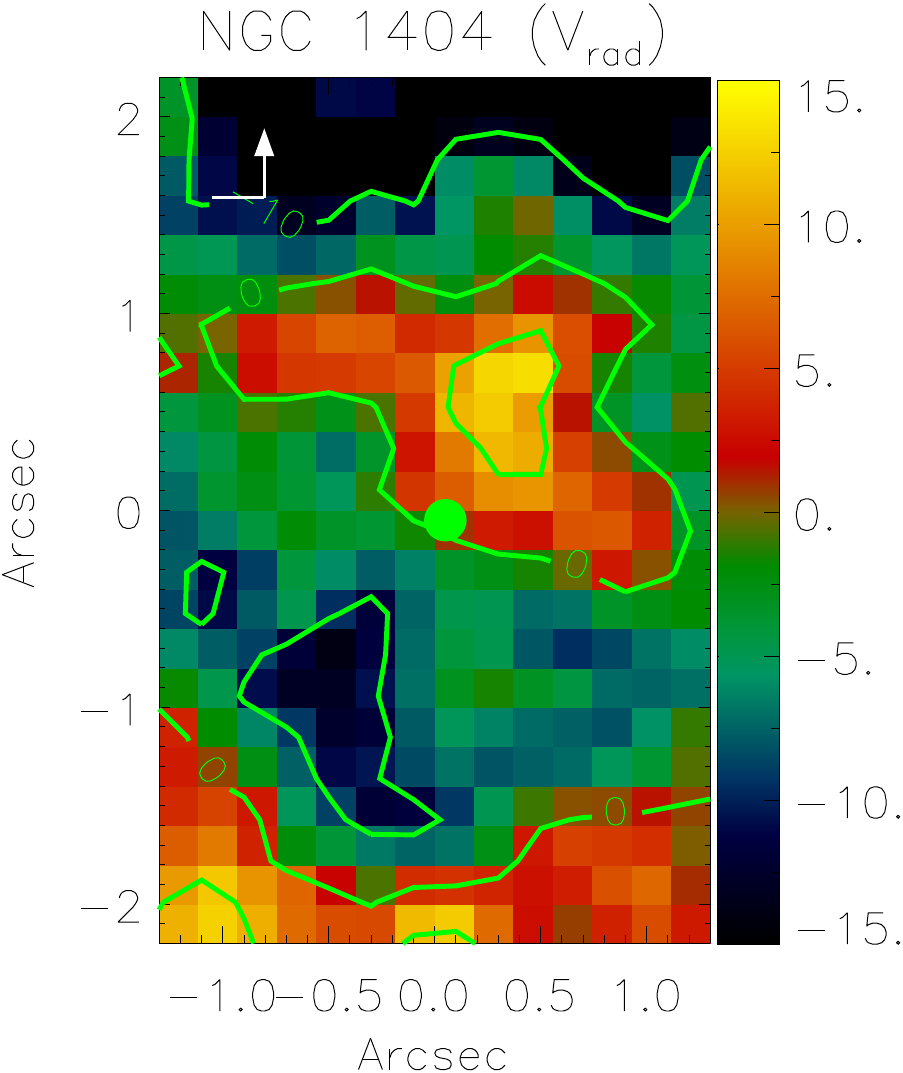}
\hspace{0.0cm}
\includegraphics[scale=0.7]{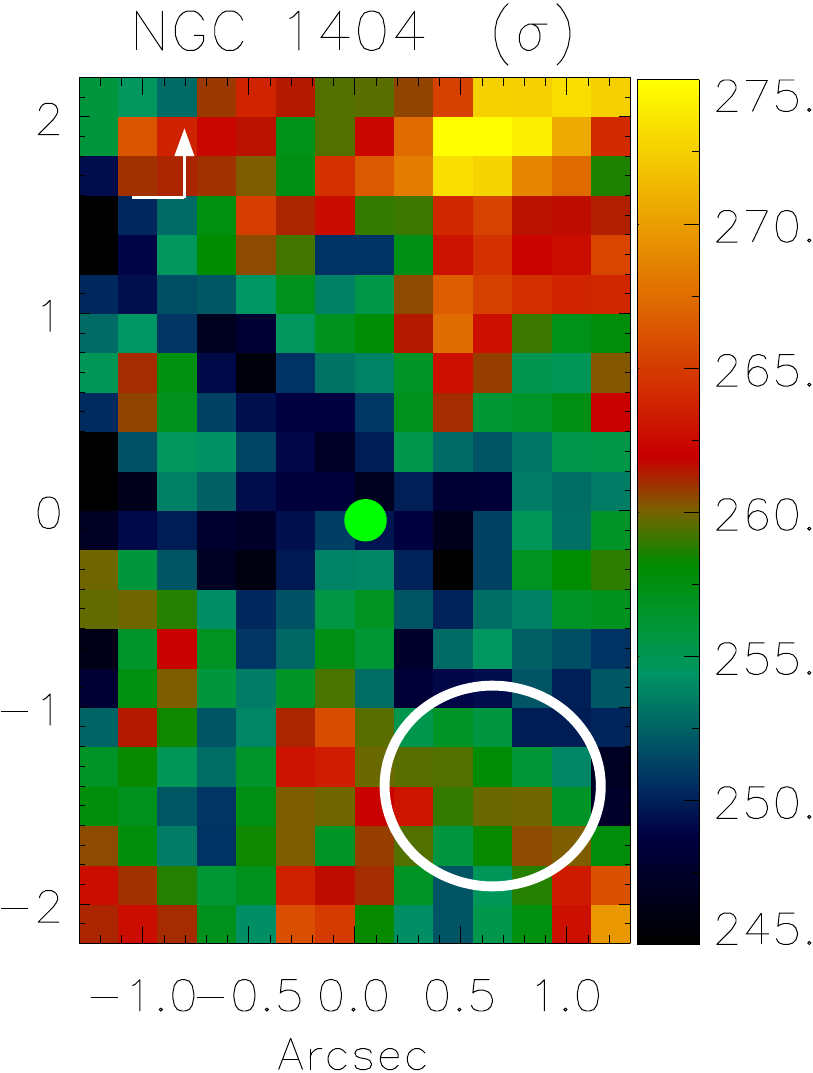}
\hspace{0.0cm}
\includegraphics[scale=0.7]{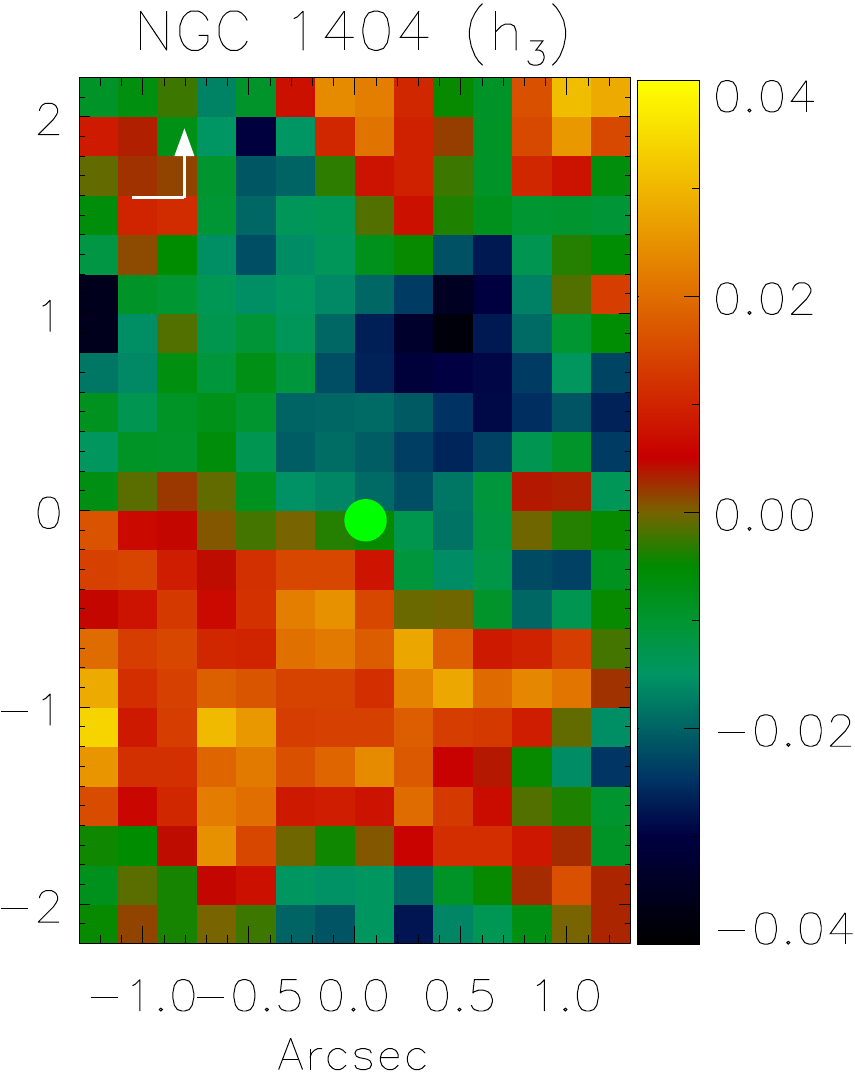}

\caption{continued \label{cinematica_estelar_2}
} 

\end{figure*}

\addtocounter{figure}{-1}

\begin{figure*}
\hspace{-1.2cm}
\vspace{1.0cm}
\includegraphics[scale=0.7]{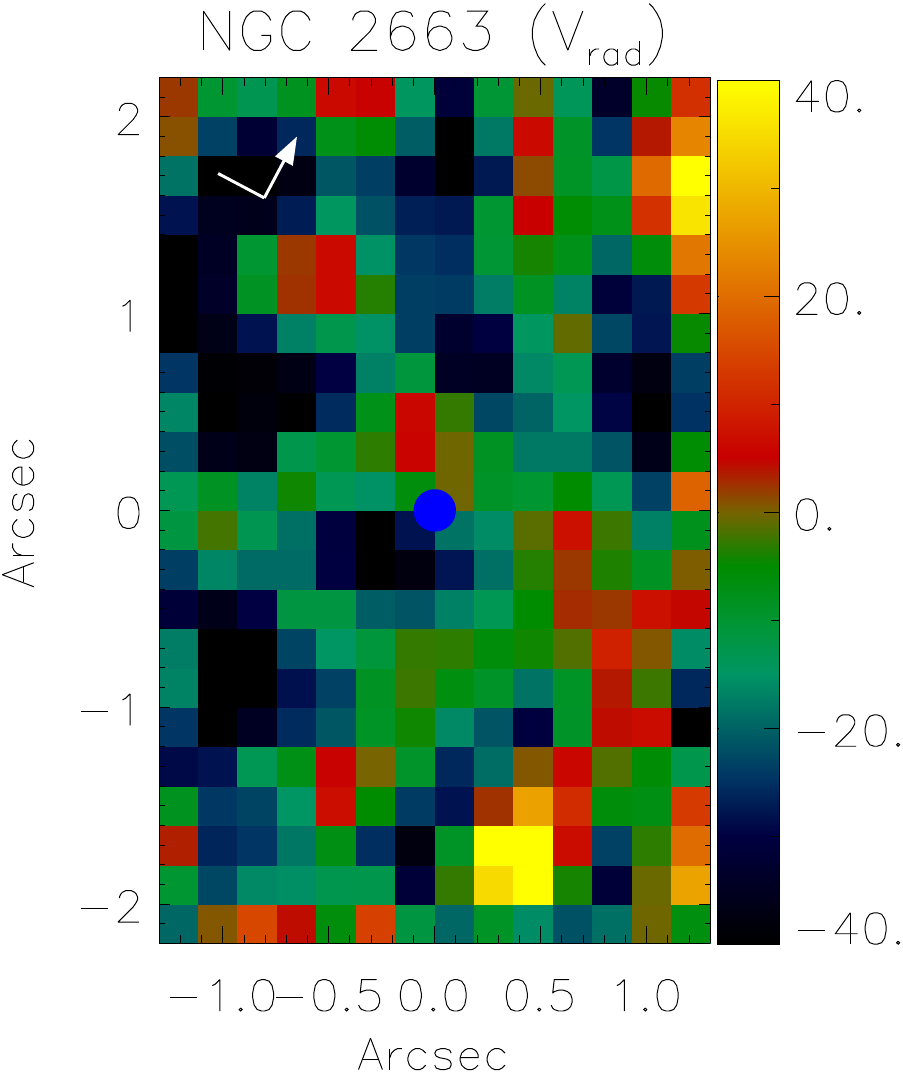}
\hspace{0.0cm}
\vspace{-0.5cm}
\includegraphics[scale=0.7]{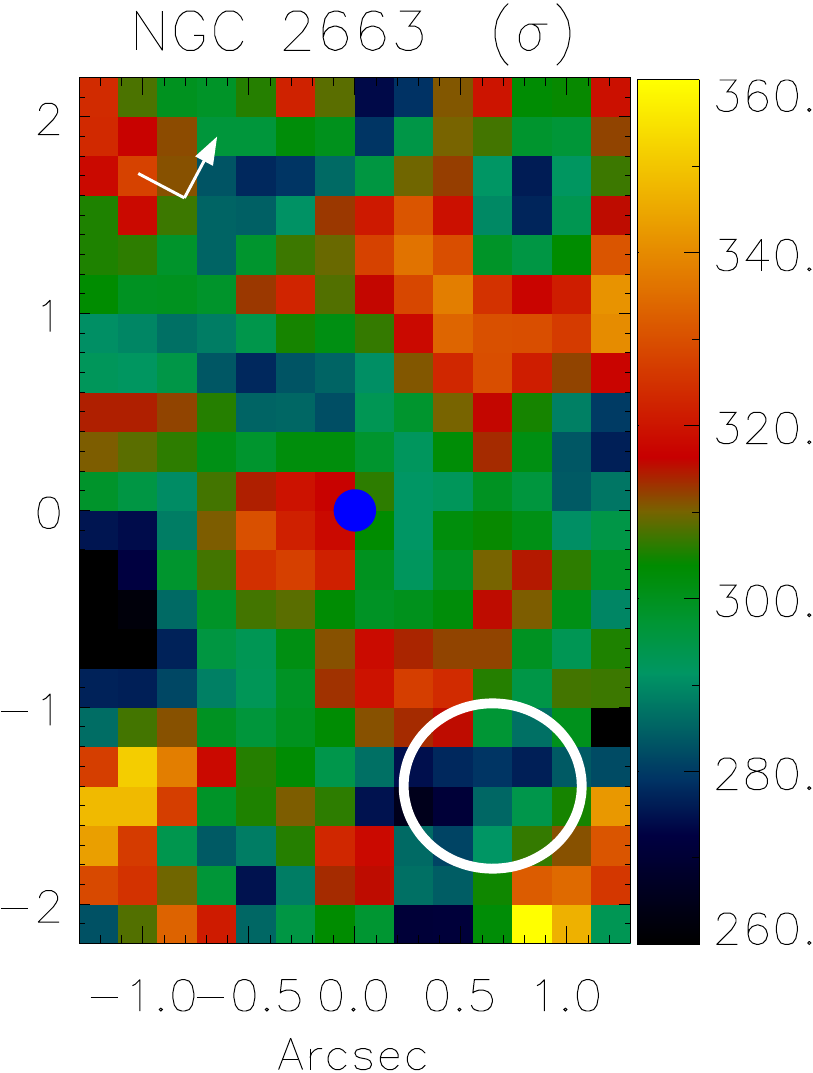}
\hspace{0.0cm}
\includegraphics[scale=0.7]{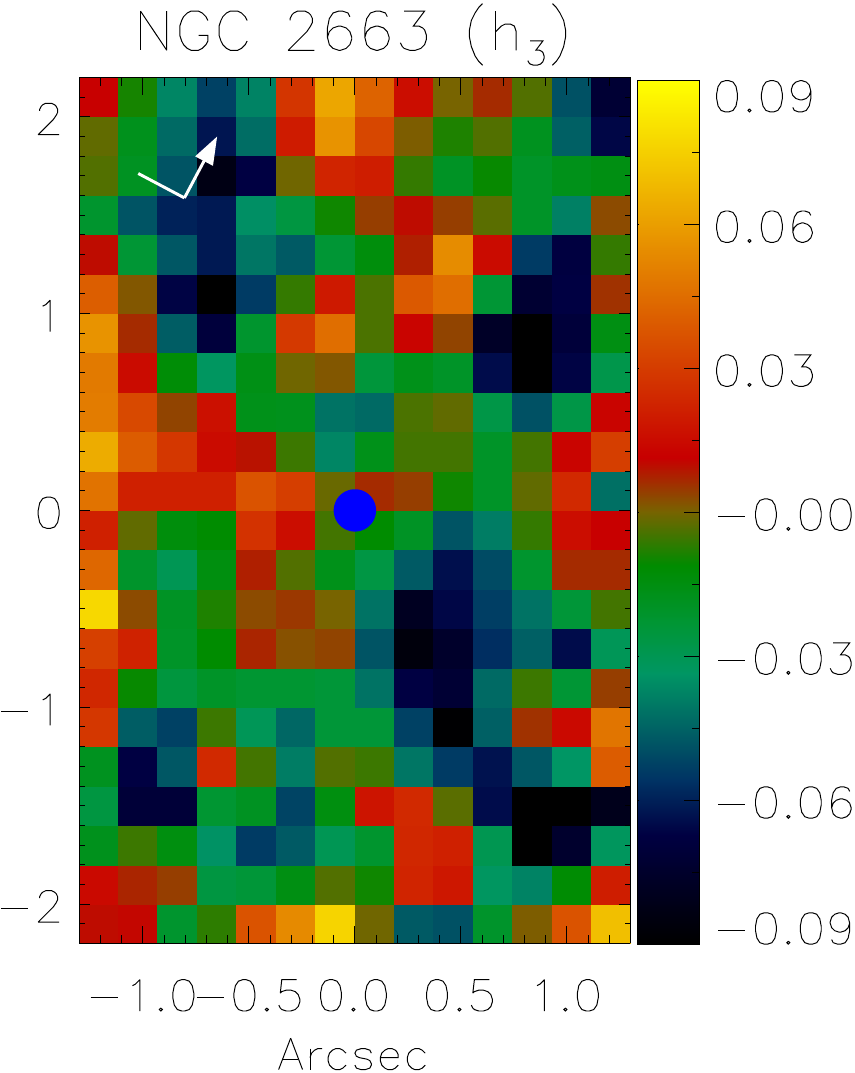} \\
\hspace{-1.4cm}
\includegraphics[scale=0.7]{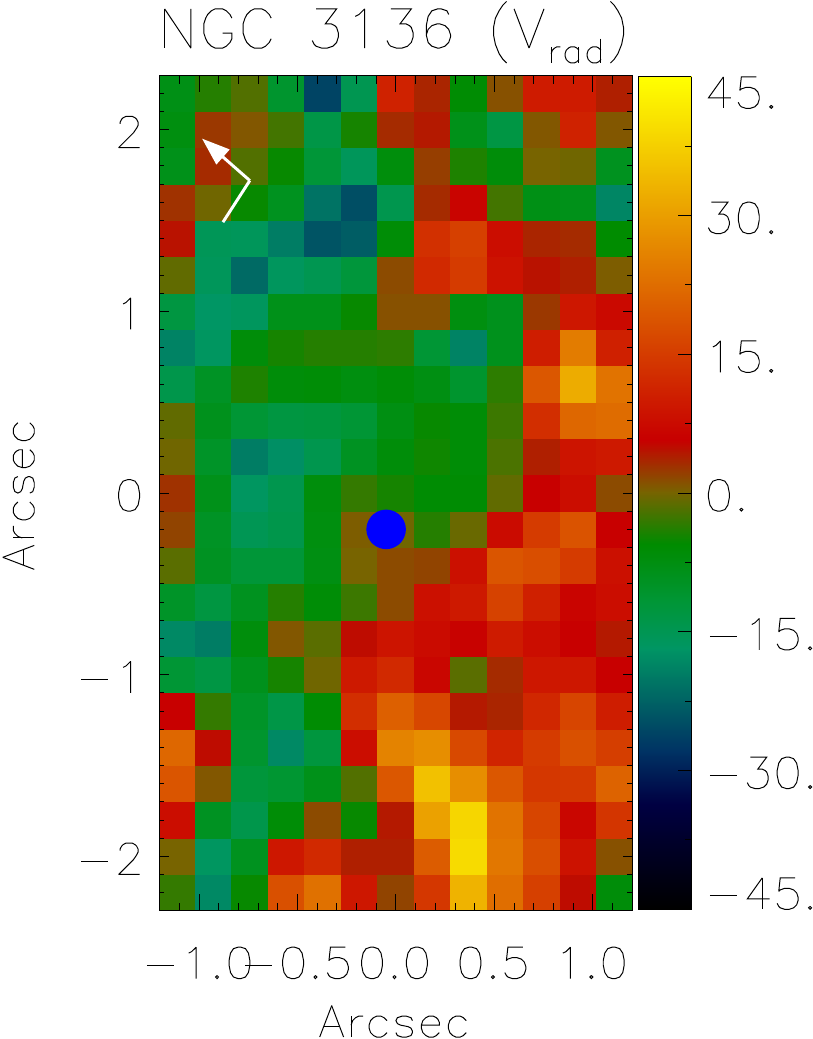}
\hspace{0.0cm}
\vspace{0.5cm}
\includegraphics[scale=0.7]{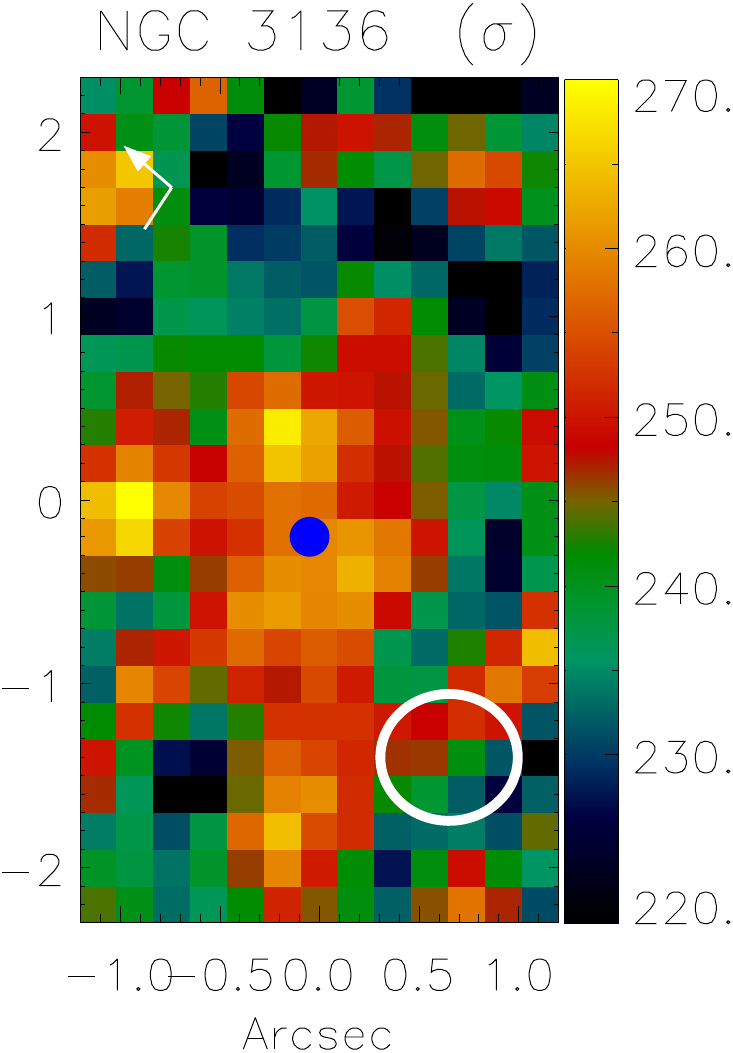}
\hspace{0.0cm}
\includegraphics[scale=0.7]{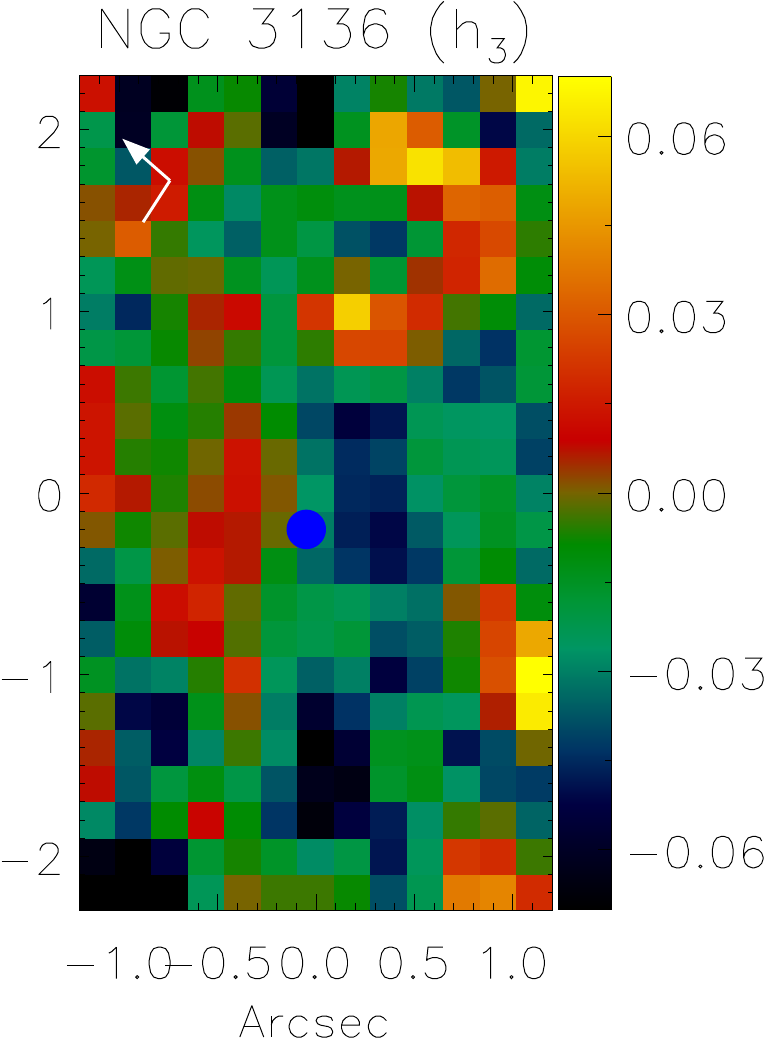} \\
\hspace{-1.4cm} 
\includegraphics[scale=0.7]{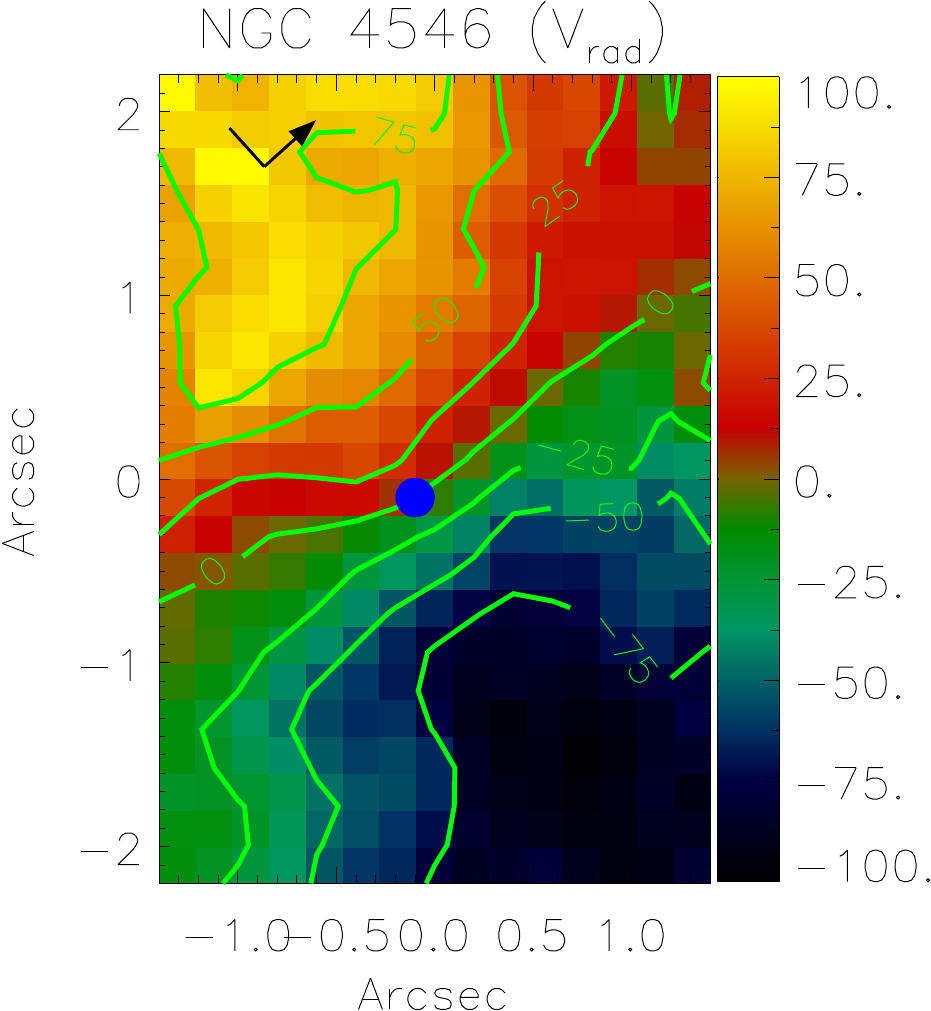}
\hspace{0.0cm}
\vspace{0.0cm}
\includegraphics[scale=0.7]{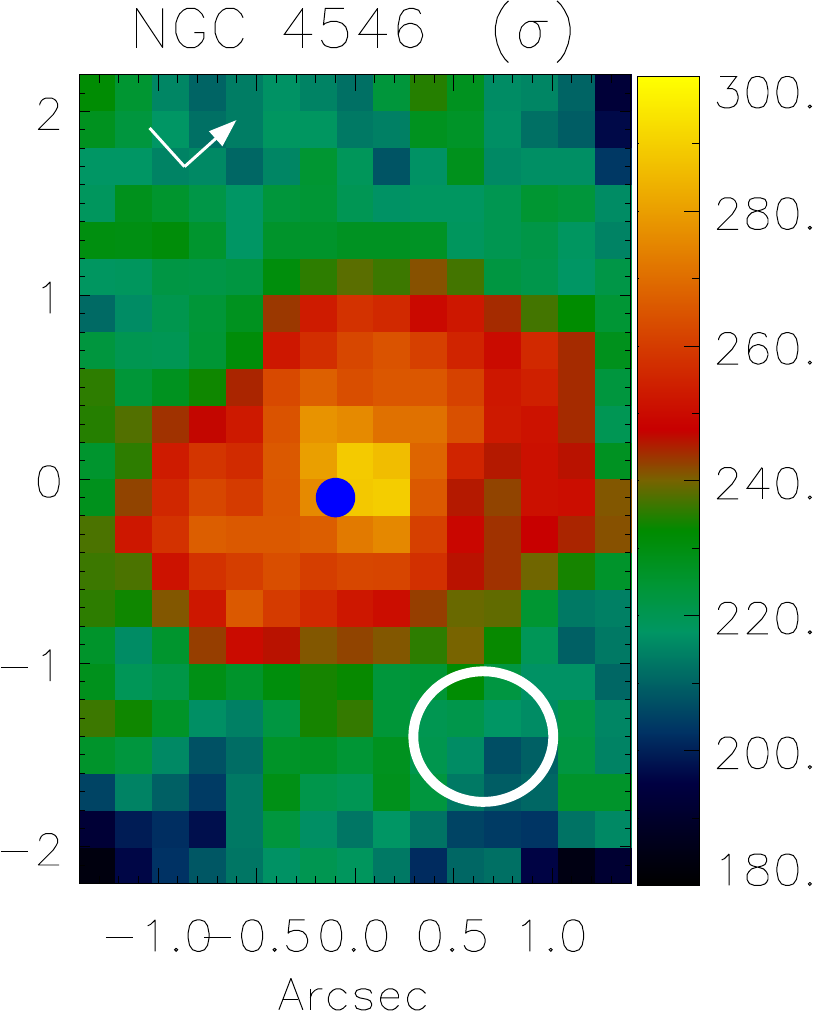}
\hspace{0.0cm}
\includegraphics[scale=0.7]{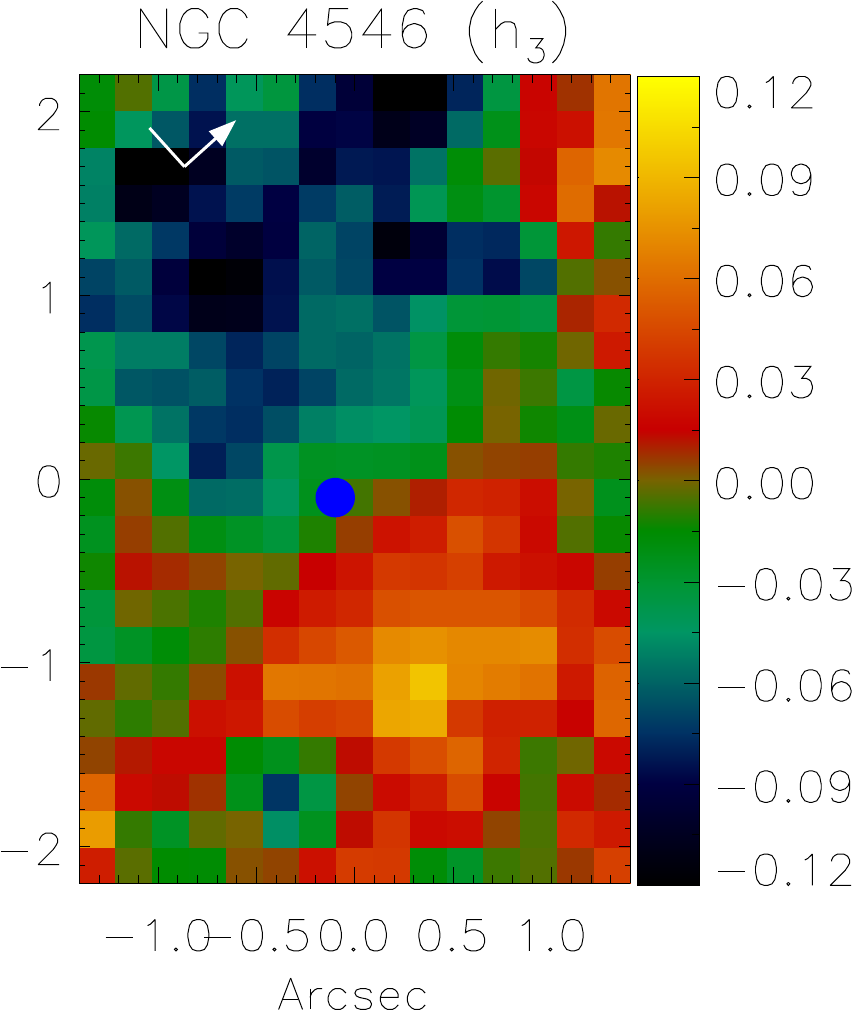}

\caption{continued \label{cinematica_estelar_3}
} 

\end{figure*}

\addtocounter{figure}{-1}

\begin{figure*}
\hspace{-1.2cm}
\includegraphics[scale=0.7]{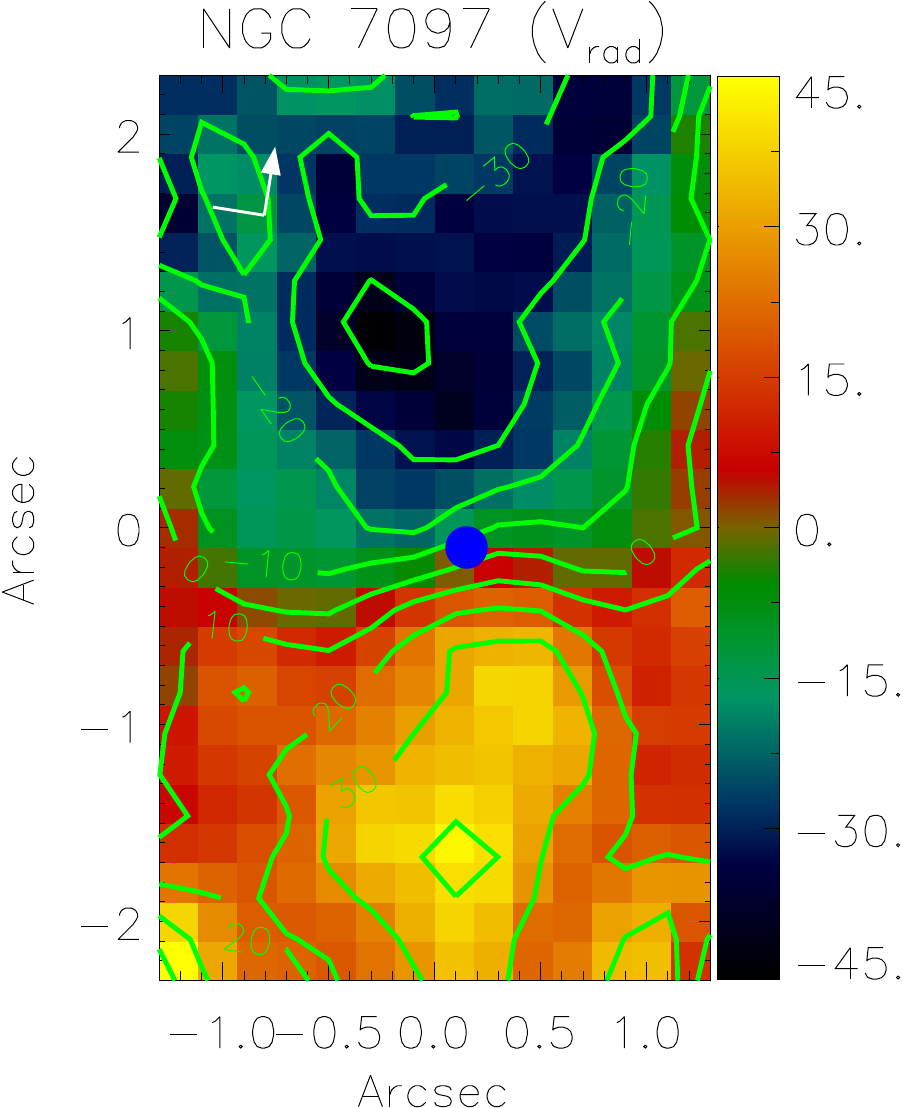}
\hspace{0.0cm}
\includegraphics[scale=0.7]{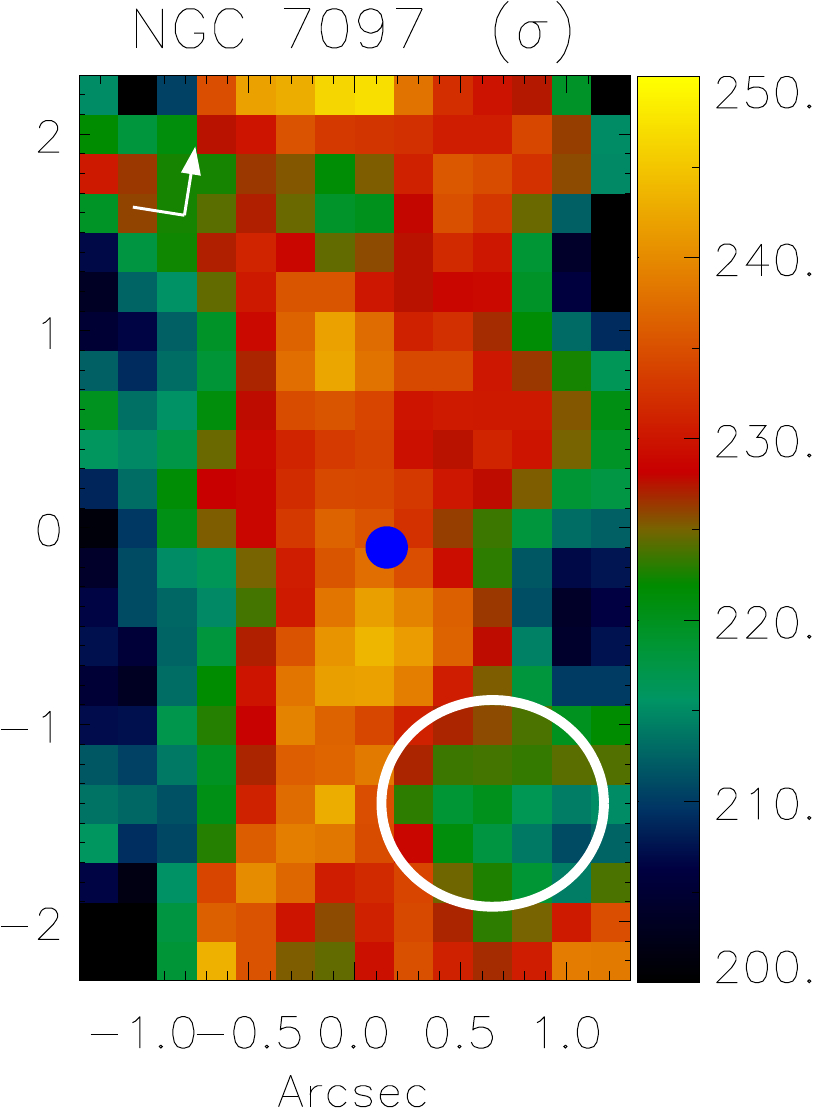}
\hspace{0.0cm}
\includegraphics[scale=0.7]{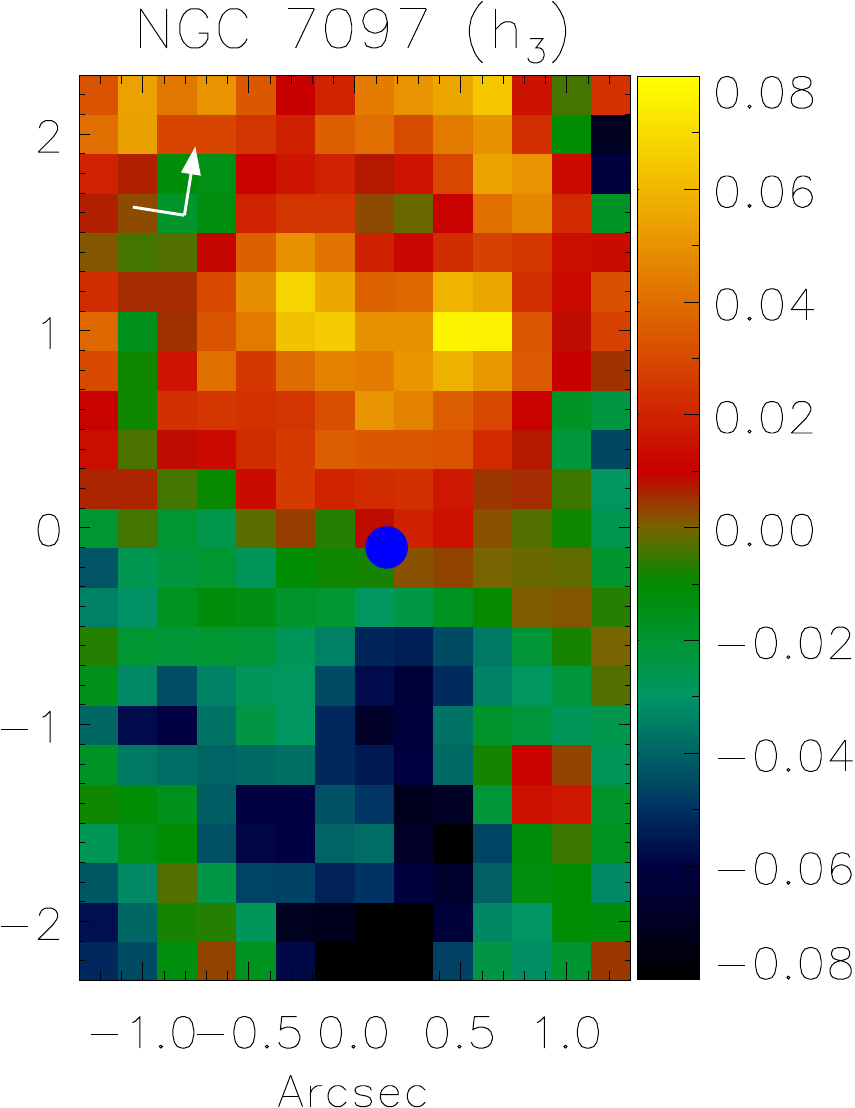}
\caption{continued \label{cinematica_estelar_4}
} 

\end{figure*}

\subsection{Radial velocity maps} \label{stellar_radial_velocities}

Radial velocity maps show that seven galaxies of the sample have a typical structure that is related to the rotation of the stellar component. Exceptions are NGC 1399, NGC 2663, and NGC 3136. We measured the position angles (PA) of these structures as the angle of the line that connects the positions of the maximum and minimum radial velocities of the stellar rotation. The kinematic centre x(V$_r$) e y(V$_r$) of the stellar rotation was determined along this line at the position where the radial velocity is equal to the average between its maximum and the minimum values. The errors associated with the PA, $x$(V$_r$), $y$(V$_r$) and also with the heliocentric radial velocity at the kinematic centre were estimated with a MC simulation, i.e., by measuring 100 times the maximum and minimum values of the radial velocity after the addition of noise to the radial velocity maps. This noise was calculated for each spaxel using the maps of the errors associated with the radial velocity measurements of the galaxies. The PA, $x$(V$_r$), $y$(V$_r$) and the heliocentric radial velocity of the kinematic centre of the galaxies are shown in Table \ref{tab_mapa_vel_radial_stellar}. In the radial velocity maps (Fig. \ref{cinematica_estelar_1}), the values of the radial velocities in each spaxel are shown with respect to the heliocentric radial velocity at the kinematic centre of the objects. For NGC 1399, NGC 2663 and NGC 3136, where we did not detect any stellar structure in rotation, the heliocentric radial velocities shown in Table \ref{tab_mapa_vel_radial_stellar} were extracted from the position of the nuclei of these galaxies. In NGC 2663, we set the position of the nucleus by the image of the red wing of the broad component of H$\alpha$ (see paper II). This component is emitted by the broad line region (BLR), which is directly associated with AGNs in galaxies. In NGC 3136, we used the image of the [O I]$\lambda$6300 line to find the nucleus. In NGC 1399, we assumed that the nucleus of the galaxy is located at the photometric centre of the stellar structure. The positions of the nuclei of these three objects are shown in Table \ref{tab_mapa_vel_radial_stellar}. The errors related to the heliocentric radial velocity of these three objects are simply the errors of the measurements of radial velocity at the spaxels that correspond to the position of their nuclei.

\begin{table*}
 \scriptsize
 \caption{Columns (1) and (2): coordinates of the kinematic centre of the sample galaxies, defined as the position where the radial velocity is equal to the average between its maximum and the minimum values along the PA of the stellar rotations.  Columns (3) and (4): coordinates of the peak of the velocity dispersion structures detected in the sample galaxies. Columns (5) and (6): coordinates of the AGNs for the sample galaxies, except for NGC 1399 and NGC 1404 (*), where the coordinates set the position of the photometric centre of the stellar structure. Column (7): position angle of the stellar structures in rotation detected in the radial velocity maps of seven galaxies of the sample. These stellar rotations are related to kinematically cold structures that are embedded in a kinematically hot stellar spheroid in these galaxies. Column (8): heliocentric radial velocity of the sample galaxies, calculated at the kinematic centre of the objects, except for NGC 1399, NGC 2663 and NGC 3136, where we measured this parameter at the position of the AGN (or at the photometric centre of the stellar structure) of the objects. \label{tab_mapa_vel_radial_stellar}
}
 \begin{tabular}{@{}lcccccccc}
  \hline
  Name & $x$(V$_r$) & $y$(V$_r$) & $x$($\sigma$) & $y$($\sigma$) & $x$$_{AGN}$ & $y$$_{AGN}$ & PA & V$_{hel}$   \\
   &  (arcsec) & (arcsec) & (arcsec) & (arcsec) & (arcsec) & (arcsec) & (degree) & (km s$^{-1}$) \\
   & (1) &(2) & (3) & (4) & (5) & (6) & (7) & (8) \\
  \hline
  ESO 208 G-21 &$-$0.2$\pm$0.2 &0.2$\pm$0.1&0.2$\pm$0.1&0.0$\pm$0.2&0.05$\pm$0.10&0.20$\pm$0.10&$-$75$\pm$6&1035$\pm$2\\
  IC 1459 &0.1$\pm$0.1 &0.0$\pm$0.1&0.1$\pm$0.1&0.0$\pm$0.1&0.05$\pm$0.05&0.00$\pm$0.05&$-$147$\pm$5&1719$\pm$2\\
  IC 5181 &0.0$\pm$0.1 &0.1$\pm$0.2&$-$0.2$\pm$0.2&0.1$\pm$0.2&0.00$\pm$0.05&0.00$\pm$0.05&90$\pm$6&2010$\pm$2\\
  NGC 1380 &0.0$\pm$0.2 &$-$0.5$\pm$0.2&$-$&$-$&0.20$\pm$0.10&$-$0.30$\pm$0.15&170$\pm$13&1858$\pm$2\\
  NGC 1399 &$-$&$-$&0.1$\pm$0.3&0.3$\pm$0.2&0.0$\pm$0.1*&0.0$\pm$0.1*&$-$&1425$\pm$5\\
  NGC 1404  &$-$0.1$\pm$0.3&0.0$\pm$0.2&$-$&$-$&0.0$\pm$0.1*&0.0$\pm$0.1*&$-$35$\pm$15&1936$\pm$1\\
  NGC 2663 &$-$ &$-$&$-$&$-$&0.00$\pm$0.05&0.00$\pm$0.10&$-$&2087$\pm$9\\
  NGC 3136  &$-$&$-$&$-$&$-$&$-$0.05$\pm$0.30&$-$0.20$\pm$0.30&$-$&1733$\pm$4\\
  NGC 4546 &0.0$\pm$0.2&$-$0.2$\pm$0.1&0.2$\pm$0.1&$-$0.2$\pm$0.1&$-$0.10$\pm$0.05&$-$0.10$\pm$0.05&80$\pm$5&1025$\pm$3\\
  NGC 7097 &$-$0.1$\pm$0.2&$-$0.1$\pm$0.2&$-$&$-$&0.15$\pm$0.05&$-$0.10$\pm$0.05&$-$162$\pm$9&2582$\pm$2\\
 
  \hline
 \end{tabular}
  
\end{table*}

We also detected the stellar structures in rotation of these seven galaxies with PCA Tomography (paper I). By comparing the PA values measured with the radial velocity maps and those measured with the tomograms, we noted that both methods match for IC 1459, NGC 1380, NGC 1404 and NGC 4546 (PA$_{tom}$ - PA$_{Vr}$ $<$ 7$^o$). In ESO 208 G-21, the difference is 18$^o$. However, one should be aware that a stellar rotation detected with PCA Tomography is more affected by dust than the radial velocity maps. Tomograms associated with stellar structures in rotation are built in such a way that the result is a combination of the anti-correlation between the red and blue wings of the stellar absorption lines and the intensity of the stellar content of the galaxies (see paper I). Since ESO 208 G-21 has a dust lane across the FOV (paper III), this structure may have affected the measurement of the PA of the stellar rotation detected in the tomogram shown in paper I. In IC 5181, the difference is 29$^o$. We showed in paper III that the gaseous disc of IC 5181 has a spiral structure, which may be the result of a non-axisymmetric potential in the central region of this galaxy (see e.g., \citealt{2006MNRAS.366.1151S}). Although we did not see any clear sign of deviation from an axisymmetric rotation in the radial velocity map, the tomogram associated with the stellar rotation of IC 5181 has more weight close to the nucleus of this galaxy. Thus, there is a small variation of the PA with radius of the stellar structure in rotation that may also be related to a non-axisymmetric potential in the circumnuclear region of IC 5181. In the case of NGC 7097, this difference is about 27$^o$. Also here, the stellar kinematics of this object seems to be peculiar. \citet{2001ApJ...546..903D} showed that NGC 7097 has a stellar structure with r $<$ 3.5 arcsec that is counter-rotating with respect to the external regions of this galaxy. Using models to describe the stellar kinematics of NGC 7097, they demonstrated that it is not possible to define a specific radial range that contains only counter-rotating orbits and another range with only corotating orbits. Indeed, they showed that in both the internal and external structures of NGC 7097, the projection of the radial velocity is the result of the combination of a large number of stellar orbits in co-rotation and a large number of orbits in counter-rotation. In other words, the counter-rotation detected within $r$ $<$ 3.5 arcsec is not associated with a compact rotating group of stars. Probably, the combination of stellar orbits found by \citet{2001ApJ...546..903D} should occur across the entire FOV of NGC 7097, leading to this peculiar stellar kinematics when compared to the other sample galaxies. 

We also built 1D profiles of the radial velocities, extracted along (parallel to the PA) and perpendicularly to the directions of the stellar rotation of the seven galaxies with such structure. Both profiles traverse the nucleus of the galaxies. We determined the position of the nuclei through the image of the red wing of the broad component of H$\alpha$ in five galaxies where we detected a BLR (paper II). In NGC 1380, we used the image of the [O I]$\lambda$6300 line to set the position of the nucleus. We assumed that the nucleus of NGC 1404 is in the photometric centre of the stellar structure (paper I). The positions of the AGN (or the stellar structure) of the galaxies are shown in Table \ref{tab_mapa_vel_radial_stellar}. The profiles are shown in Fig. \ref{perfil_stellar_kin_1}, where the radial velocities are shown with respect to the heliocentric radial velocity at the kinematic centre of the galaxies. 

\begin{figure*}

\begin{center}
\includegraphics[scale=0.30]{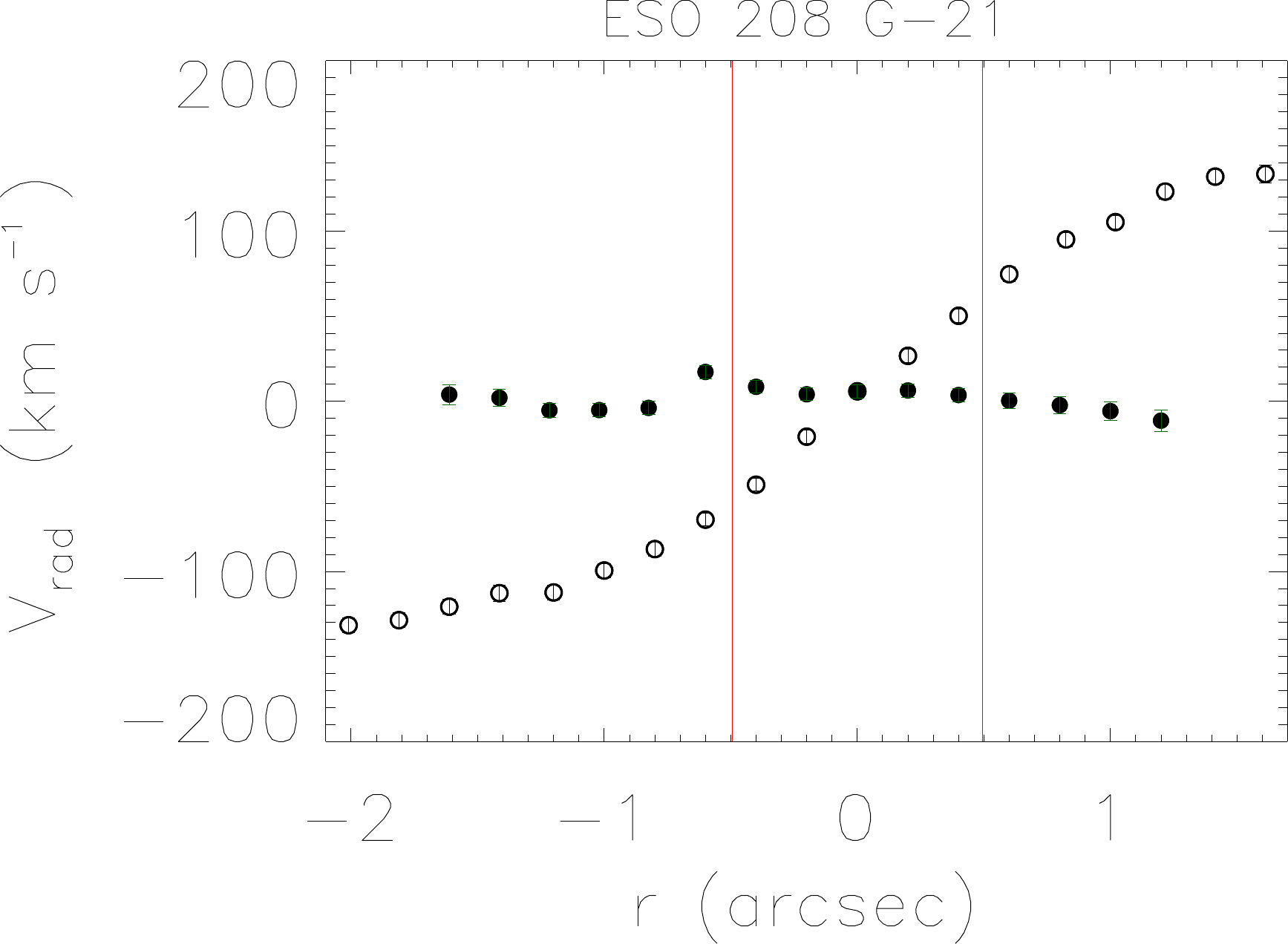}
\hspace{-0.0cm}
\vspace{0.3cm}
\includegraphics[scale=0.30]{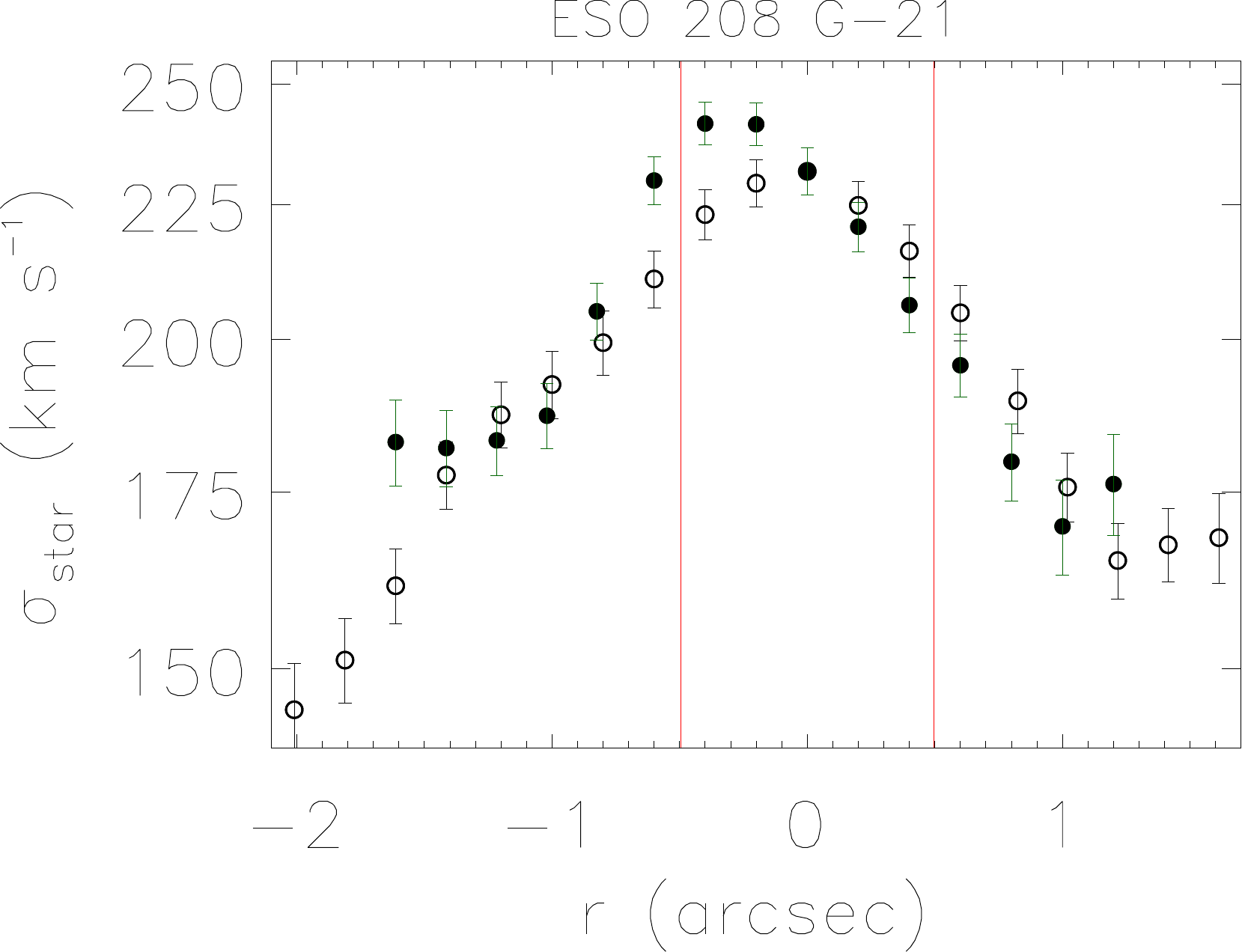}
\hspace{-0.0cm}
\includegraphics[scale=0.30]{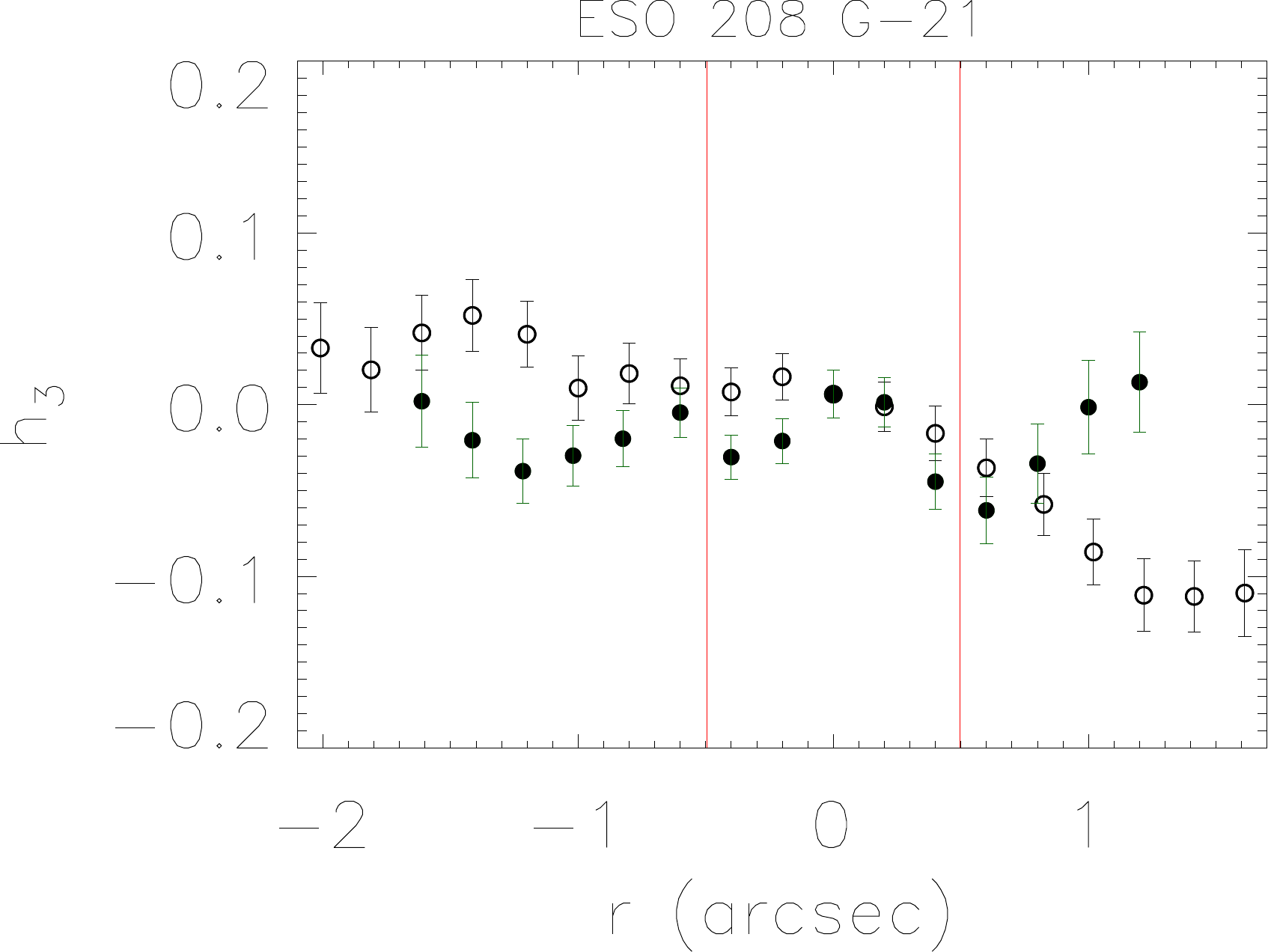}

\includegraphics[scale=0.30]{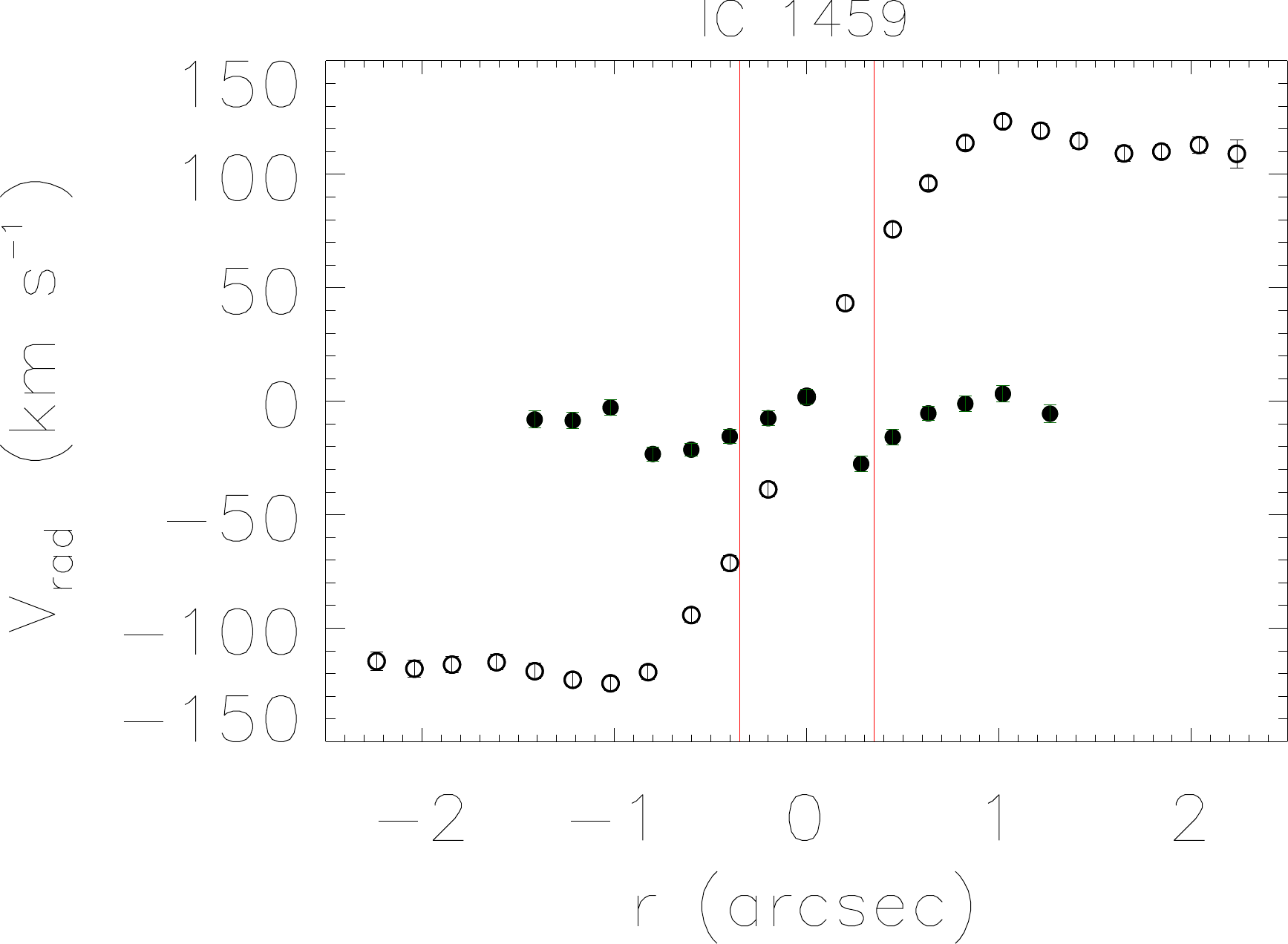}
\hspace{-0.0cm}
\vspace{0.3cm}
\includegraphics[scale=0.30]{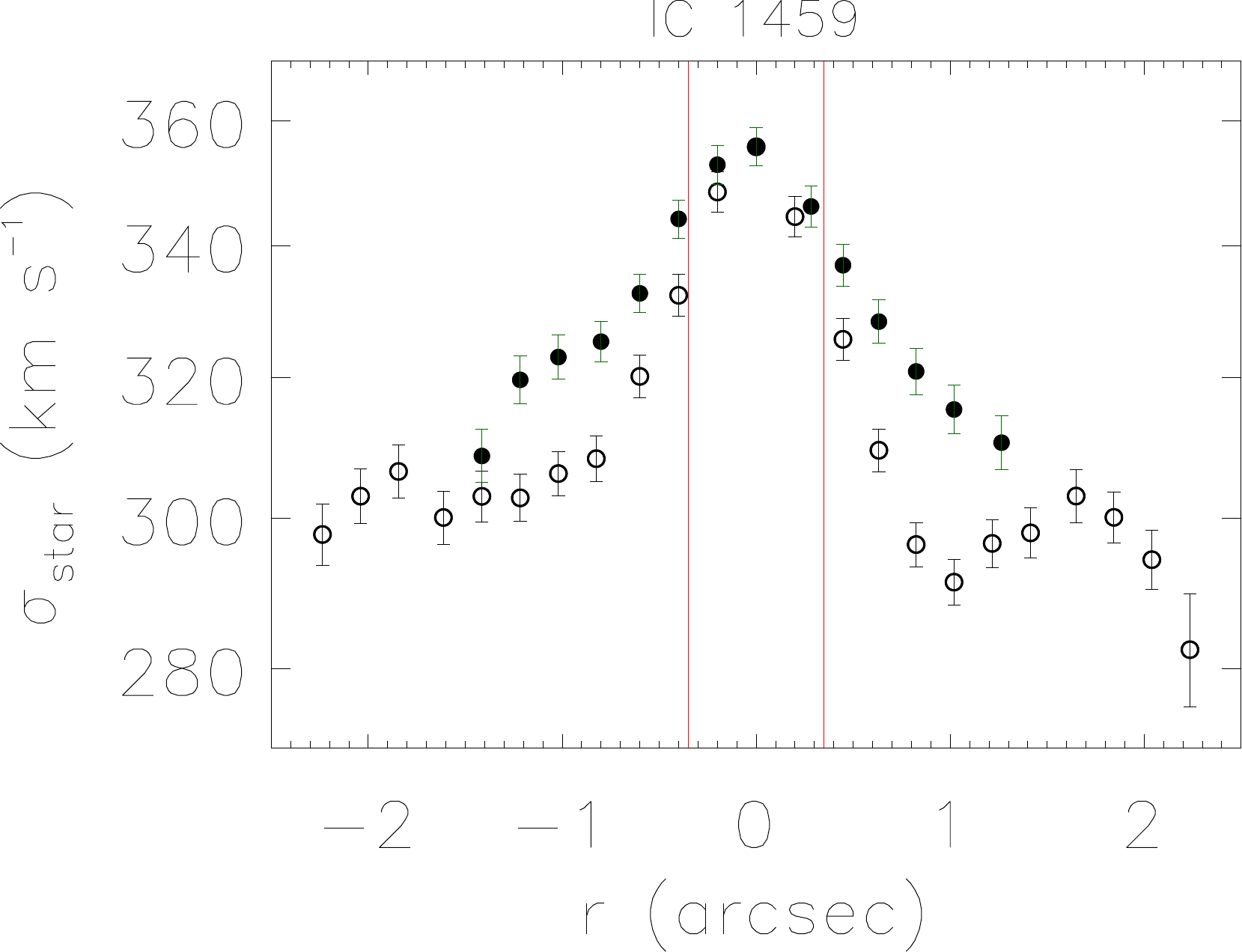}
\hspace{-0.0cm}
\includegraphics[scale=0.30]{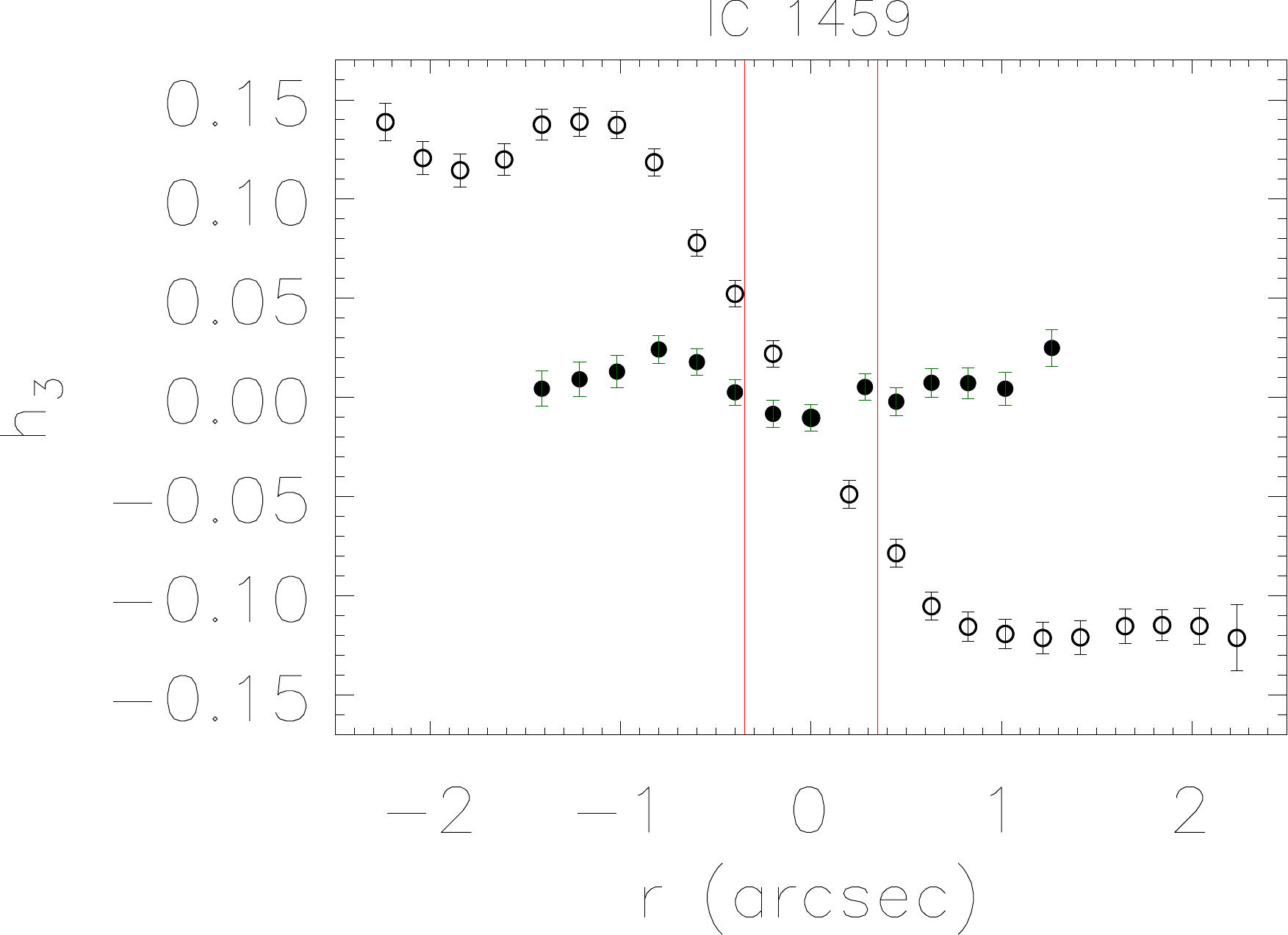}

\includegraphics[scale=0.30]{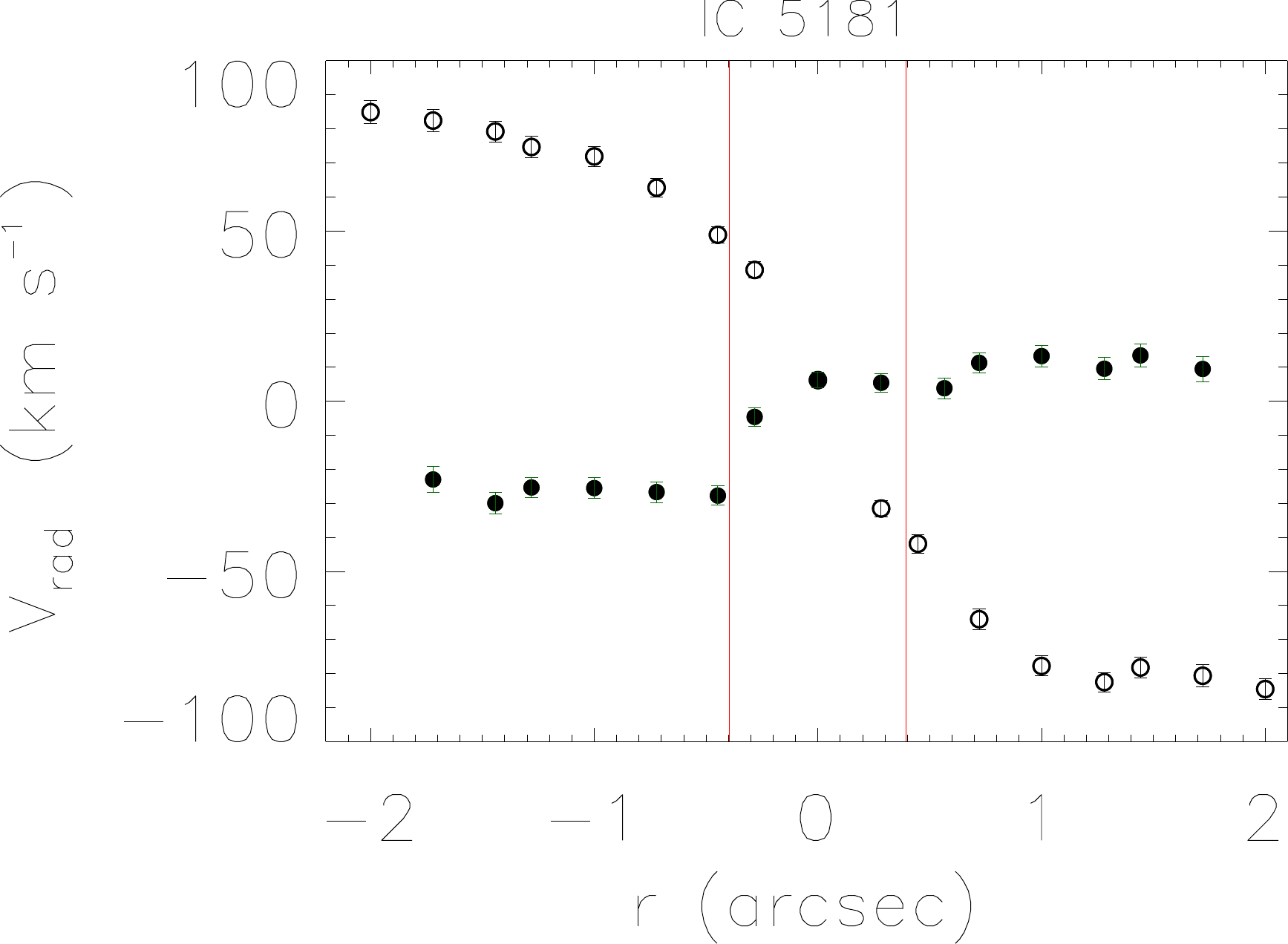}
\hspace{-0.0cm}
\vspace{0.3cm}
\includegraphics[scale=0.30]{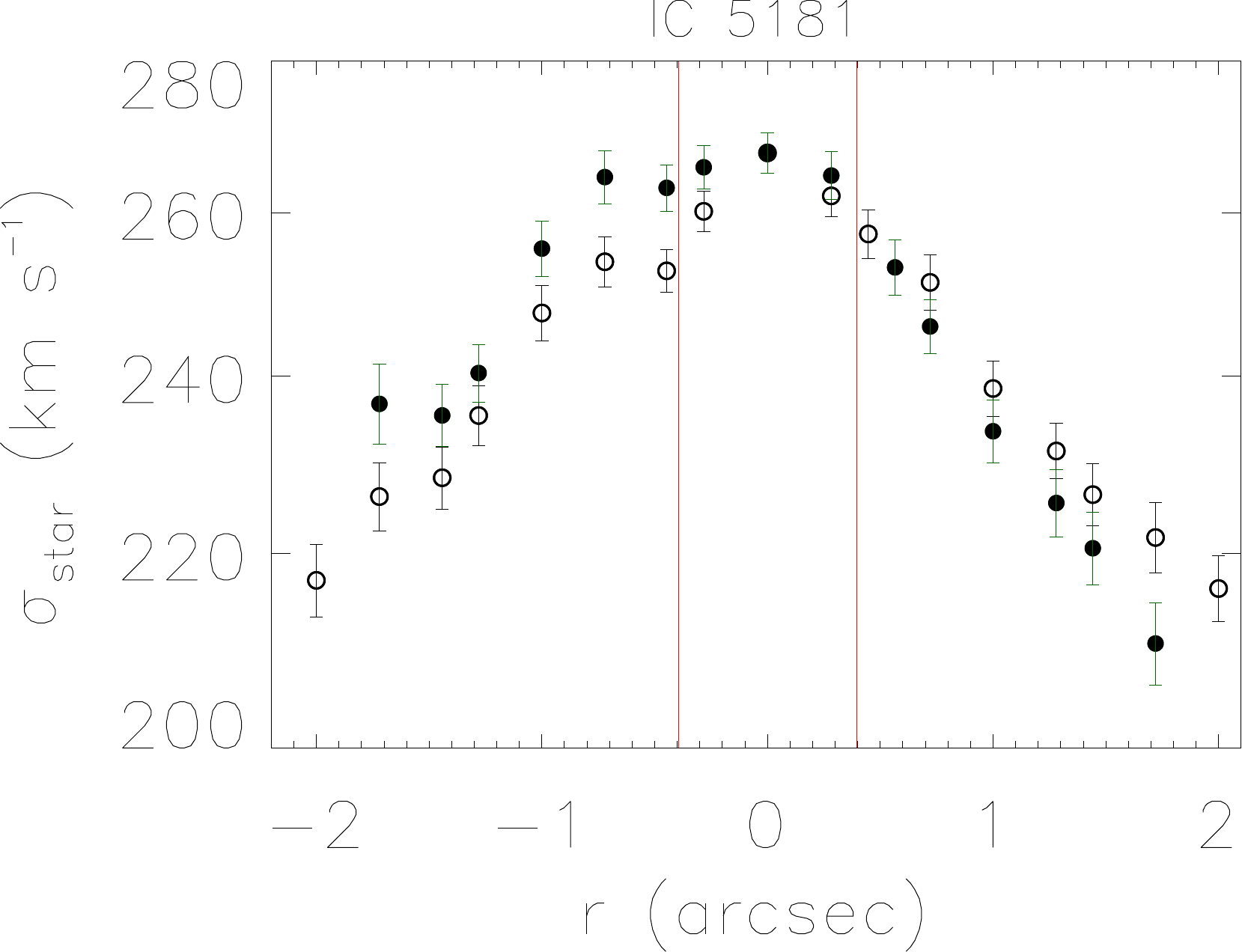}
\hspace{-0.0cm}
\includegraphics[scale=0.30]{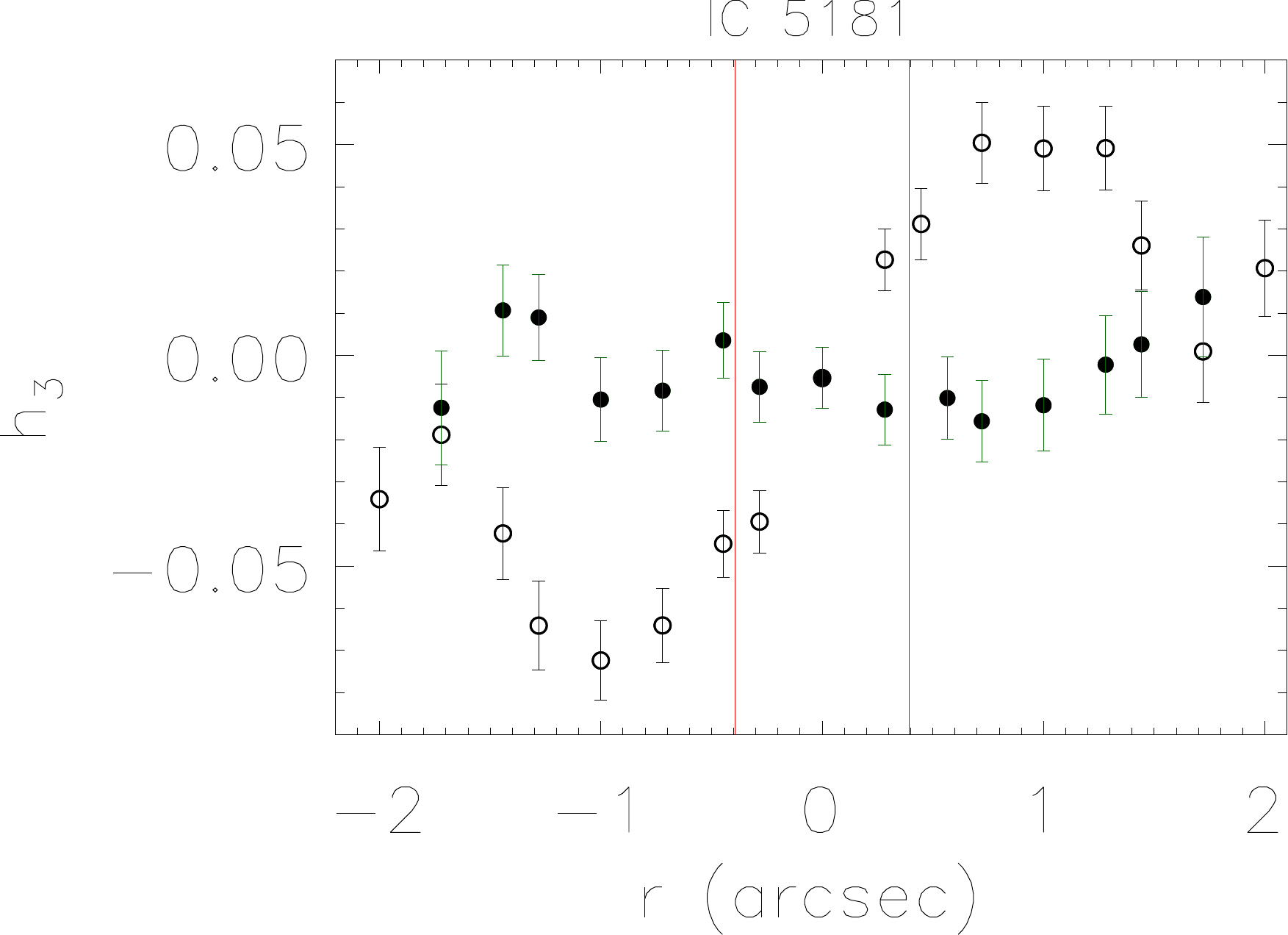}

\includegraphics[scale=0.30]{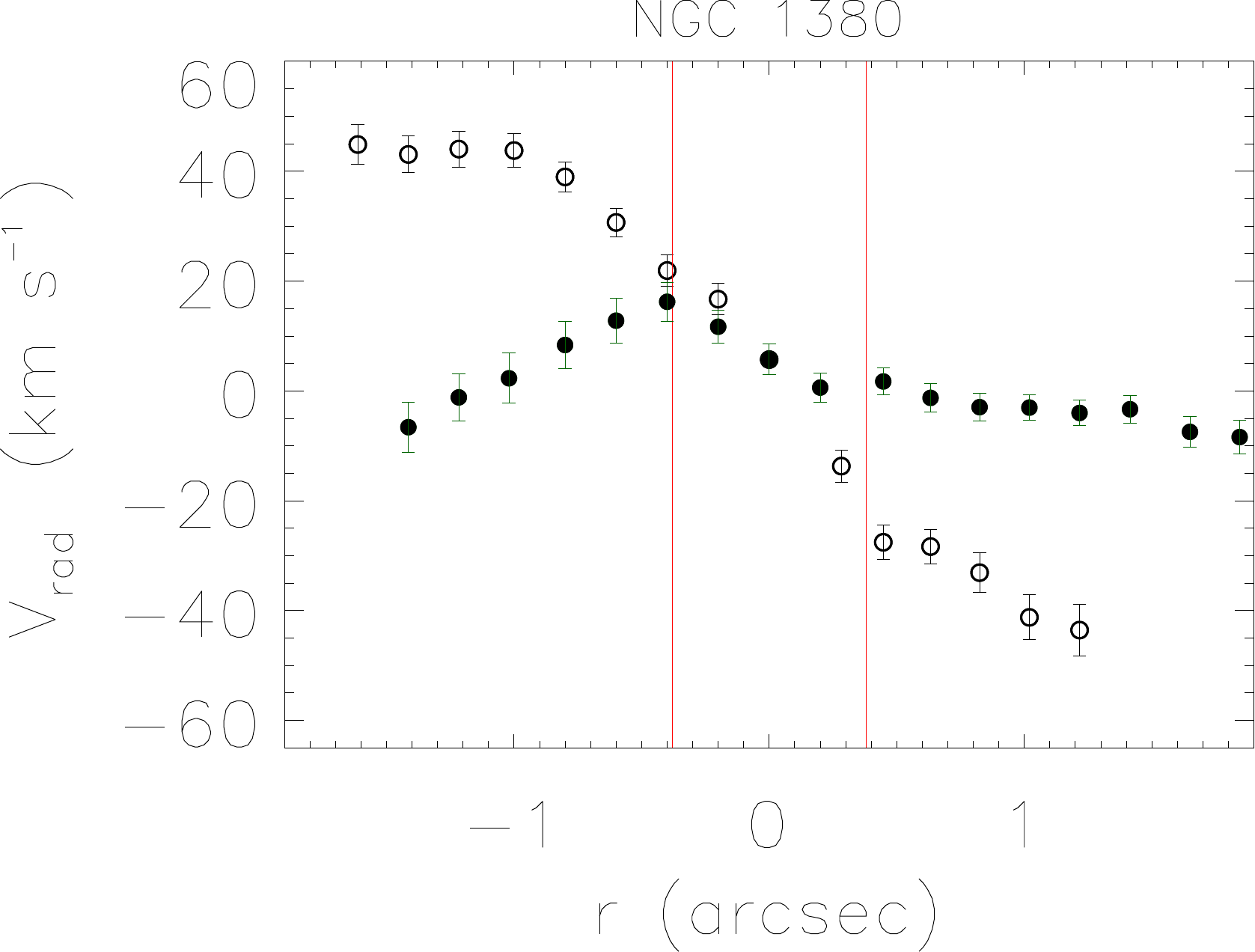}
\hspace{-0.0cm}
\vspace{0.3cm}
\includegraphics[scale=0.30]{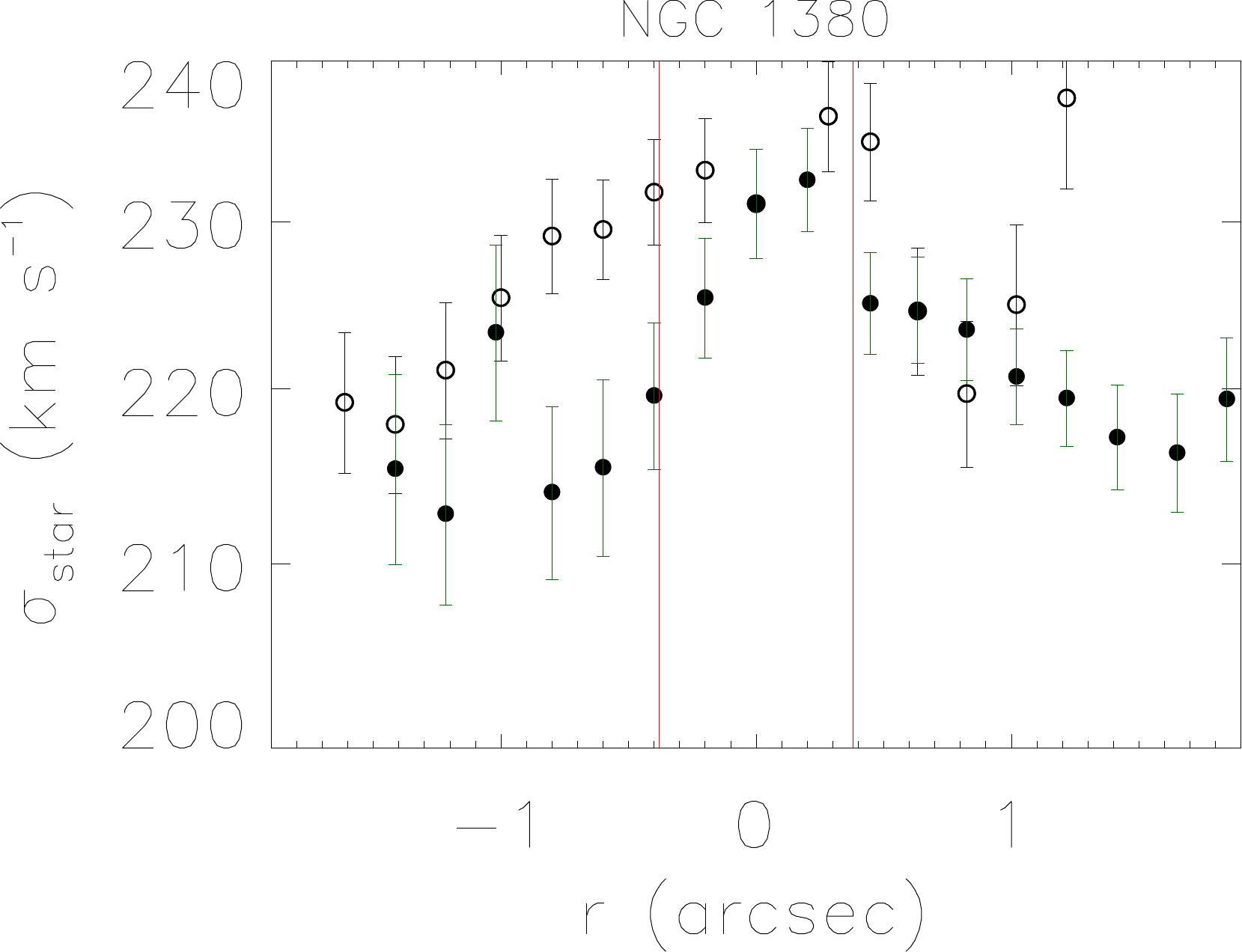}
\hspace{-0.0cm}
\includegraphics[scale=0.30]{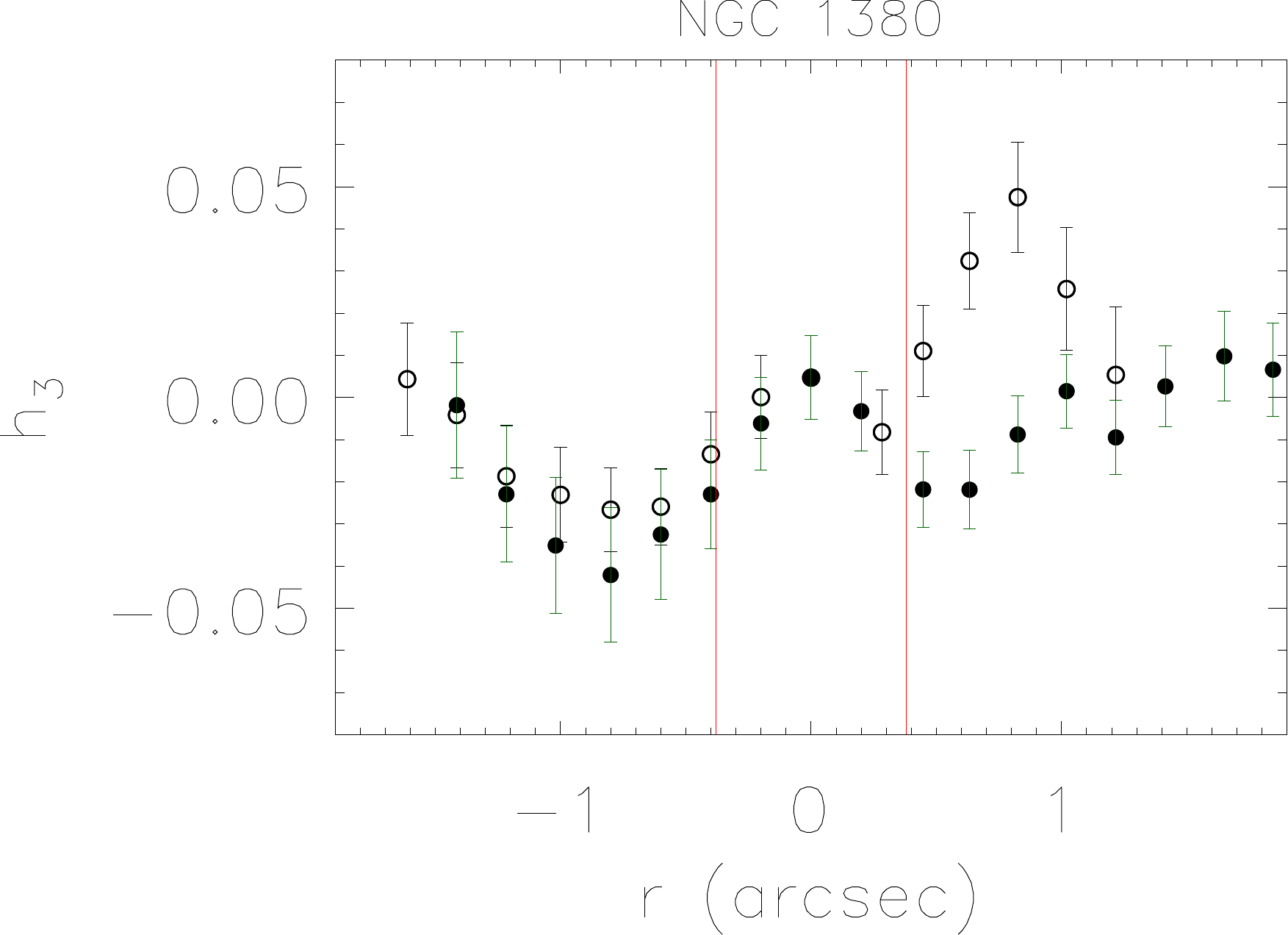}
\caption{1D profiles of the kinematic maps. The hollow black circles were extracted along the direction of the stellar rotations. The filled green circles were extracted perpendicularly to the stellar rotations. The red lines delimit the FWHM of the PSF of the data cubes (see papers I and II).   \label{perfil_stellar_kin_1}}

\end{center}

\end{figure*}

\addtocounter{figure}{-1}

\begin{figure*}
\begin{center}

\includegraphics[scale=0.30]{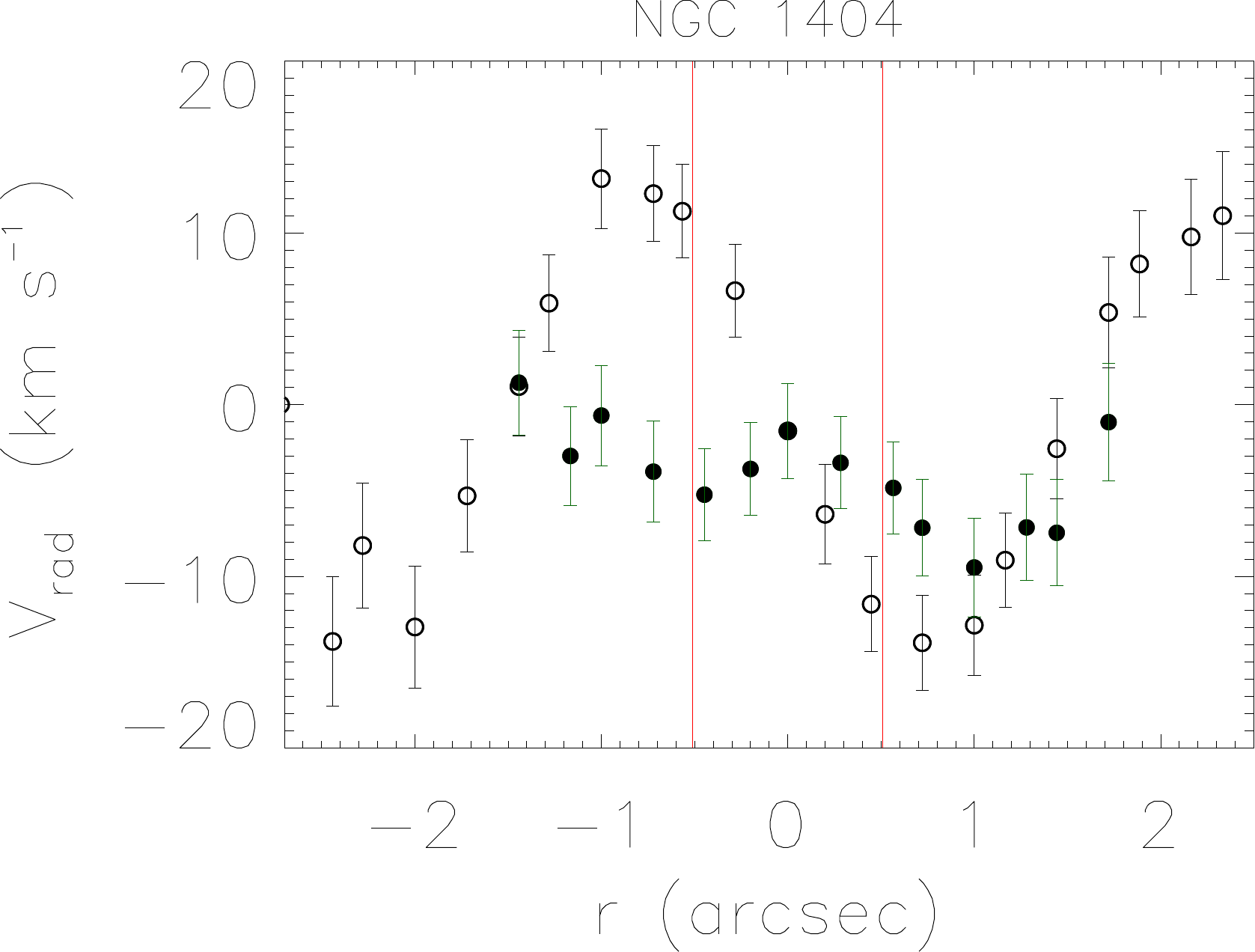}
\hspace{-0.0cm}
\vspace{0.3cm}
\includegraphics[scale=0.30]{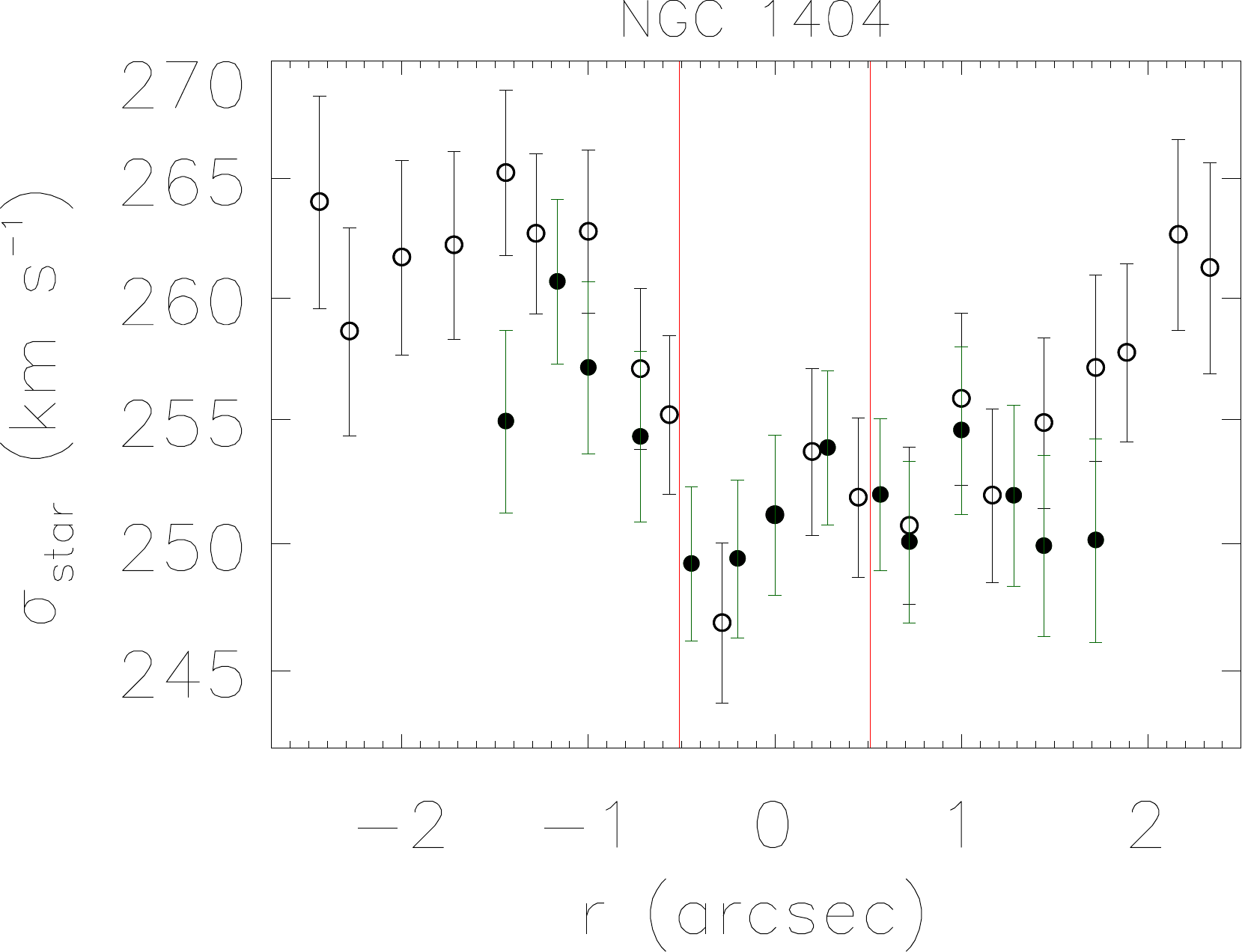}
\hspace{-0.0cm}
\includegraphics[scale=0.30]{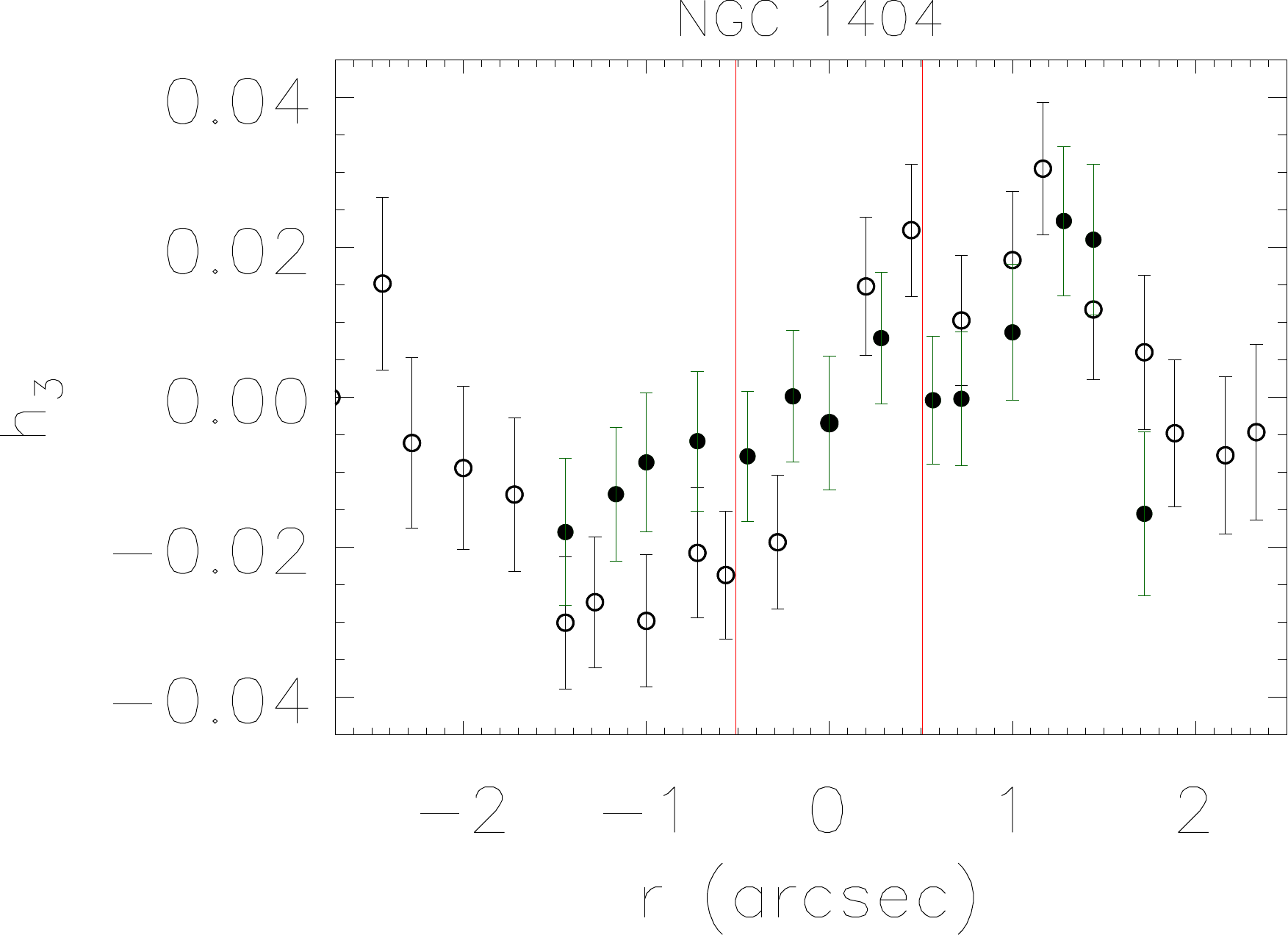}

\includegraphics[scale=0.30]{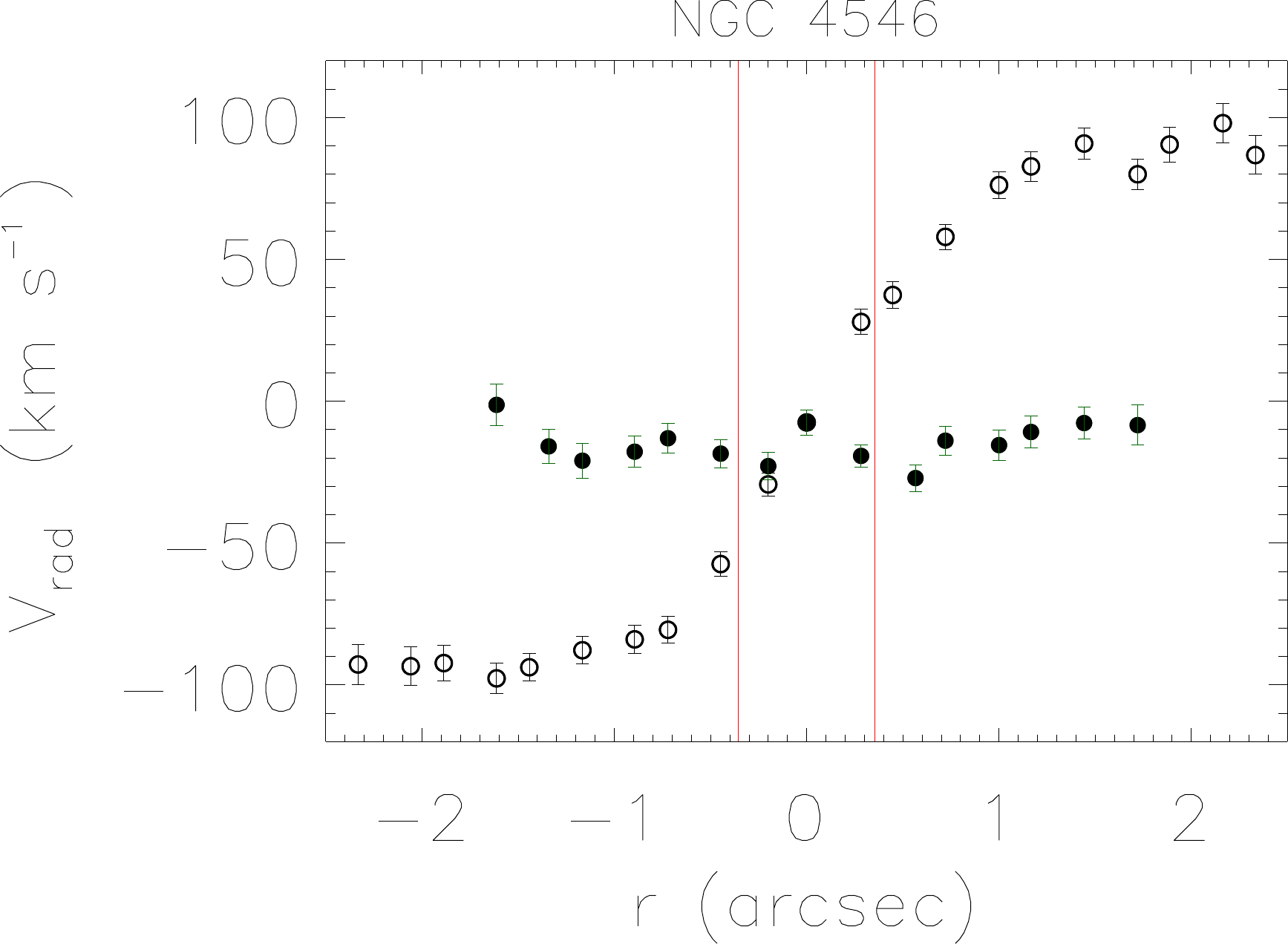}
\hspace{-0.0cm}
\vspace{0.3cm}
\includegraphics[scale=0.30]{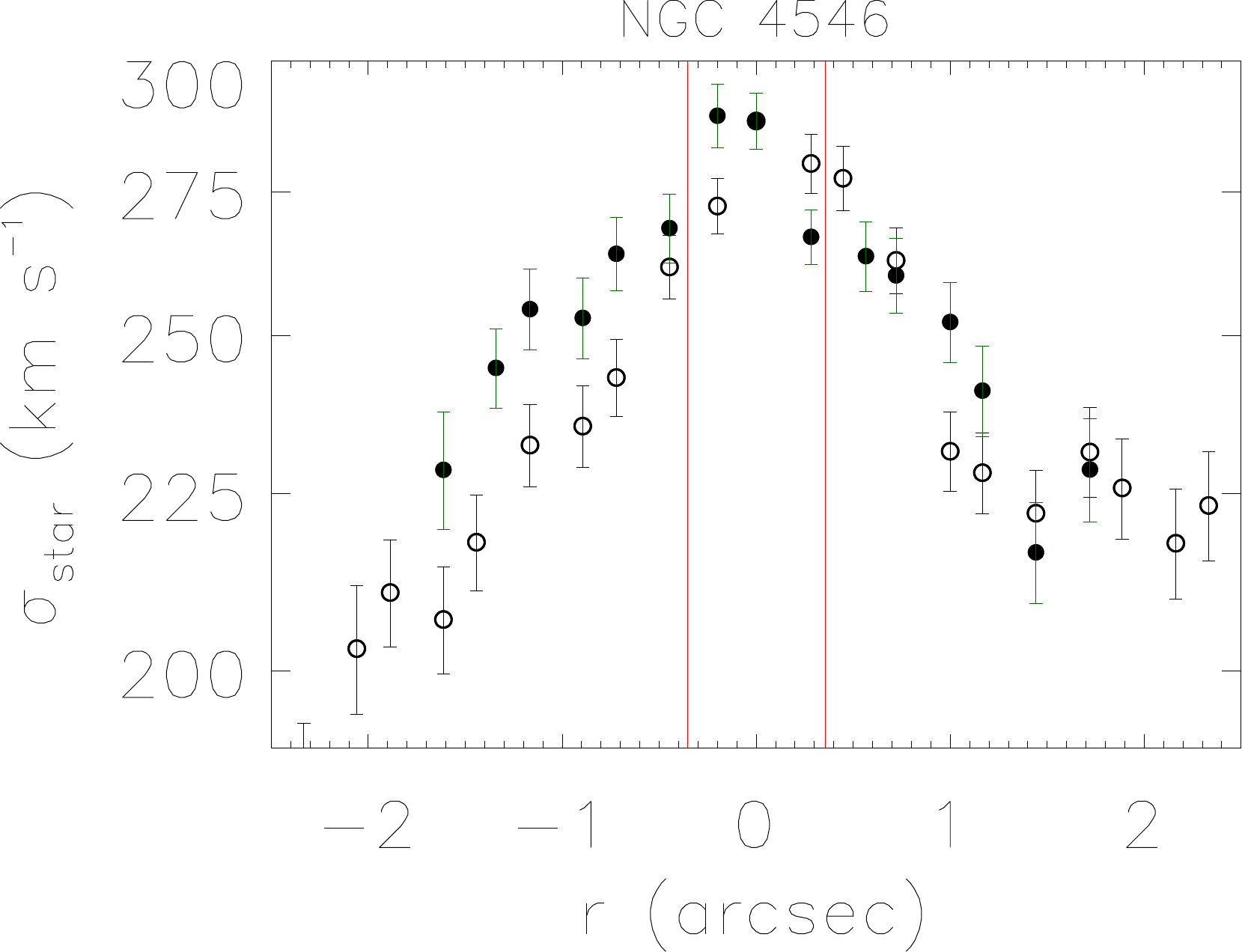}
\hspace{-0.0cm}
\includegraphics[scale=0.30]{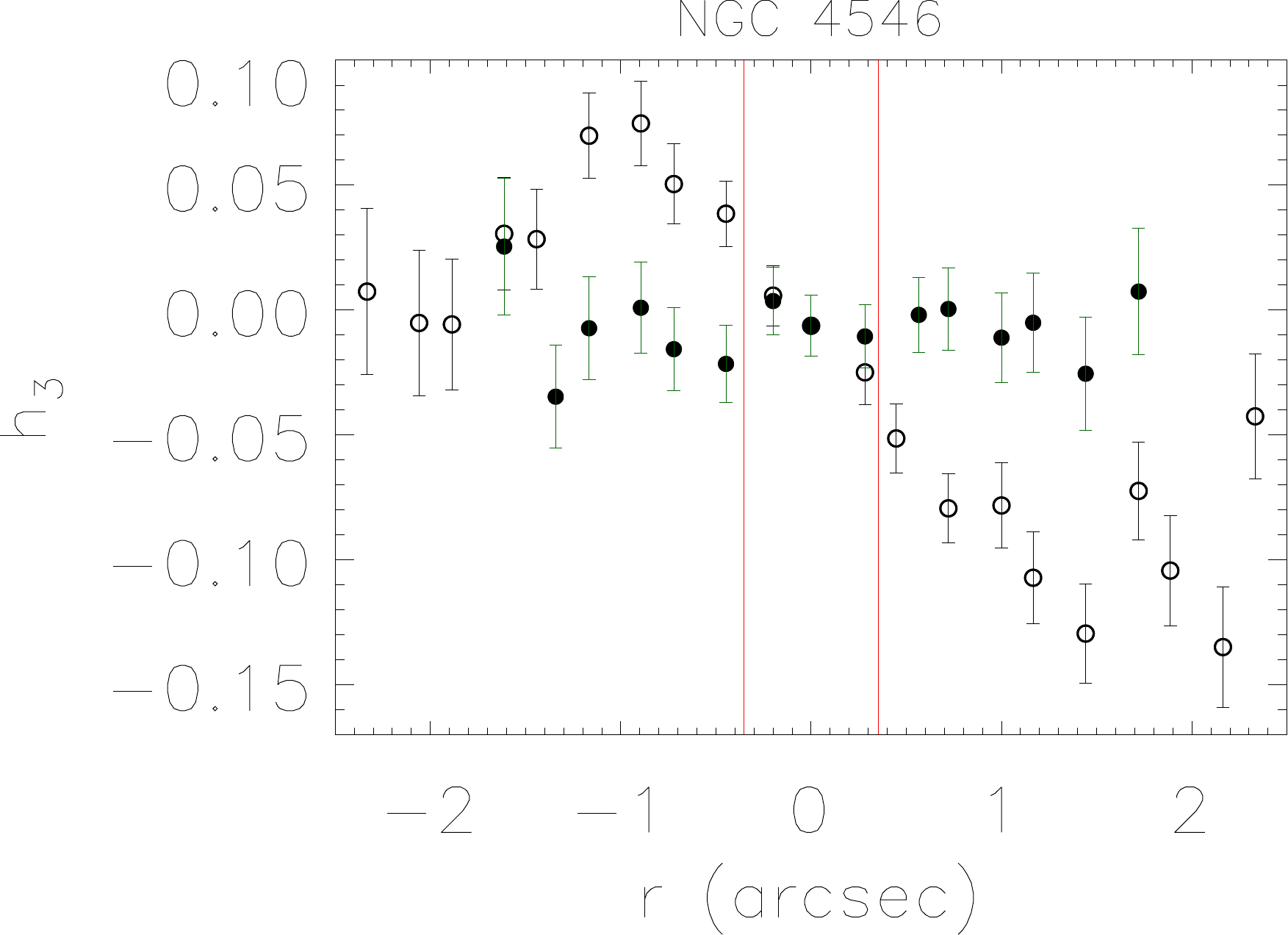}

\includegraphics[scale=0.30]{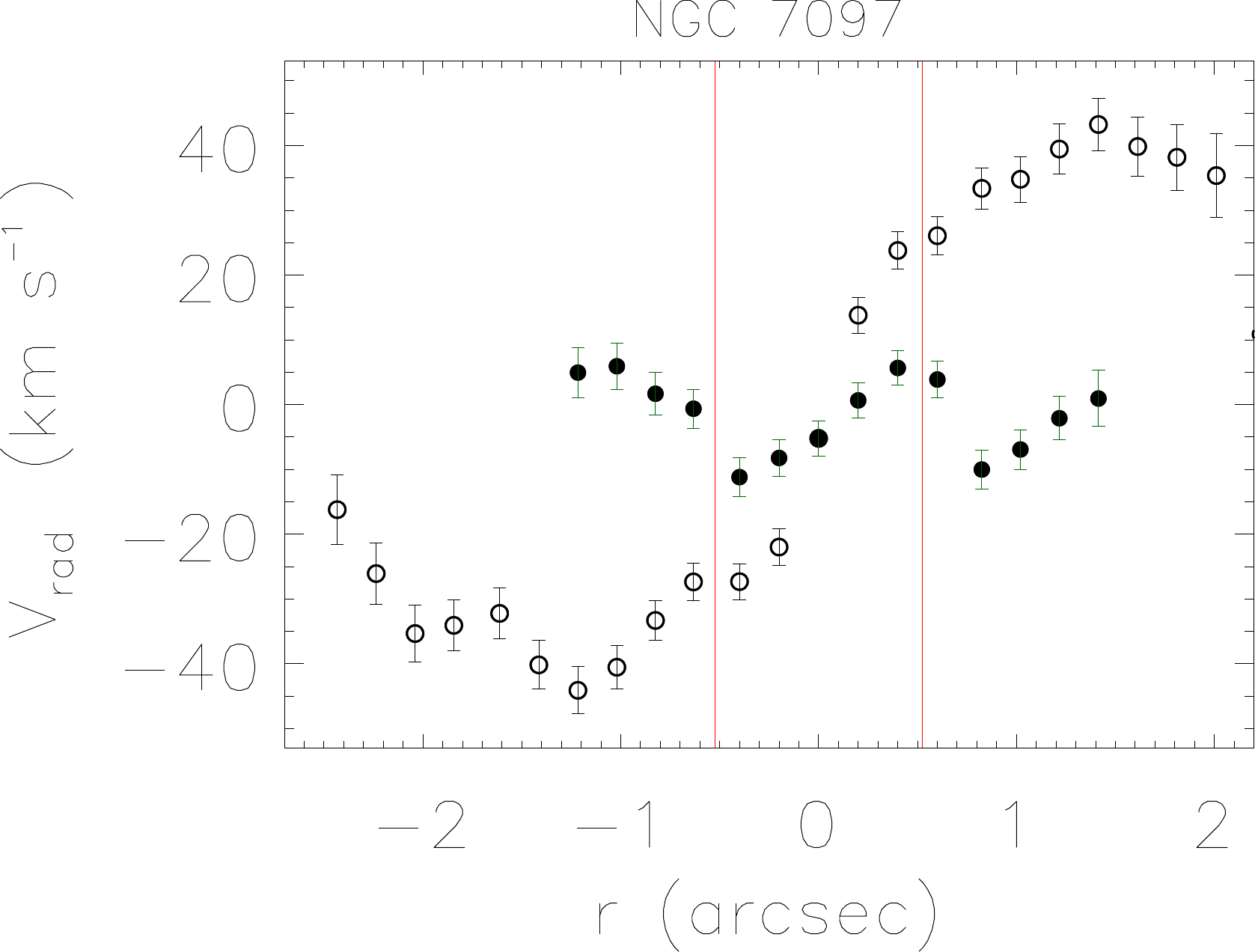}
\hspace{-0.0cm}
\vspace{0.3cm}
\includegraphics[scale=0.30]{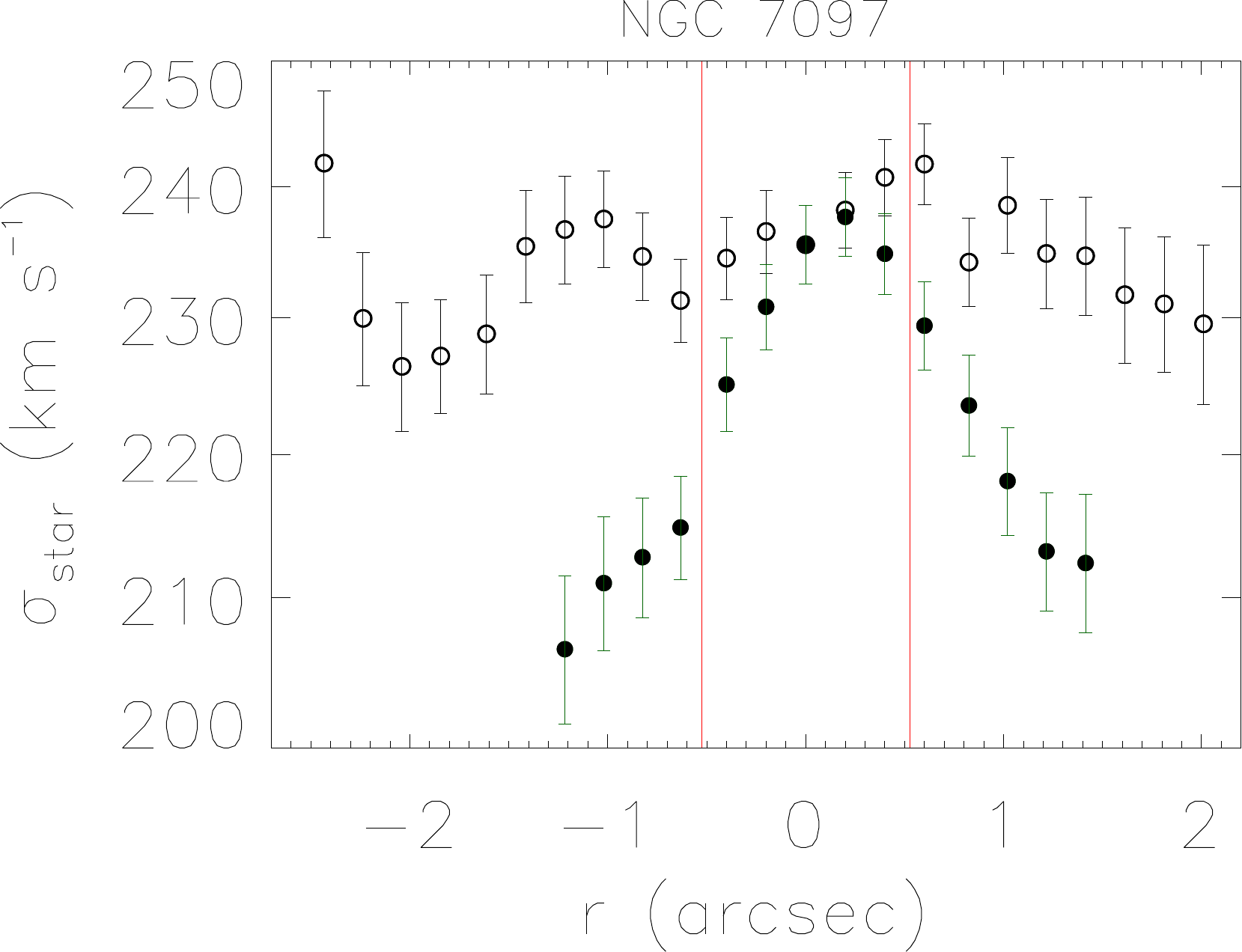}
\hspace{-0.0cm}
\includegraphics[scale=0.30]{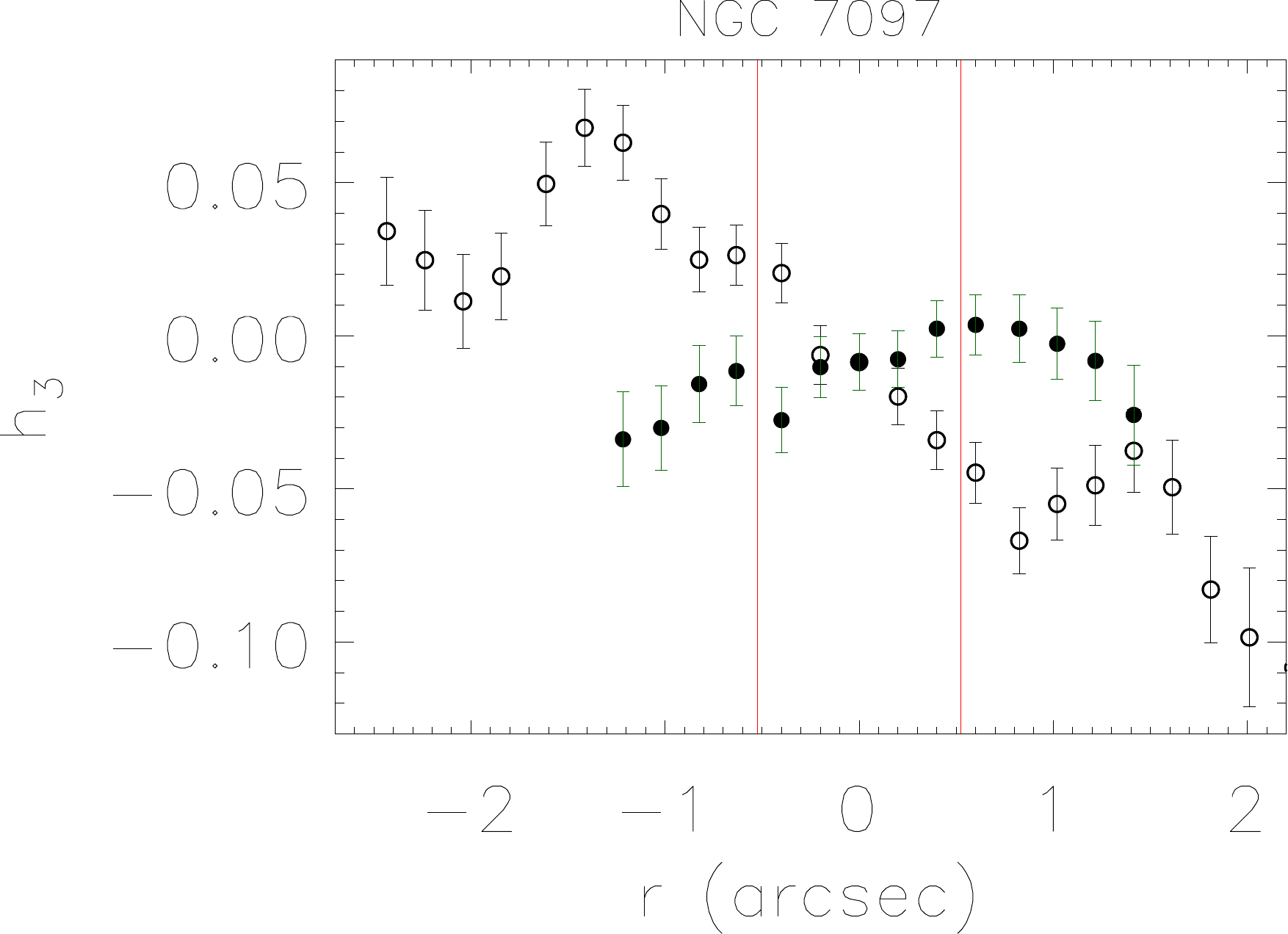}

\caption{continued. \label{perfil_stellar_kin_2}}

\end{center}

\end{figure*}

We proposed in paper I that these seven galaxies have stellar structures in rotation. Both the maps and the profiles of, at least, six objects suggest a rotating structure. Note that there is a well-defined rotation in the 1D profile along the rotating stellar structures, while in the perpendicular direction, the radial velocities are nearly constant near the nucleus. In NGC 1404, the 1D profile along the PA shows a sinusoidal behaviour. By comparing the radial velocity map of this object, shown in Fig. \ref{perfil_stellar_kin_2}, with the radial velocity map of \citet{2014MNRAS.441..274S}, which has a larger FOV (25 $\times$ 28 arcsec$^2$), we note that NGC 1404 has a kinematic decoupled core (KDC) counter-rotating with respect to the external structure of this object. This should explain the sinusoidal behaviour of the 1D profile along the PA For IC 5181, note that the profile along the perpendicular direction is not zero along the region where $V_{rad}$ $>$ 0 km s$^{-1}$. This is possibly related to the twist detected in the stellar structure in rotation of this galaxy. In the internal region of NGC 7097, it should be noted that there may be both corotating and counter-rotating orbits. The peculiarity of the stellar kinematics of this object is also reflected in the 1D profile along the perpendicular direction of the stellar rotation, whose V$_{rad}$ values are not constant along the radius. 

\subsection{Velocity dispersion maps} \label{stellar_velocity_dispersion}

\begin{figure*}
	\centering
		\includegraphics[scale=0.3]{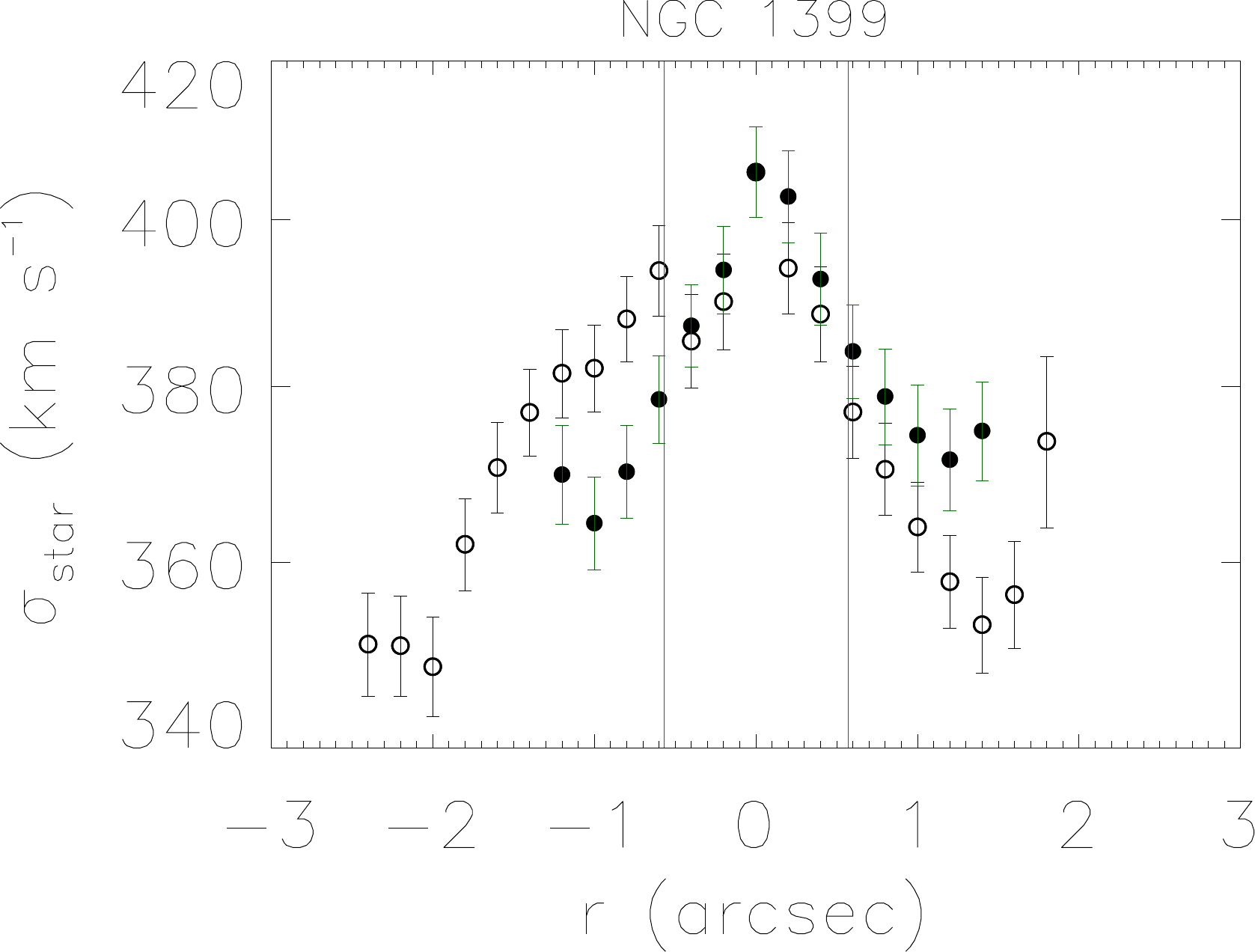}
		\includegraphics[scale=0.3]{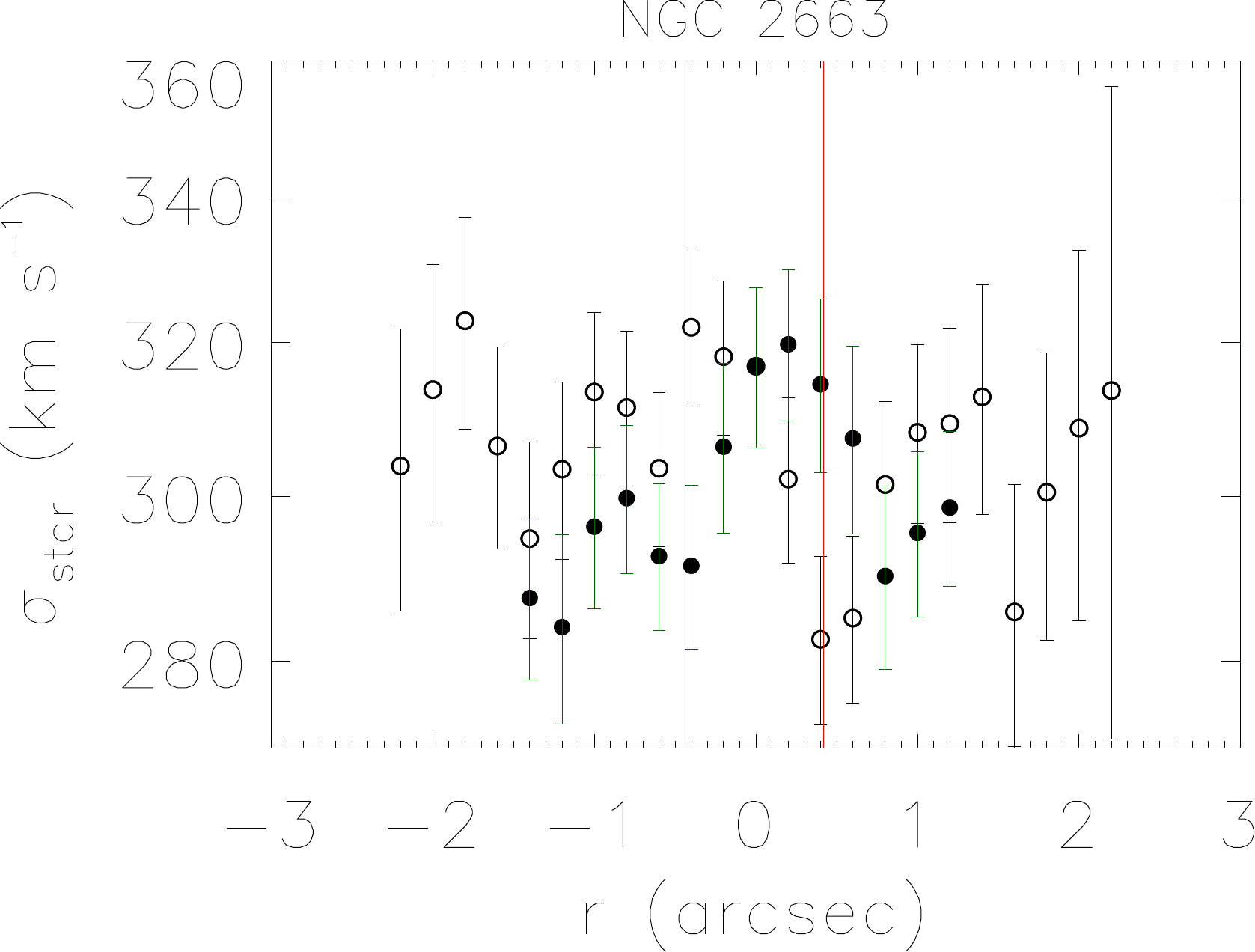}
		\includegraphics[scale=0.3]{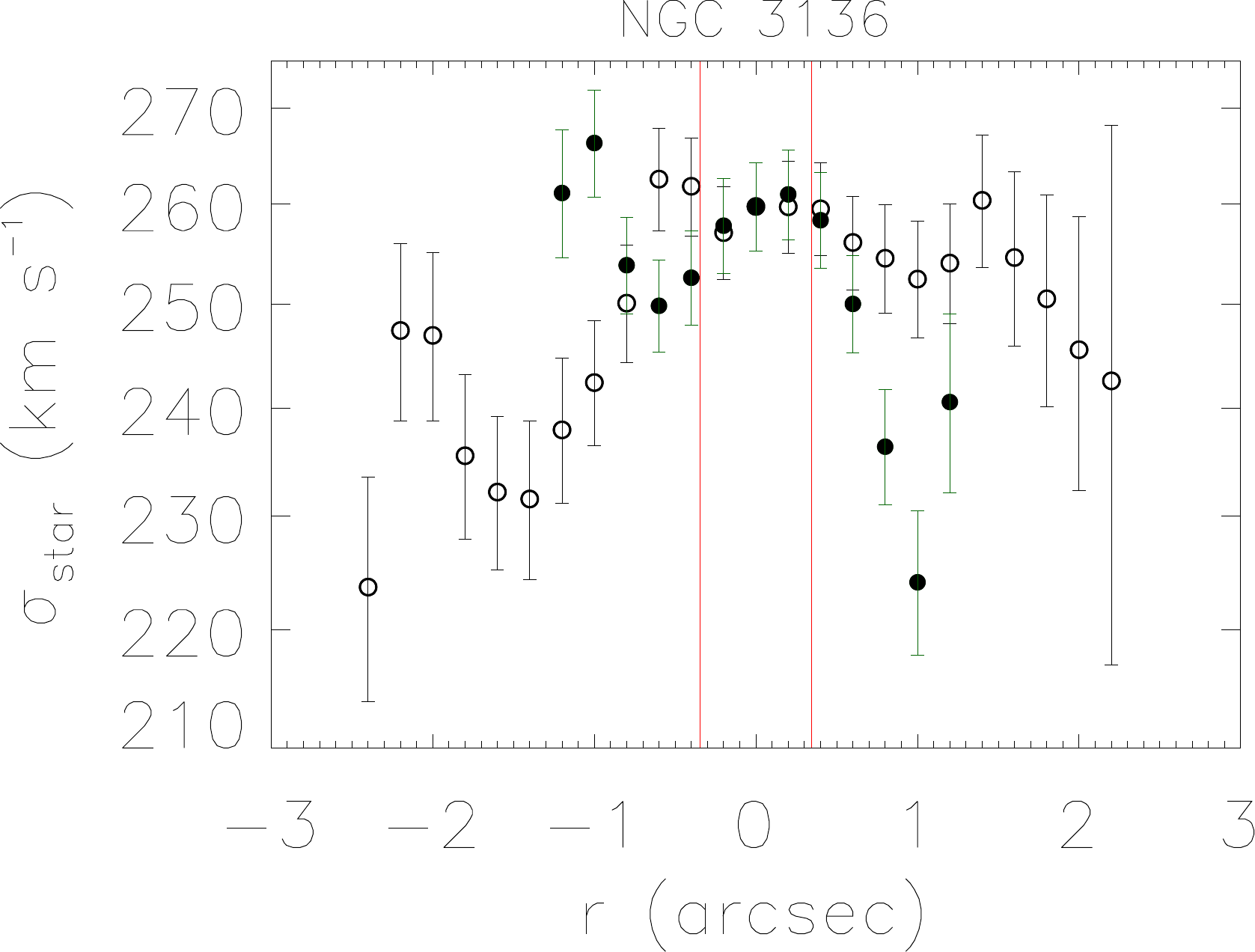}
	\caption{1D profiles of the velocity dispersion for NGC 1399, NGC 2663, and NGC 3136. We did not detect any evidence of ordered stellar rotation in these galaxies. In this case, the hollow black circles are related to the profile extracted along the longest spatial dimension of the data cubes while the filled green circles correspond to the profile extracted along the shortest spatial dimension. \label{fig:perfil_disp_N2663_N1399}}
	
\end{figure*}

The velocity dispersion maps are shown in Fig. \ref{cinematica_estelar_1}. Their 1D profiles, shown in Fig. \ref{perfil_stellar_kin_1}, were also extracted along and perpendicularly to the directions of the stellar rotations detected in the radial velocity maps. For NGC 1399, NGC 2663, and NGC 3136, where we did not detect any evidence of ordered stellar rotation, the 1D profiles were extracted along the longest and the shortest spatial dimension of the data cubes. They are shown in Fig. \ref{fig:perfil_disp_N2663_N1399}. In these three galaxies, both profiles traverse their nuclei.  

We noticed that the velocity dispersion has higher values in the nuclear region of the galaxies ESO 208 G-21, IC 1459, IC 5181, NGC 4546, and also NGC 1399. The positions of these nuclear peaks of velocity dispersion are given in Table \ref{tab_mapa_vel_radial_stellar}. We do not see any evidence of peaks in the velocity dispersion maps of NGC 1380, NGC 2663, and NGC 3136. In NGC 1404, we detected a sigma-drop in the central region. This feature has already been reported by \citet{1998A&AS..133..325G}. We will revisit the sigma-drop detected in NGC 1404 in Section \ref{ngc1404_case}.

The nuclear peaks detected in ESO 208 G-21, IC 1459, IC 5181, and NGC 4546 should correspond, in part, to unresolved stellar rotation. This interpretation is not new at all (see e.g., \citealt{1987IAUS..127...17K,1988ApJ...327L..55F}). First, the peaks of velocity dispersion agree, within the errors, with the positions of the kinematic centre and also with the positions of the AGNs of these four objects. Also, note that in IC 1459, IC 5181 and NGC 4546, the shape of the peaks are elongated perpendicularly to the direction of the rotation. The 1D profiles confirm this statement, since the velocity dispersion decreases slowly along this direction. A possible interpretation for this characteristic is that it is caused by a stellar structure in rotation. Along the plane of rotation, the sigma peak has a width limited by the PSF. On the perpendicular direction it may not be limited by the PSF as the size of the  structure may go beyond it.

In NGC 7097, the elongated structure detected in the velocity dispersion map is a possible consequence of the complexity of the stellar kinematics discussed in Section \ref{stellar_radial_velocities}. Also the elongation detected in IC 1459, IC 5181, and NGC 4546 may be the result of a more complex setup of their stellar kinematics. For NGC 1380, one would expect a peaky structure, on the basis of the radial velocity map, somewhat similar to those of ESO 208-G21, IC 5181, IC 1459 or NGC 4546. Instead, for unknown reasons, it shows an irregular structure with a tendency to increase towards the centre. 

\citet{2007ApJ...671.1321G}, using data from the \textit{Hubble Space Telescope} \textit{(HST)}, analysed the stellar kinematics of NGC 1399, where we detected a nuclear peak, but no sign of stellar rotation. They showed that the stellar orbits are predominantly tangential near the SMBH. However, our data do not have enough spatial resolution to detect the decrease of the observed velocity dispersion that results from this anisotropy in the nuclear region. 

\subsection{Circumnuclear velocity dispersion of the sample galaxies} \label{central_velocity_dispersion_section}

To obtain a representative value of the circumnuclear velocity dispersion of the sample galaxies, we calculated $\sigma_{*}$ within a radius $R_{*}$ as
\begin{equation}
	\sigma_{*} \equiv \frac{\sum_{i=1}^{N_p}{I_i\sqrt{\sigma_i^2+V_i^2}}}{\sum_{i=1}^{N_p}{I_i}},
	\label{central_velocity_dispertion_equation}
\end{equation}
where $I_i$ is the intensity of the stellar continuum in the spaxel $i$ and $N_p$ is the total number of spaxels within $R_{*}$. The values of $\sigma_{*}$ and $R_{*}$ are shown in Table \ref{tab_disp_vel_stellar}. The statistical errors for $\sigma_{*}$ were estimated with a MC simulation and are $\sim$ 1 km s$^{-1}$. 

\begin{table*}
 \scriptsize
   \caption{Column (1): velocity dispersion within $R_{*}$ of the sample galaxies, calculated with equation (\ref{central_velocity_dispertion_equation}). The errors were estimated with a Monte Carlo simulation and are $\sim$ 1 km s$^{-1}$. Column (2): $R_{*}$ in arcsec. Column (3): $R_{*}$ in units of the effective radius. The values of $R_e$ for the sample galaxies are shown in paper I. \label{tab_disp_vel_stellar}
}
 \begin{tabular}{@{}lccc}
 \hline
  Name & $\sigma_{*}$ (km s$^{-1}$)& $R_{*}$ (arcsec)& $R_{*}$ ($R_e$)   \\
   & (1) & (2)& (3)  \\
  \hline
  ESO 208 G-21 & 205$\pm$1 &1.6 &0.06   \\
  IC 1459 &325$\pm$1 &1.4 &0.04   \\
  IC 5181 &251$\pm$1 &1.6 &0.14   \\
  NGC 1380 &224$\pm$1 &1.6 &0.04   \\
  NGC 1399 &378$\pm$1 &1.4 &0.04   \\
  NGC 1404 &253$\pm$1 &1.4 &0.06  \\
  NGC 2663 &304$\pm$1 &1.4 &0.02   \\
  NGC 3136 &250$\pm$1 &1.4 &0.04   \\
  NGC 4546 &254$\pm$1 &1.6 &0.06   \\
  NGC 7097 &230$\pm$1 &1.4 &0.08   \\

  \hline
 \end{tabular}

\end{table*}

\subsection{Maps of the Gauss Hermite $h_3$ moment} \label{stellar_hermite_h3_moment}

The maps of the $h_3$ moment are shown in Fig. \ref{cinematica_estelar_1}. The profiles extracted along and perpendicularly to the directions of the stellar rotations detected in the radial velocity maps are shown in Fig. \ref{perfil_stellar_kin_1}.

The most remarkable feature of the $h_3$ maps is the anti-correlation between this parameter and the radial velocities in the galaxies where we detected stellar structures in rotation. LOSVDs with signatures of asymmetry are related to rotating stellar structures embedded in the spheroidal structure, which is common for some E galaxies \citep{1994MNRAS.270..325C,2010gfe..book.....M}. These asymmetric profiles have already been detected for IC 1459 by \citet{1988ApJ...327L..55F}. Indeed, these authors interpreted this result as a kinematically cold disclike component that is embedded in a kinematically hot structure, although they described the LOSVD profile as a sum of two Gaussian functions.

In NGC 1404, the anti-correlation between $h_3$ and the radial velocity is probably a consequence of the KDC also being embedded in the spheroidal structure of this object. 

\subsection{Comparison of the GMOS-IFU results with other data sets} \label{comparison_results_previous_studies}
\subsubsection{Comparison with long-slit data} \label{comp_longslit}

\begin{figure*}

\begin{center}
\includegraphics[scale=0.30]{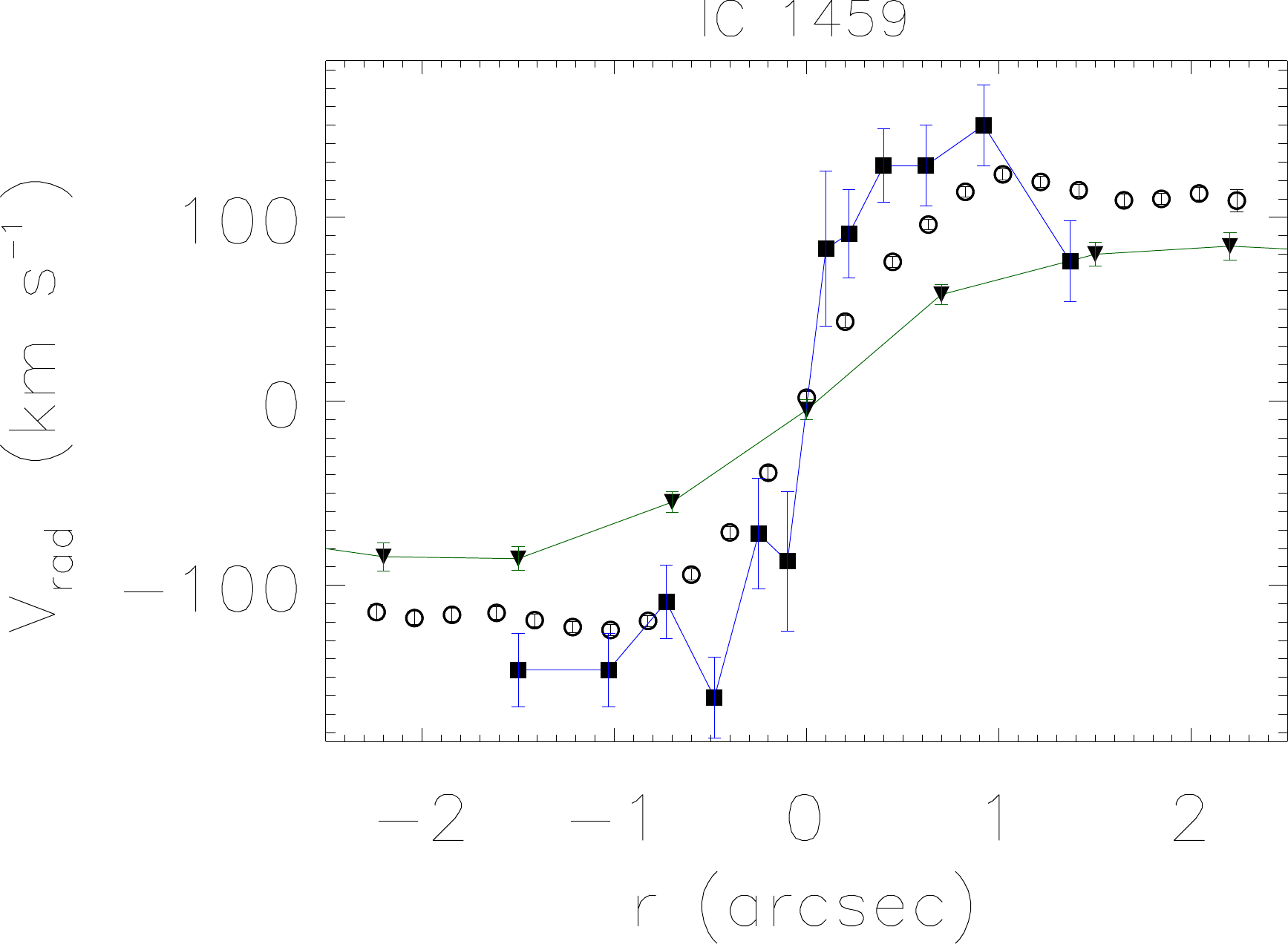}
\hspace{-0.0cm}
\vspace{0.3cm}
\includegraphics[scale=0.30]{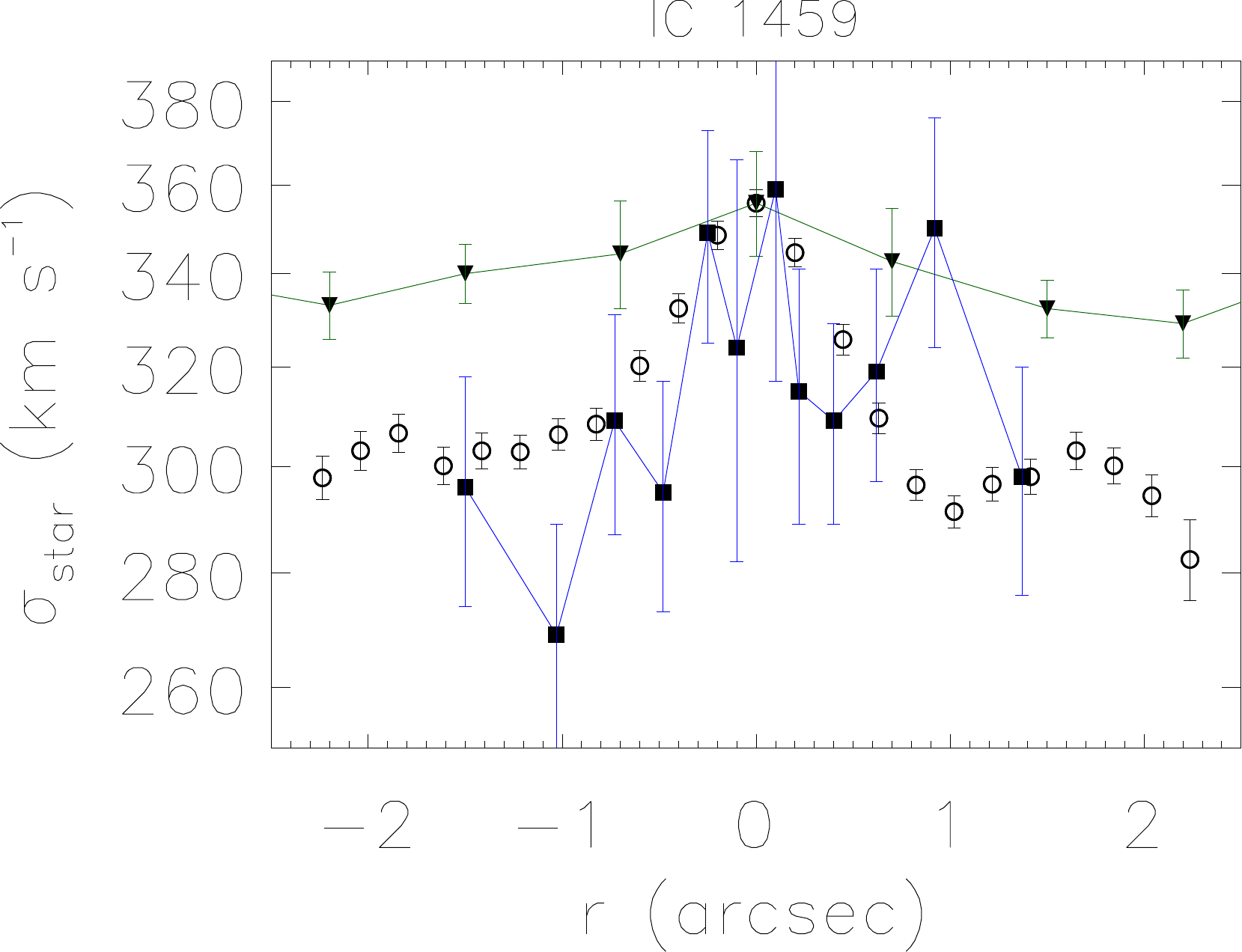}
\hspace{-0.0cm}
\includegraphics[scale=0.30]{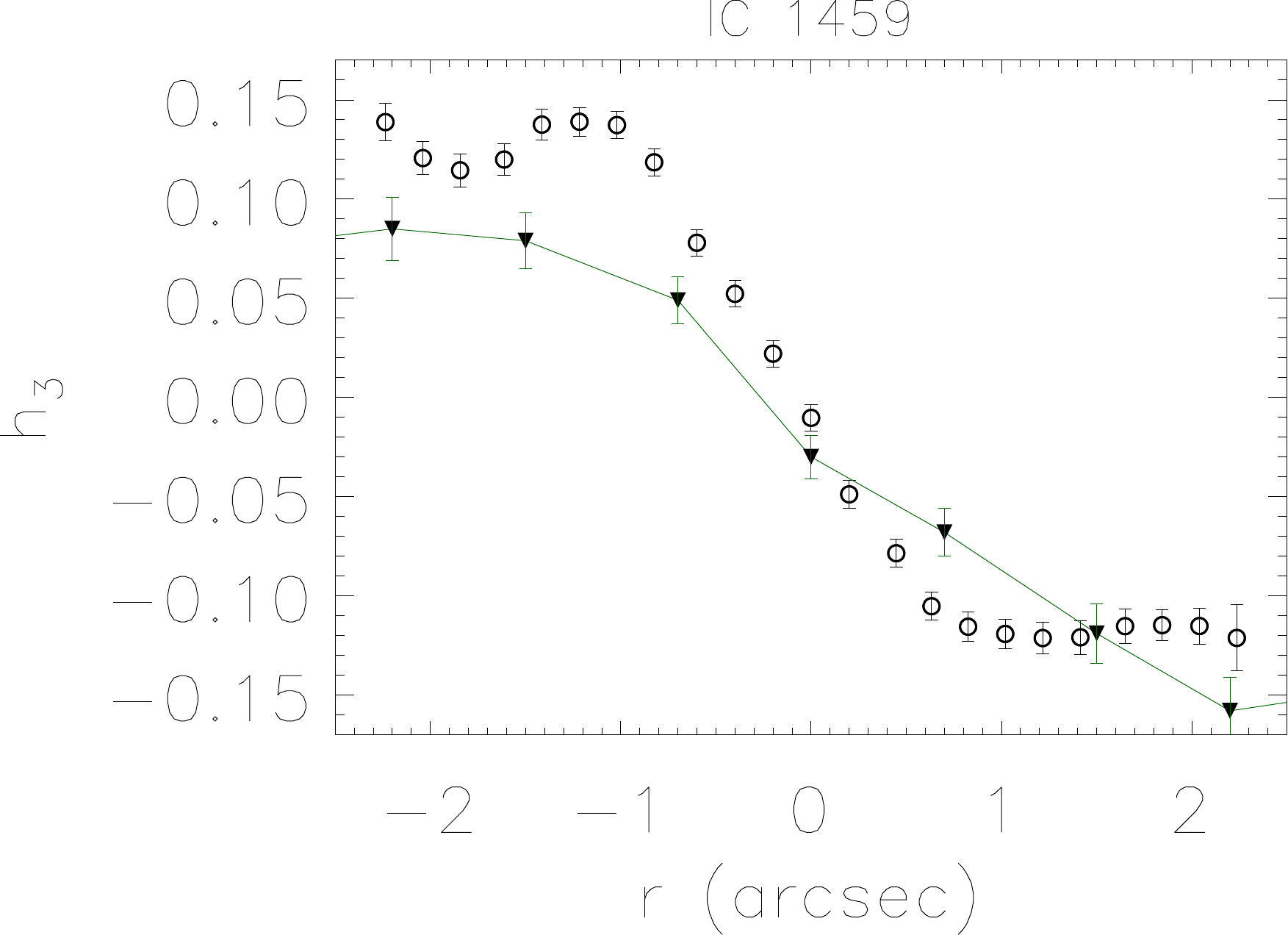}

\caption{Comparison of the GMOS results (hollow black circles) with long-slit results from the \textit{HST} (filled black squares with blue lines) and from ground-based spectroscopy (filled black triangles with green lines) for the galaxy IC 1459. All profiles were extracted from the major-axis of this object. Both long-slit results were taken from \citet{2002ApJ...578..787C}. For the $h_3$ moment, only ground-based results were available. \label{perfil_comparacao_ic1459}}

\end{center}

\end{figure*}

In order to check the consistency of our results, we compared the stellar kinematics extracted from the GMOS data cubes with other data sets presented in the literature. The kinematics of the galaxy IC 1459 were previously analysed by \citet{2002ApJ...578..787C} using long-slit data from the \textit{HST} and from ground-based spectroscopy. In Fig. \ref{perfil_comparacao_ic1459}, we show 1D profiles of the radial velocity, velocity dispersion and $h_3$ moment (only ground-based results) along the major-axis of IC 1459 extracted from the GMOS data cube and from both long-slit spectrographs. One may see that the results from GMOS seem to be consistent with the data presented by \citet{2002ApJ...578..787C}. The differences are probably caused by the fact that these measurements have distinct spatial resolutions.

\begin{figure*}

\begin{center}

\includegraphics[scale=0.30]{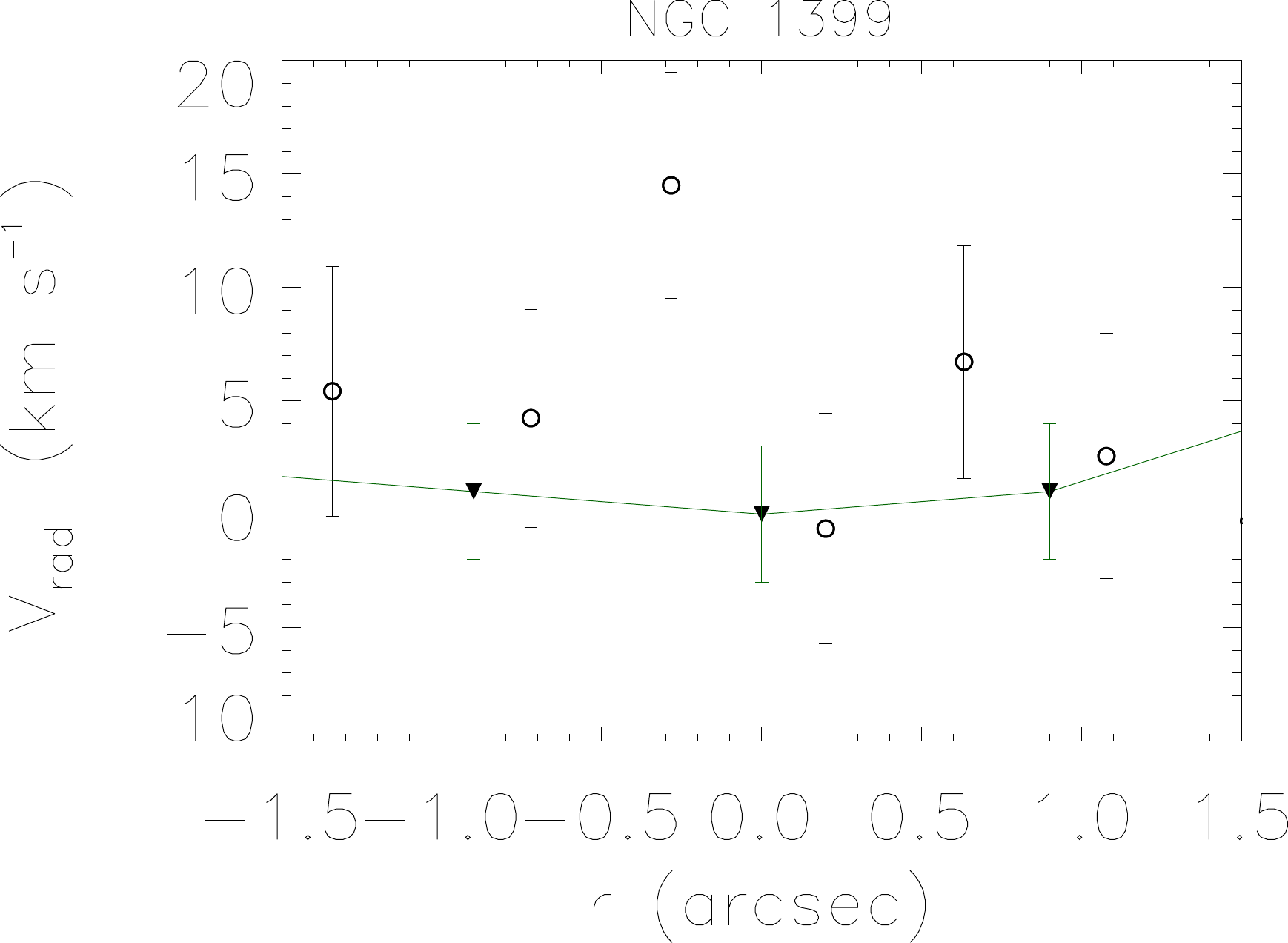}
\hspace{-0.0cm}
\vspace{0.3cm}
\includegraphics[scale=0.30]{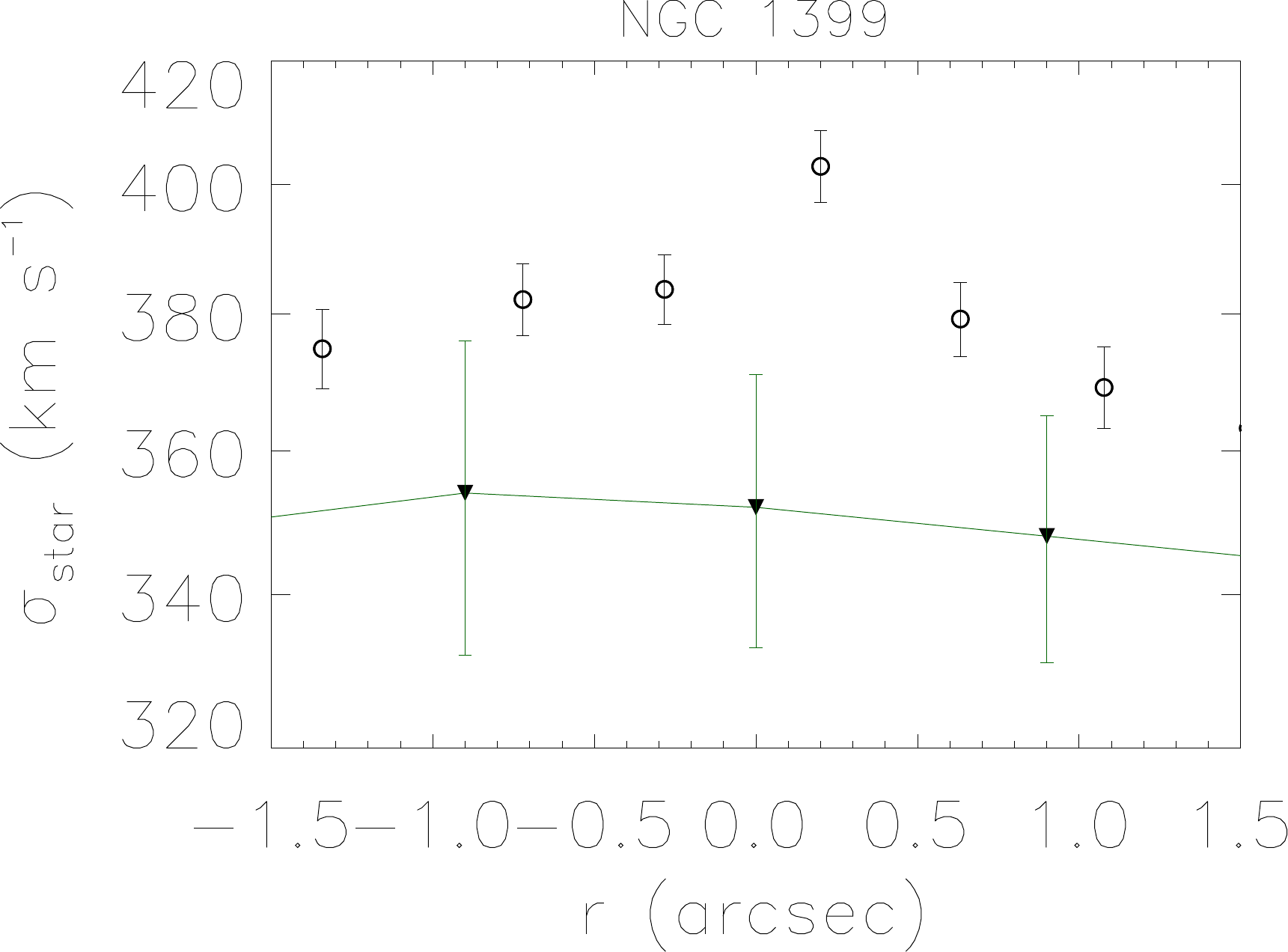}
\hspace{-0.0cm}

\includegraphics[scale=0.30]{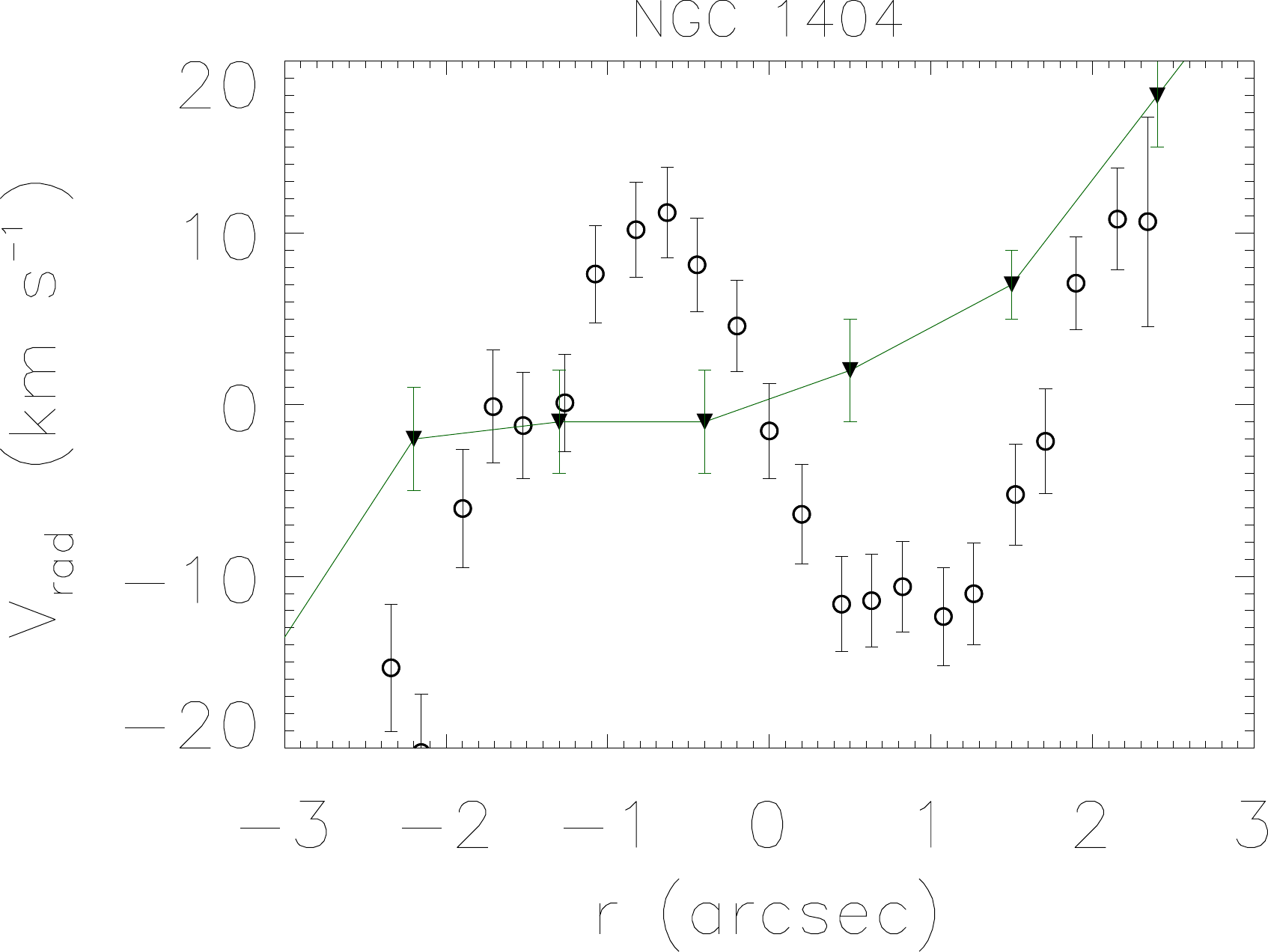}
\hspace{-0.0cm}
\vspace{0.3cm}
\includegraphics[scale=0.30]{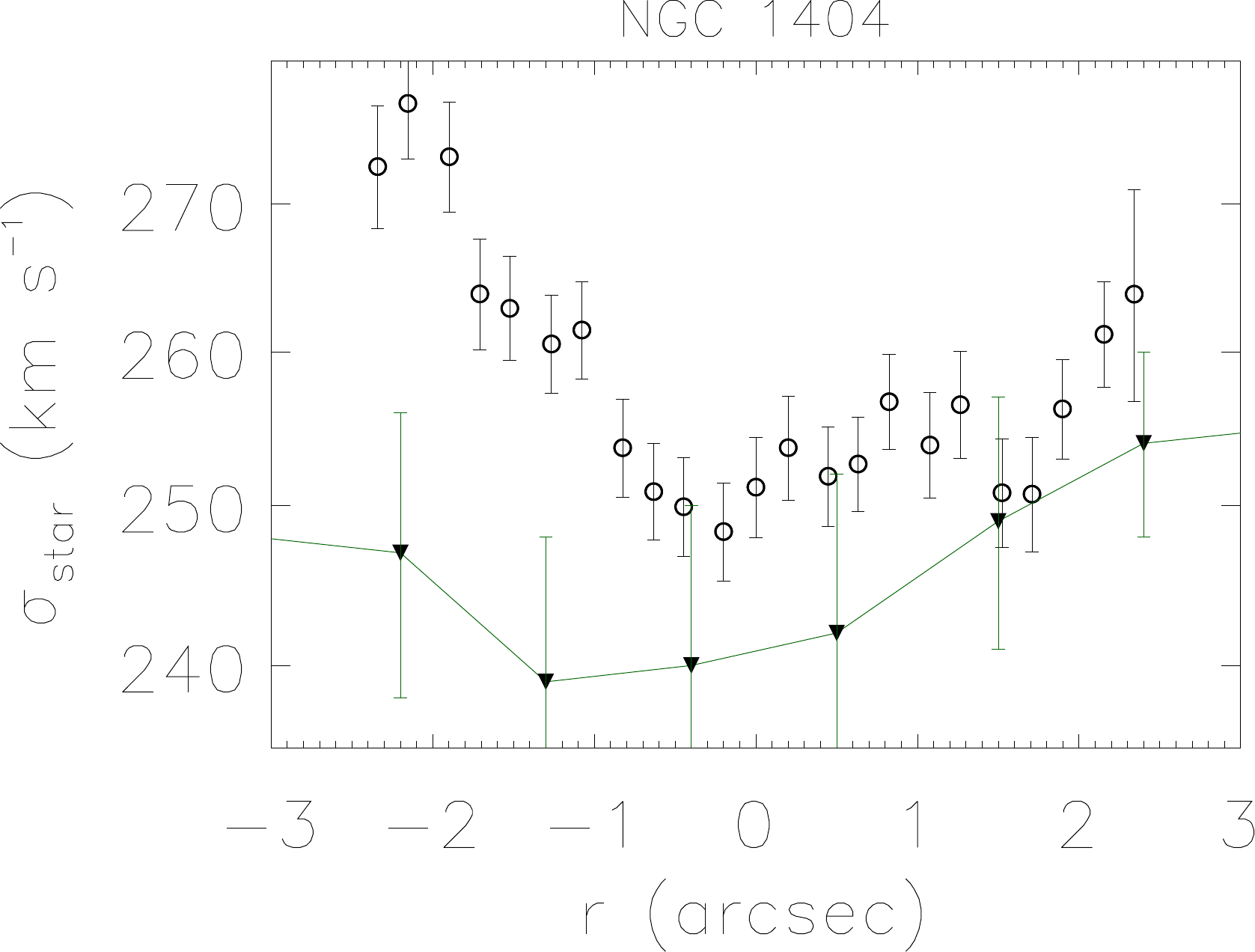}
\hspace{-0.0cm}
\caption{Comparison of the GMOS results (hollow black circles) with long-slit results from ground-based spectroscopy (filled black triangles with green lines) for the galaxies NGC 1399 and NGC 1404. All profiles were extracted from the major-axis of this object. Both long-slit results were taken from \citet{1998A&AS..133..325G}   \label{perfil_comp_fornax}}

\end{center}

\end{figure*}

For NGC 1399 and NGC 1404, we obtained ground-based long-slit data from \citet{1998A&AS..133..325G} along the major-axis of these objects. Fig. \ref{perfil_comp_fornax} shows the 1D profiles of the kinematic parameters extracted from GMOS and from the long-slit data. Again, one may see a consistency between the GMOS and long-slit measurements. Moreover, the GMOS results of both galaxies match the stellar kinematics IFU data presented by \citet{2014MNRAS.441..274S}. A direct comparison between Fig. \ref{perfil_comp_fornax} and fig. A2 from \citet{2014MNRAS.441..274S} confirms this statement. It is worth mentioning that both long-slit data and the IFU data from \citet{2014MNRAS.441..274S} do not have enough spatial resolution to resolve the KDC in NGC 1404. In fact, the differences between all results seem to be caused by distinct spatial resolutions among the data sets.

\subsubsection{Comparison with SAURON data cubes} \label{comp_sauron}

\begin{figure*}

\begin{center}

\includegraphics[scale=0.30]{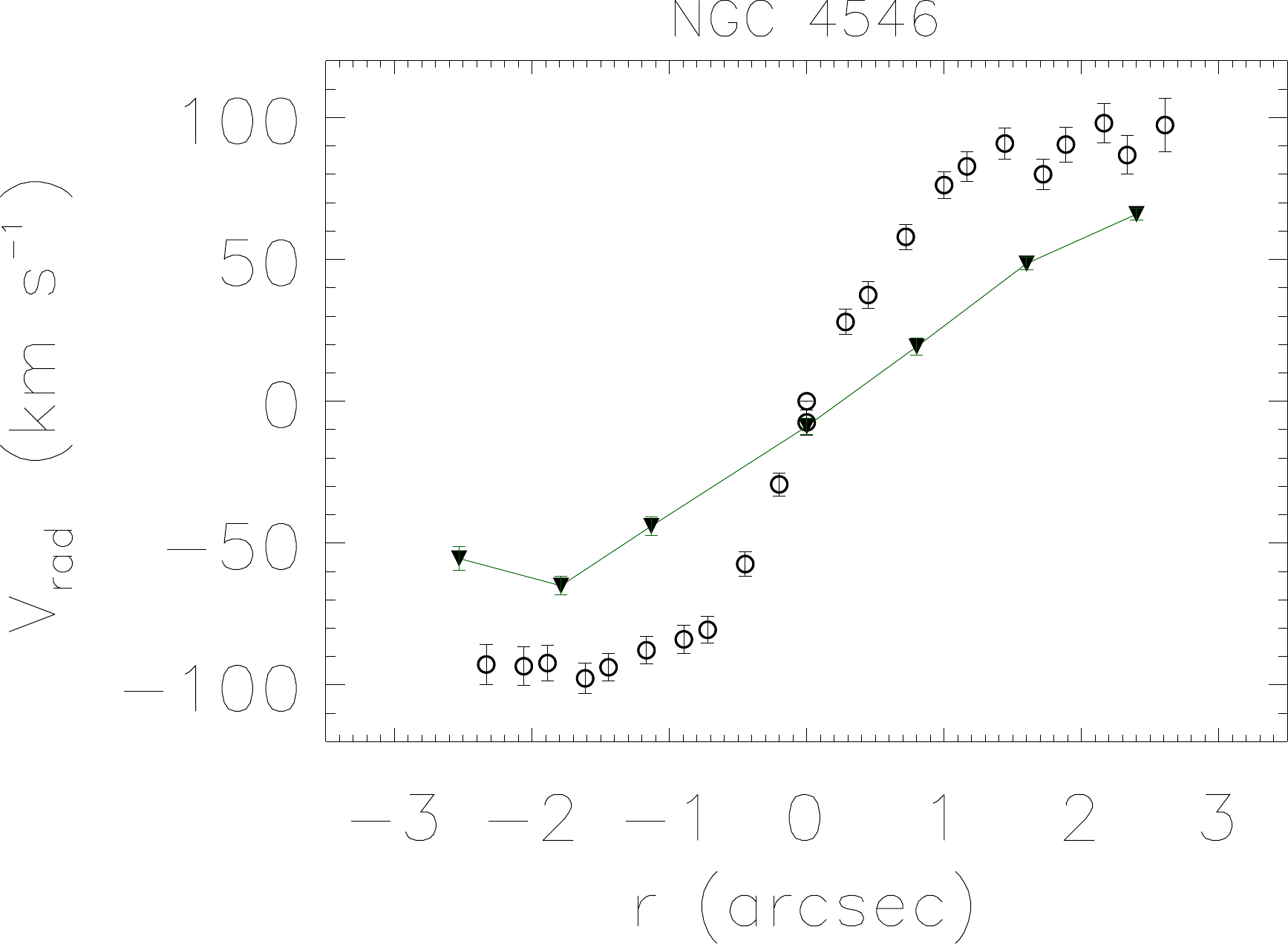}
\hspace{-0.0cm}
\vspace{0.3cm}
\includegraphics[scale=0.30]{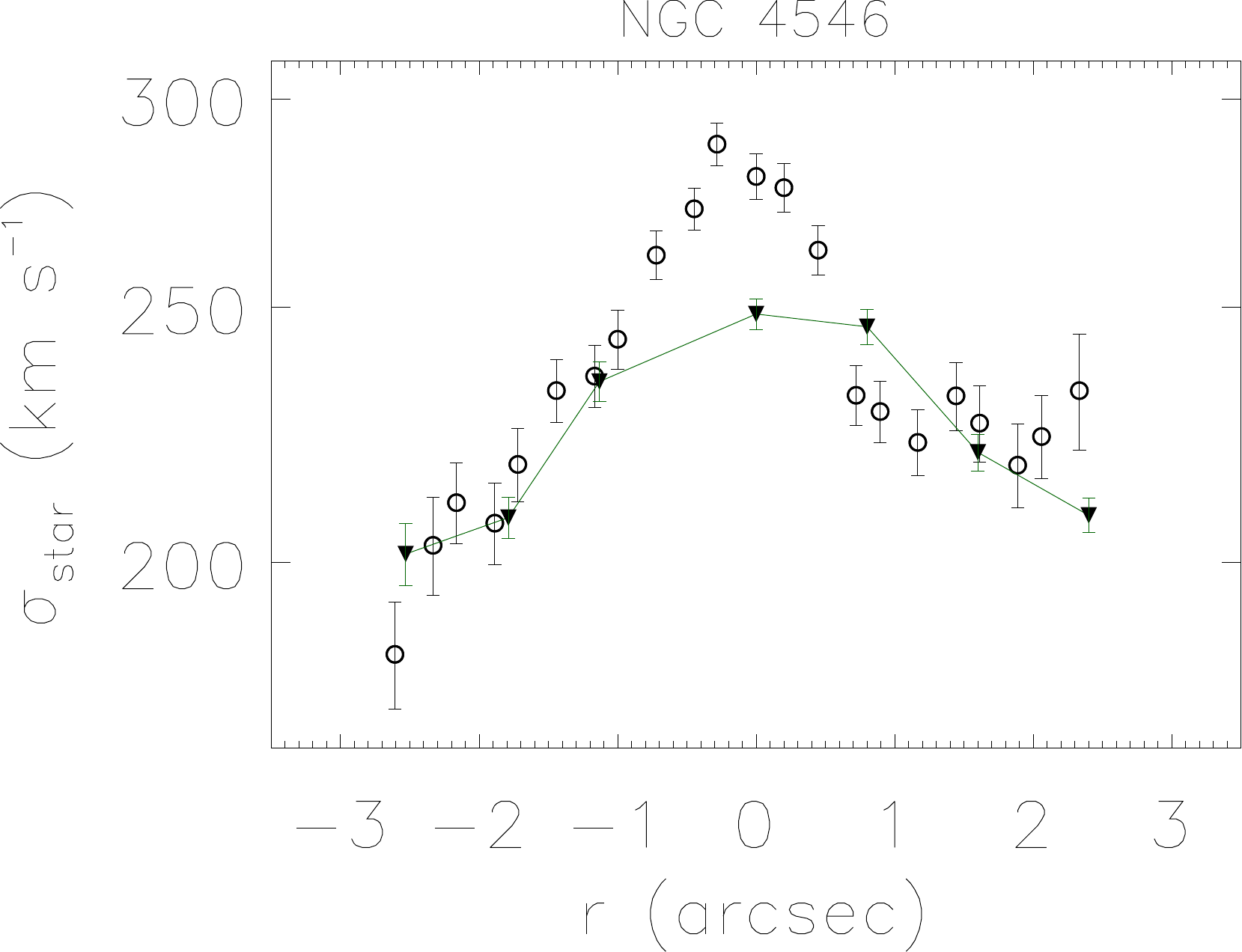}
\hspace{-0.0cm}
\includegraphics[scale=0.30]{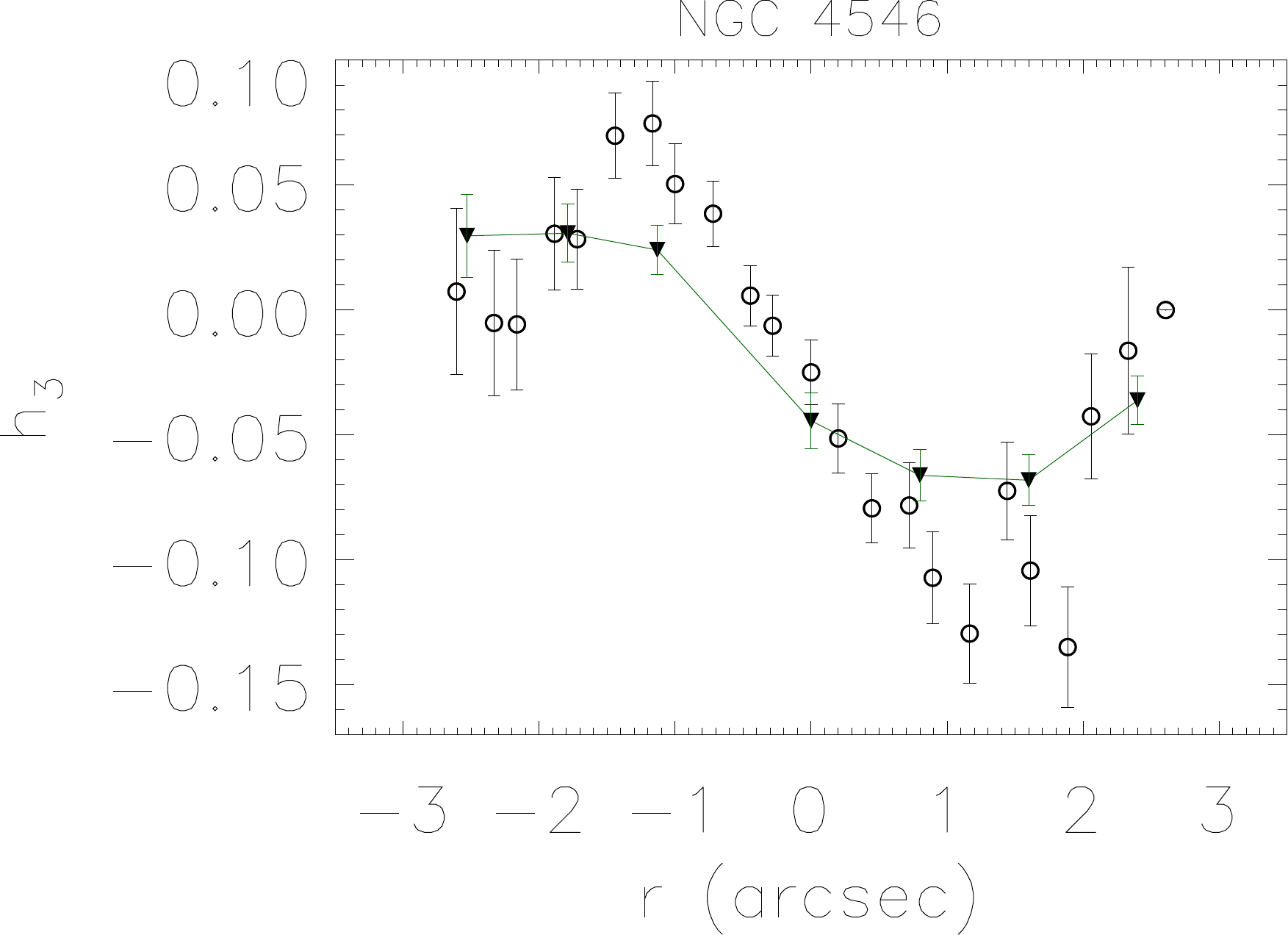}

\caption{Comparison of the GMOS results (hollow black circles) with results extracted from the SAURON data cube (filled black triangles with green lines) for the galaxy NGC 4546. In this case, both GMOS and SAURON results were extracted using exactly the same procedures described in Section \ref{ppxf_meth}. \label{perfil_comparacao_atlas}}

\end{center}

\end{figure*}

The galaxy NGC 4546 is also part of the SAURON sample and of the ATLAS\textsuperscript{3D} project. So, we decided to analyse their data cube of this object \citep{2011MNRAS.413..813C}\footnote{The data cube was downloaded from the ATLAS\textsuperscript{3D} project website at http://www-astro.physics.ox.ac.uk/atlas3d/} mainly to compare the circumnuclear velocity dispersion measurements. We built kinematic maps for this data cube using exactly the same procedures described in Section \ref{ppxf_meth}. We assured that the kinematic centres of both data cubes have the same radial velocity. In Fig. \ref{perfil_comparacao_atlas}, we compare the 1D profile of the kinematic maps taken from the GMOS data cubes and from the SAURON data cubes along the direction of the stellar rotations. We used the same PA from Table \ref{tab_mapa_vel_radial_stellar} for this extraction. Again, the differences that are seen in the 1D profiles are probably caused by the fact that the data sets do not have the same spatial resolution. Indeed, the 1D profile of the velocity dispersion shows that both measurements agree for $r$ $>$ 1.0 arcsec, where the effects of the seeing of the GMOS data cube are less pronounced. Also the $h_3$ results seem to reasonably match each other. 

We may also compare if the values measured for $\sigma_{*}$ agree for both data cubes. For the SAURON data cube, we found $\sigma_{*}$ = 241$\pm$12 km s$^{-1}$ (assuming that the systematic error for the SAURON data cubes is $\sim$ 5\%, as stated by \citealt{2006MNRAS.366.1126C}), while we found $\sigma_{*}$ = 254$\pm$1 km s$^{-1}$ using the GMOS data cube. Although both measurements agree within the errors, we explore some possibilities to explain this difference between them: since the spatial sampling of data cubes from the ATLAS\textsuperscript{3D} project is 0.8 arcsec, we resampled the GMOS data cubes to the same spaxel size. We built kinematic maps using such a configuration and we obtained $\sigma_{*}$ = 253$\pm$1 km s$^{-1}$. Also the spectral resolutions of both data cubes are different ($\sigma_{inst}$ = 30 km s$^{-1}$ in $\lambda$ = 5577\AA\ for this GMOS data cube and, from \citealt{2004MNRAS.352..721E}, $\sigma_{inst}$ = 108 km s$^{-1}$ for the SAURON data cube). So, we modified the spectral resolution of the GMOS data cubes to $\sigma_{inst}$ = 108 km s$^{-1}$, but maintaining the spaxels with sizes of 0.2 arcsec. With this configuration, $\sigma_{*}$ = 243$\pm$1 km s$^{-1}$. Using spaxels with sizes of 0.8 arcsec and a spectral resolution $\sigma_{inst}$ = 108 km s$^{-1}$ for the GMOS data cubes, we calculated $\sigma_{*}$ = 257$\pm$1 km s$^{-1}$. Although the values of $\sigma_{*}$ for the GMOS data cube with spaxels of 0.2 arcsec and $\sigma_{inst}$ = 108 km s$^{-1}$ and for the SAURON data cube match each other, the spectral resolution does not seem to be, alone, the cause of the systematic differences for the GMOS data cubes. The point here is that the kinematic maps for the GMOS data cubes with such a configuration is noisier than the same results with any other configuration. Moreover, the GMOS data cube with exactly the same instrumental configuration as the SAURON data cube provides $\sigma_{*}$ = 257$\pm$1 km s$^{-1}$, which is almost the same result found for the original GMOS configuration. Another parameter is the seeing of the observation, but it does not seem to affect the measurements made with GMOS. We found $\sigma_{*}$ = 251$\pm$1 km s$^{-1}$ using the GMOS data cube in its original configuration, but with a FWHM of the PSF degraded to 1.4 arcsec (the original PSF has a FWHM = 0.71 arcsec). 

One possible source of systematic errors in the $\sigma_{*}$ measurements using GMOS data cubes is the wavelength range where {\sc ppxf} is applied. To test this, we fitted each spectrum of the GMOS data cube of NGC 4546 with the same wavelength range as the SAURON data cube, which is from $\sim$ 4800 to $\sim$ 5250\AA. We found $\sigma_{*}$ = 243$\pm$1 km s$^{-1}$. We also applied {\sc ppxf} from $\sim$ 5250 to $\sim$ 6100\AA. For this setup, we obtained $\sigma_{*}$ = 223$\pm$1 km s$^{-1}$. These differences between the values may come from the fact that GMOS observations made with the R831-G5322 grating is noisier in the blue region of the spectra. Also, the spectral resolution is constant in wavelength but not in velocity. It seems that, depending on the wavelength range, the values for the $\sigma_{*}$ parameter may vary by $\sim$ 10\%, at least for NGC 4546. For IC 1459, for example, which was observed with the B600-G5323 grating (that provides a better response on the blue region of the spectra), we obtained $\sigma_{*}$ = 319$\pm$1 km s$^{-1}$ for the $\sim$ 4800\AA\ to $\sim$ 5250\AA\ range and 322$\pm$1 km s$^{-1}$ for the $\sim$ 5250\AA\ to $\sim$ 6100\AA. Both measurements are very close to $\sigma_{*}$ = 325$\pm$1 km s$^{-1}$, obtained when we fitted all the wavelength range of the data cube. It is worth mentioning that only these two gratings were used in all sample galaxies (see paper I). A systematic investigation about this issue will be revisited in a future paper, but, for now, we suggest that the errors caused by applying {\sc ppxf} to different wavelength ranges may be as large as 10\%.

\section{The case of NGC 1404} \label{ngc1404_case}

NGC 1404 was classified as a fast rotator by \citet{2014MNRAS.441..274S}. They used an IFU with a spaxel of 1 arcsec$^2$ and a FOV of 25 $\times$ 28 arcsec$^2$. By comparing our results with those obtained by \citet{2014MNRAS.441..274S}, we conclude that NGC 1404 has a KDC with an extension of $\sim$ 200 pc that is counter-rotating with respect to the external structure of the galaxy. Indeed, the presence of a decoupled structure in NGC 1404 was suggested by \citet{2014MNRAS.441..274S} due to the detection of two peaks in their velocity dispersion map. \citet{2008IAUS..245..303G} have also reported this counter-rotating KDC. 

\citet{2006MNRAS.373..906M} found KDCs with diameters of about one-hundred pc in some ETGs. These features were detected in fast rotating galaxies and are counter-rotating with respect to their external stellar components. According to these researchers, the host galaxies of these compact KDCs are among the youngest ETGs of the SAURON sample. Moreover, these KDCs are usually (but not always) younger than the overall population of their host galaxies. Although this result might suggest a link between counter-rotation and enhanced star formation, \citet{2006MNRAS.373..906M} showed that it is more likely to detect compact KDCs in young galaxies due to the effect of luminosity-weighting. By means of a simple model, they demonstrated that when a KDC of a galaxy gets older, its mass-to-light ratio increases and it becomes harder to distinguish the KDC from the background rotation field. The results of their model revealed that KDCs older than $\sim$ 5 Gyr are barely seen in the radial velocity maps. In other words, it is easier to detect a younger KDC, but they may have a broad range of ages. Indeed, \citet{2006MNRAS.373..906M} showed in their fig. 16 that, among six galaxies with KDCs with intrinsic sizes of $\sim$ 100 pc, in five the central luminosity-weighted ages of these objects are between 0.5 and 4 Gyr. Only one galaxy has such a compact structure with a luminosity-weighted age of $\sim$ 18 Gyr.  Probably, NGC 1404 would be similar to these objects studied by \citet{2006MNRAS.373..906M}, since the estimated age of the central region of NGC 1404 is about 5 $-$ 8 Gyr \citep{2008IAUS..245..303G}.

We detected a $\sigma$-drop in the velocity dispersion map of NGC 1404. Moreover, the velocity dispersion profile rises more steeply along the direction of the stellar rotations. This may be a consequence of the superposition of the two stellar components in counter-rotation. Indeed, \citet{2007MNRAS.379..418C}, using three-integral Schwarzschild models, showed that such a configuration predicts two peaks in the velocity dispersion along the rotation of both stellar components. This result for NGC 1404 resembles that of the 2$\sigma$ galaxies (those objects that shows two off-centred peaks in the velocity dispersion maps along the major axis of the galaxy), as defined by \citet{2011MNRAS.414.2923K}, although it is worth mentioning that the definition of such a class of objects given by these authors claims that the separation of the peaks of the maximum values of the velocity dispersion should be, at least, half of the effective radius. However, the data from the ATLAS\textsuperscript{3D} project, analysed by \citet{2011MNRAS.414.2923K}, does not have enough spatial resolution to detect velocity dispersion peaks with a separation of a few arcsec. Thus, a systematic study with seeing-limited data would be of great value in order to better understand objects with the same characteristics.

Another hypothesis is that the $\sigma$-drop is not related to the KDC. In this case, we may speculate the existence a star-forming region that is kinematically cold and dominates the stellar light within a projected radius smaller than the PSF of the data cube of this object ($\sim$ 1 arcsec, see paper I). According to \citet{2006MNRAS.369..853W}, kinematically cold gas is able to form stars in the centre of barred late-type galaxies. They showed that a star formation rate of $\sim$ 1 M$_\odot yr^{-1}$ is enough to maintain kinematically cold stars, and thus a $\sigma$-drop. The reason is that, although there are mechanisms that make stars kinematically hot (e.g., gravitational disturbances), the nuclear region is fueled again with kinematically cold stars. However, NGC 1404 is a non-barred ETG. One possible scenario suggested by \citet{2006MNRAS.369..853W} is that the star formation rate is quite low, possibly undetectable. In this case, the $\sigma$-drop may fade away, but over a long time-scale ($\sim$ 100 Myr). Therefore, an in-depth study of the stellar populations of NGC 1404 is necessary to verify this hypothesis (\citealt{2008IAUS..245..303G} do not give much detail of their analysis of the stellar populations of NGC 1404). In addition, a detailed modelling of the stellar kinematics would be interesting, since it may confirm if this is either related to a kinematically cold component or associated with the KDC. 

\section{Summary and Conclusions} \label{conclusions_section}

Stellar kinematics may provide information about the formation history of galaxies. Fast and slow rotators, for example, have different formation paths \citep{2014MNRAS.444.3357N}. However, to assert a global characteristic for the stellar kinematics, data cubes with a FOV that englobes, at least, the region within one effective radius of the galaxies are necessary. The problem is that such data cubes have a spatial resolution that is limited by the size of the fibre (diameters of $\sim$ 2 arcsec). Observations using seeing-limited IFU data are, thus, very important in order to access information about the circumnuclear (scales of 100 pc) properties of the stellar kinematics. There are features that are only seen in such scales, as in the case of NGC 1404 [but see also \citet{2006MNRAS.373..906M}]. The point here is that an analysis with seeing-limited data cubes may provide complementary information about the stellar kinematics of a galaxy as a whole. Another interesting issue is that, with such data cubes, one may look for off-centred AGNs in a reliable way. Comparing the position of the kinematic centre with the position of the AGNs and also with the position of the photometric centre of the stellar structure may reveal that the SMBH is not always at the centre of a galaxy (see e.g. \citealt{2014ApJ...796L..13M}). Thus, the main subject of the present paper is to provide stellar kinematics results from seeing-limited data cubes for the sample galaxies. A more complete analysis in order to search for general characteristics of the stellar kinematics from the circumnuclear region of massive ETGs demands more data and is not possible with the present sample.

Below we present the main findings of the present paper.

\begin{itemize}
	\item We detected stellar structures in rotation embedded in a kinematically hot stellar structure (spheroid) in the circumnuclear region of seven galaxies of the sample. The other three objects do not seem to possess any sign of stellar rotation within a radius of 100 pc from the nucleus.

	\item The galaxy IC 5181 seems to have a small variation of the PA of the stellar structure in rotation with radius. This may be a consequence of a non-axisymmetric potential within the central region of the galaxy. It is worth mentioning that an evidence of a non-axisymmetric potential for this galaxy is also seen in the gas kinematics (paper III).

	\item In six galaxies, we detected a nuclear peak in the velocity dispersion maps. A fraction of this peak in four galaxies is likely to be associated with the stellar rotation in a region that is not fully resolved. In NGC 1380, NGC 2663, NGC 3136, and NGC 7097, the velocity dispersion maps are more complex and require a careful analysis for a reliable interpretation. For NGC 1399, a nuclear peak is also seen, but no stellar rotation was detected. Also here, a proper modelling is needed to explain this result.
		
	\item NGC 1404 has a counter-rotating KDC with an extension of $\sim$ 200 pc. Moreover, we detected a $\sigma$-drop in the nuclear region. One possibility is that it is related to both stellar components in counterrotation, with a configuration that is very similar to the 2$\sigma$ galaxies. Another hypothesis is that this $\sigma$-drop is associated with kinematically cold stars being formed at a very low star formation rate ($<$ 1 M$_\odot yr^{-1}$). If this is the case, then the $\sigma$-drop is not related to the kinematics of the KDC.

\end{itemize}

\section*{Acknowledgements}

This paper is based on observations obtained at the Gemini Observatory, which is operated by the Association of Universities for Research in Astronomy, Inc., under a cooperative agreement with the NSF on behalf of the Gemini partnership: the National Science Foundation (United States), the National Research Council (Canada), CONICYT (Chile), the Australian Research Council (Australia), Minist\'{e}rio da Ci\^{e}ncia, Tecnologia e Inova\c{c}\~{a}o (Brazil) and Ministerio de Ciencia, Tecnolog\'{i}a e Innovaci\'{o}n Productiva (Argentina). {\sc iraf} is distributed by the National Optical Astronomy Observatory, which is operated by the Association of Universities for Research in Astronomy (AURA) under cooperative agreement with the National Science Foundation. We also acknowledge the usage of the HyperLeda data base (http://leda.univ-lyon1.fr). This research has made use of the NASA/IPAC Extragalactic Data base (NED), which is operated by the Jet Propulsion Laboratory, California Institute of Technology, under contract with the National Aeronautics and Space Administration.

T.V.R, J.E.S and R.B.M. also acknowledge Funda\c{c}\~{a}o de Amparo \`a Pesquisa do Estado de S\~{a}o Paulo (FAPESP) for the financial support under grants 2008/06988-0 (T.V.R.), 2012/21350-7 (T.V.R.), 2011/51680-6 (J.E.S.) and 2012/02262-8 (R.B.M.). We also thank the referee Dr. Eric Emsellem for valuable suggestions that improved the quality of this paper. 

\bibliographystyle{mn2e}
\bibliography{bibliografia}

\appendix

\section{Examples of the fitting procudure done with {\sc ppxf}} \label{starlight_results}

In this appendix, we present some results obtained with {\sc ppxf}. The idea is to show the behaviour of the fitting procedure in spectra with both high and low signal-to-noise ratios. In Fig. \ref{starlight_graf_1}, we present the spectrum, the fitted result and the residual of a representative spaxel of the galaxies' centre and of a spaxel located at the upper-left region of the FOVs, $\sim$ 2 arcsec away from the galaxies' centre. It is worth mentioning that all emission lines was previously known (papers II and III) and they were all masked in the fitting procedure. We also masked the Na I$\lambda\lambda$5890,5896 absorption lines, since they are affected by the interstellar medium of the objects, and the spectral regions that are affected by the gaps between the three CCDs that are installed in the GMOS.

\begin{figure*}
\includegraphics[width=70mm,height=55mm]{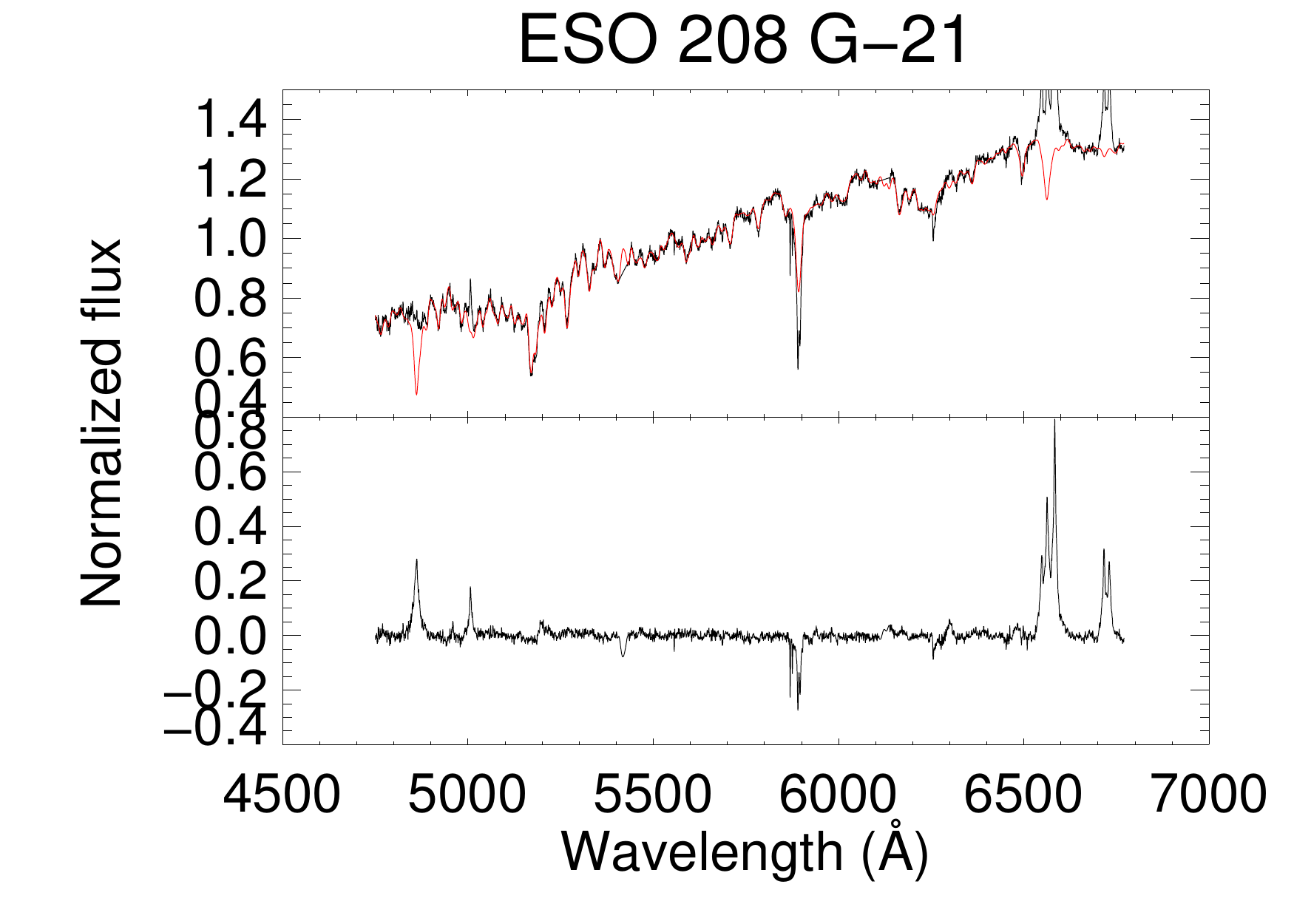}
\includegraphics[width=70mm,height=55mm]{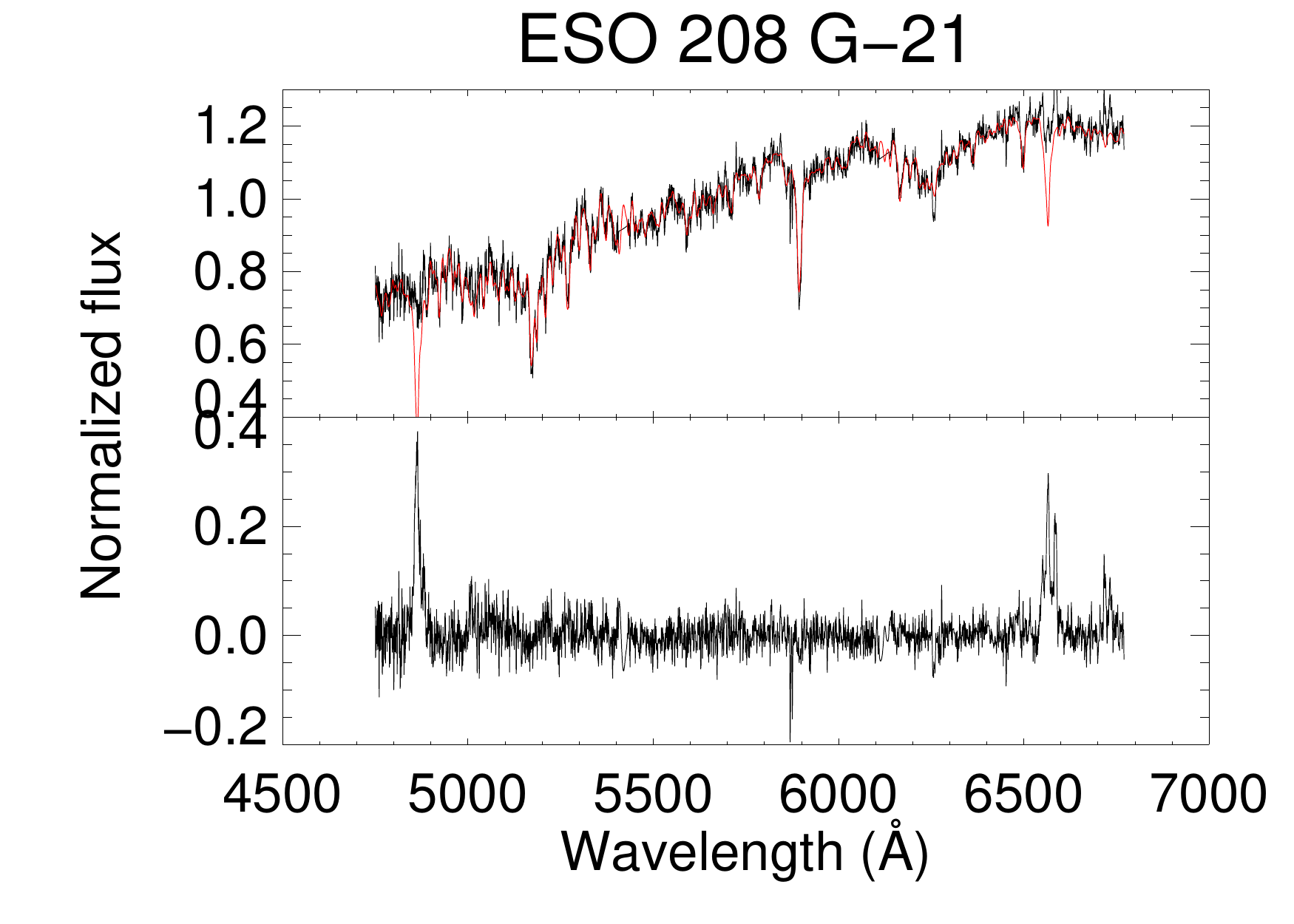}
\includegraphics[width=70mm,height=55mm]{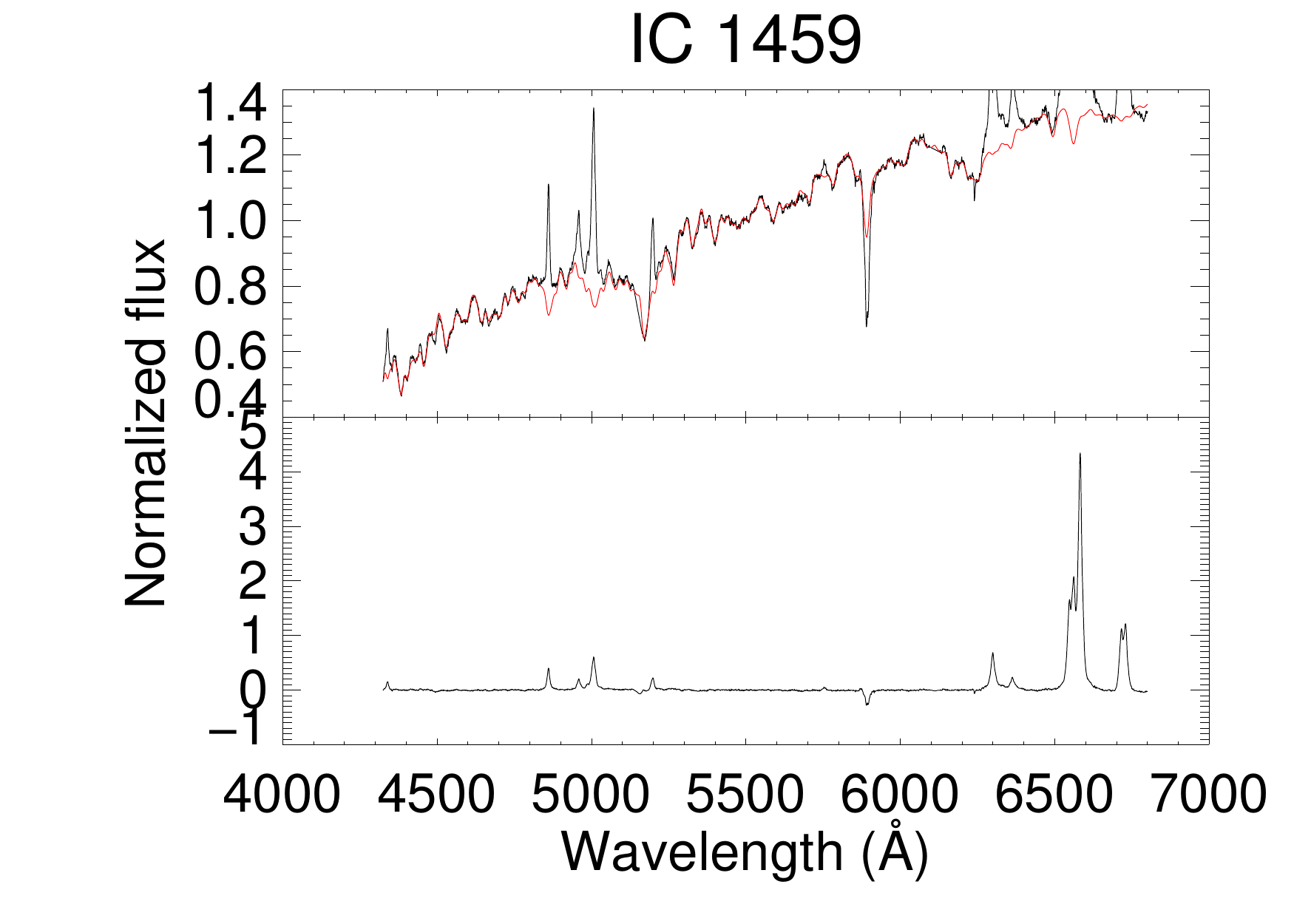}
\includegraphics[width=70mm,height=55mm]{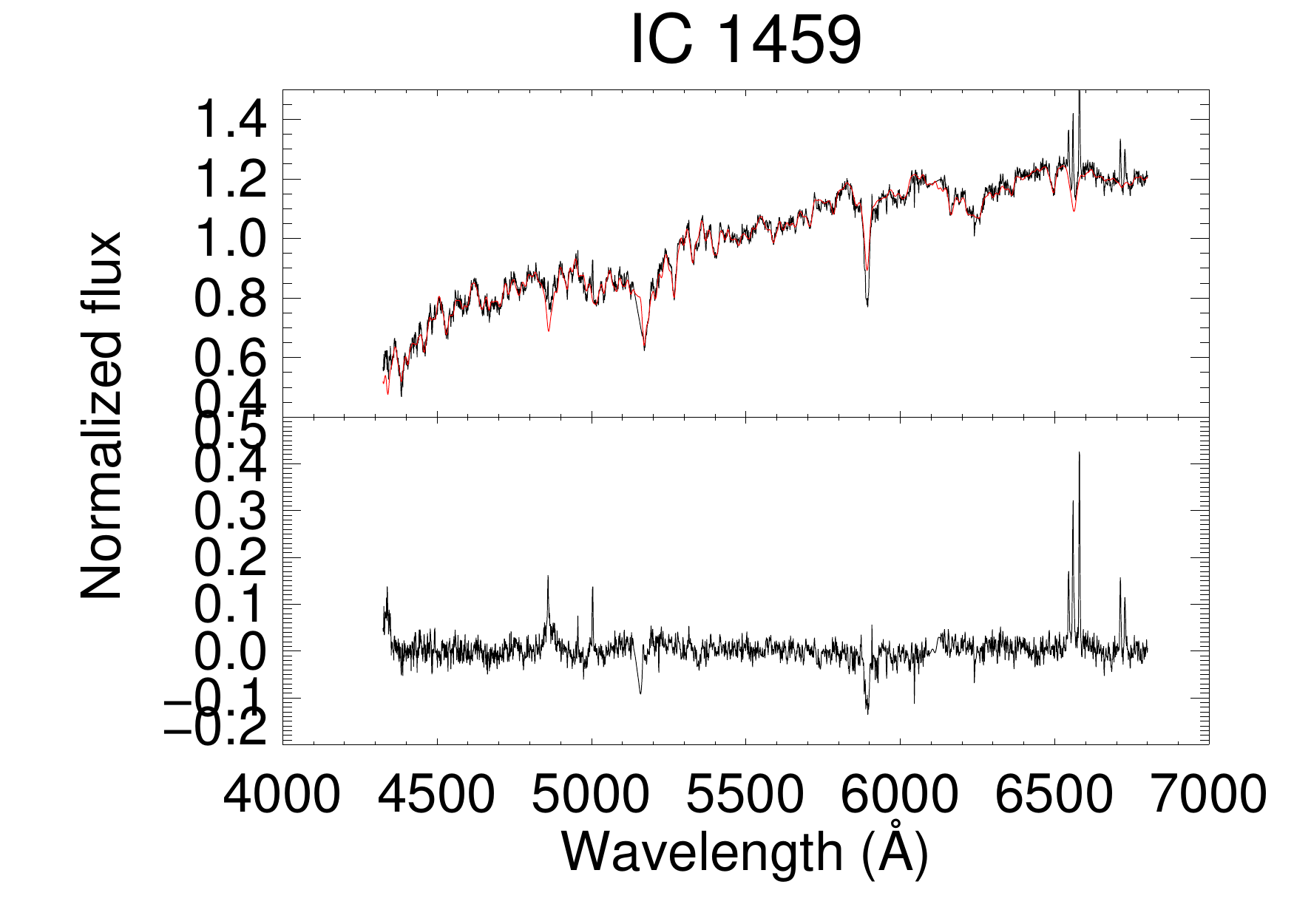}
\includegraphics[width=70mm,height=55mm]{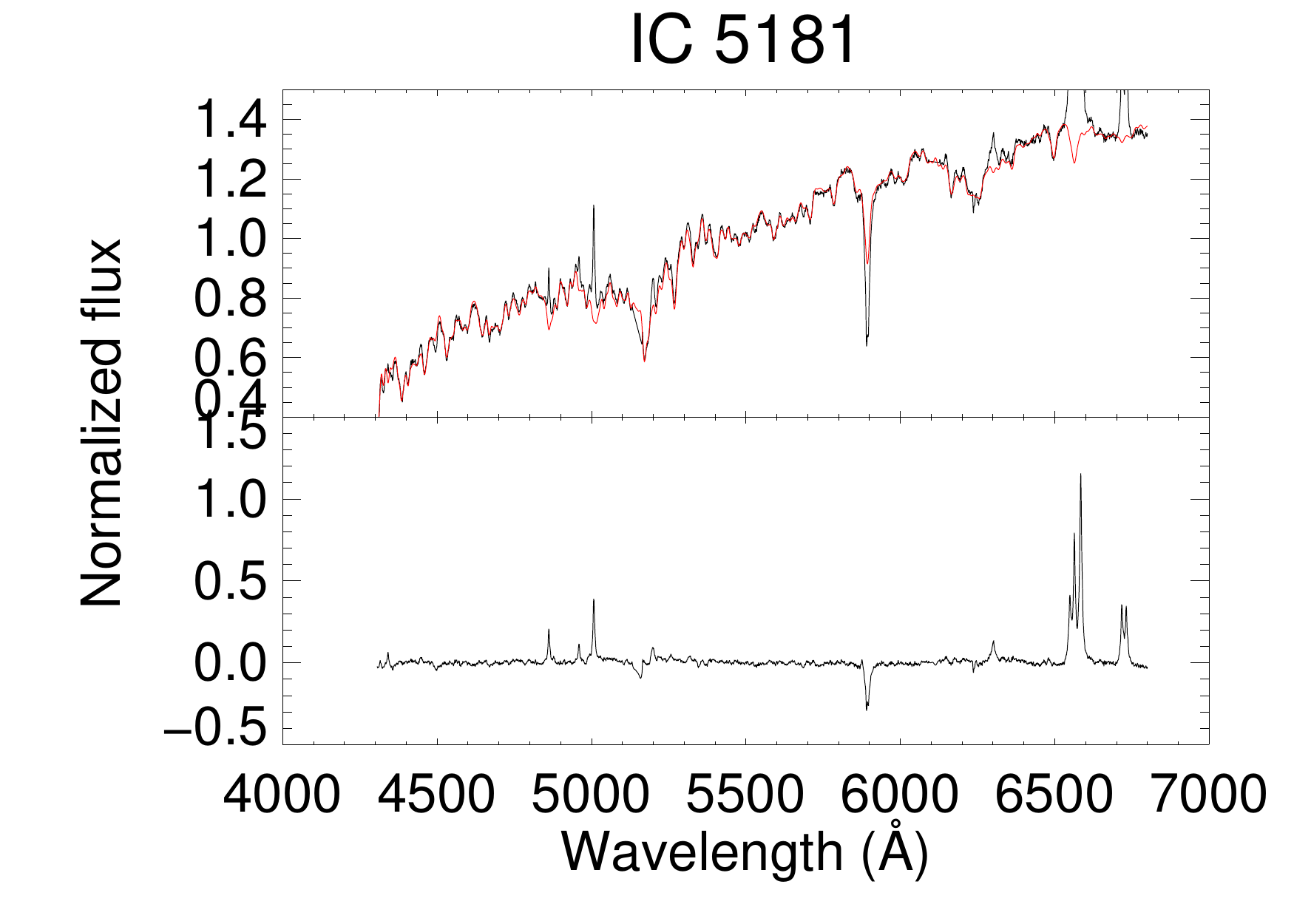}
\includegraphics[width=70mm,height=55mm]{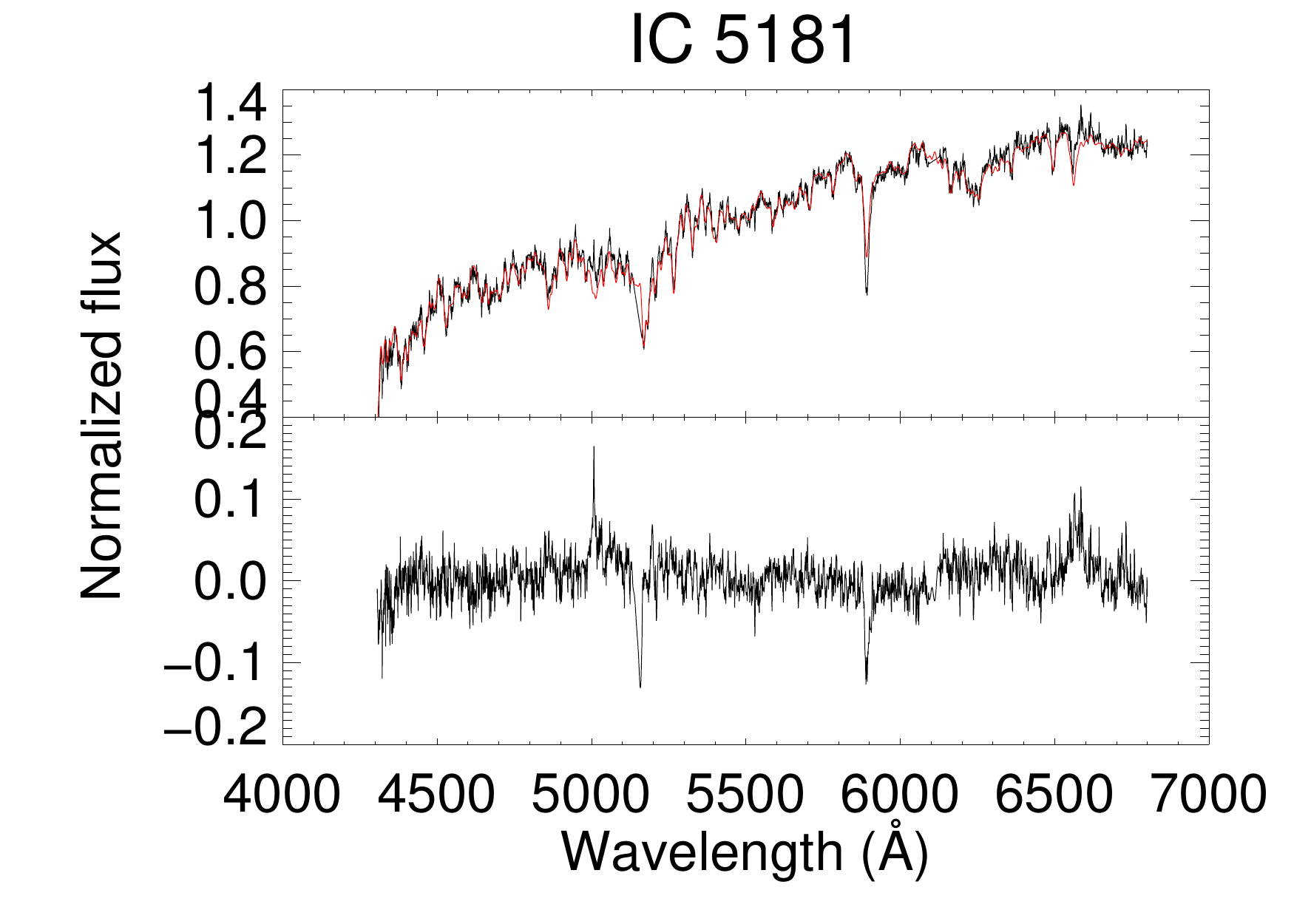}
\includegraphics[width=70mm,height=55mm]{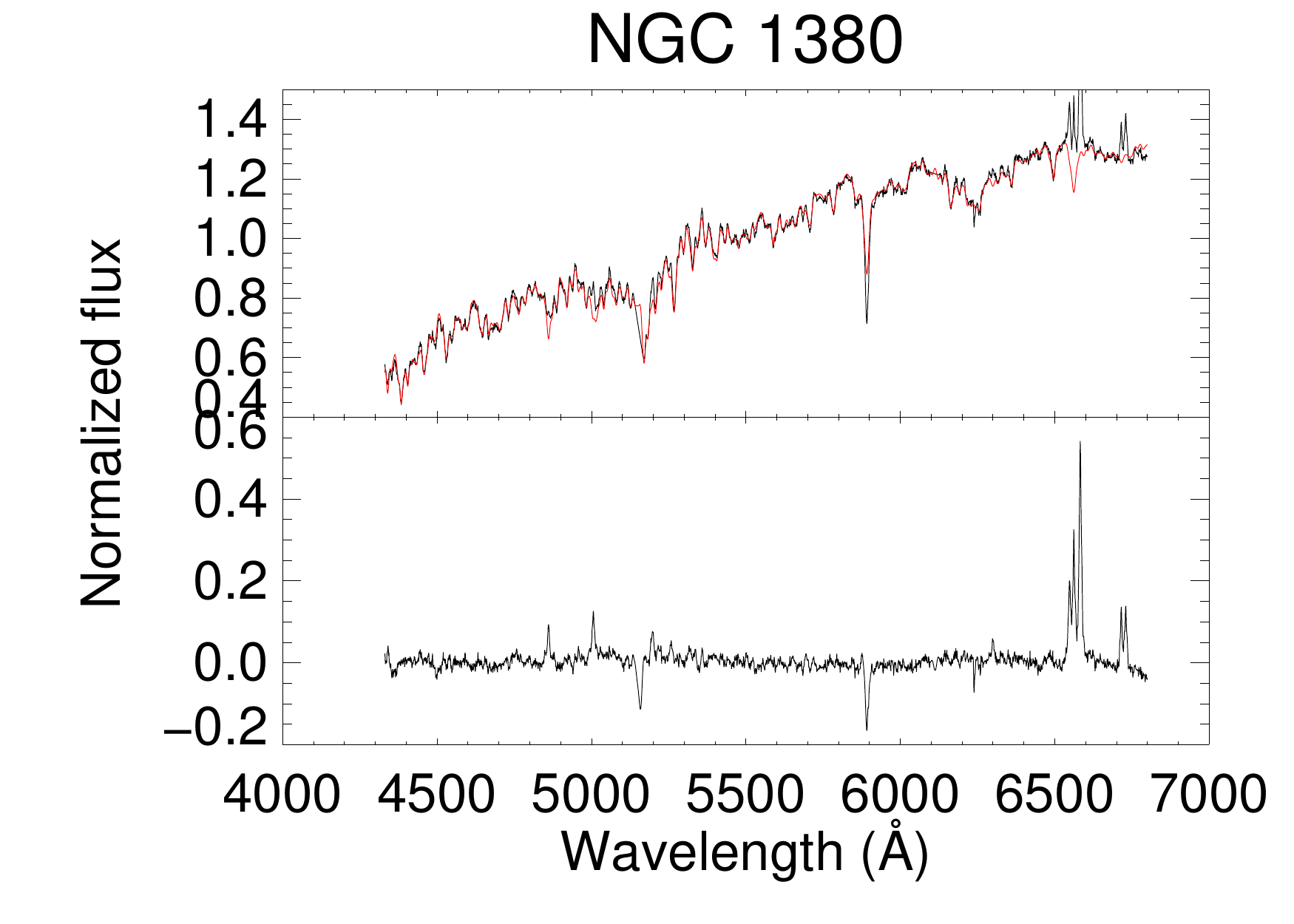}
\includegraphics[width=70mm,height=55mm]{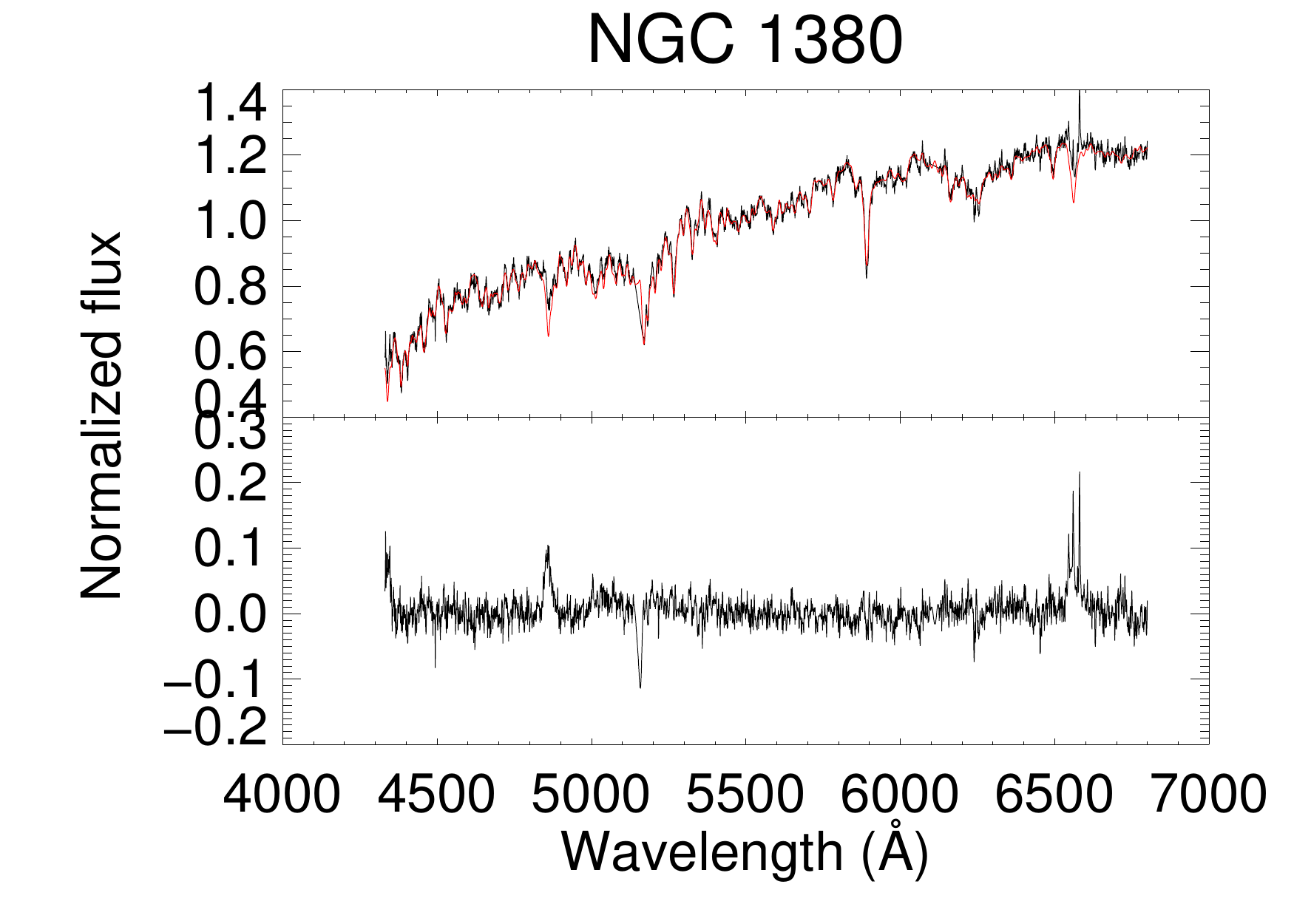}

\caption{Left: results obtained with {\sc ppxf} from a representative spaxel of the central region (high S/N ratio) of the sample galaxies. Right: results from a spaxel located at the upper-left position of the FOV, $\sim$ 2 arcsec away from the galaxies' centre (low S/N ratio). For each galaxy, we show the spectrum (black) and the fitted result (red) at the top and the residuals at the bottom. \label{starlight_graf_1}
}
\end{figure*}

\addtocounter{figure}{-1}

\begin{figure*}
\includegraphics[width=70mm,height=55mm]{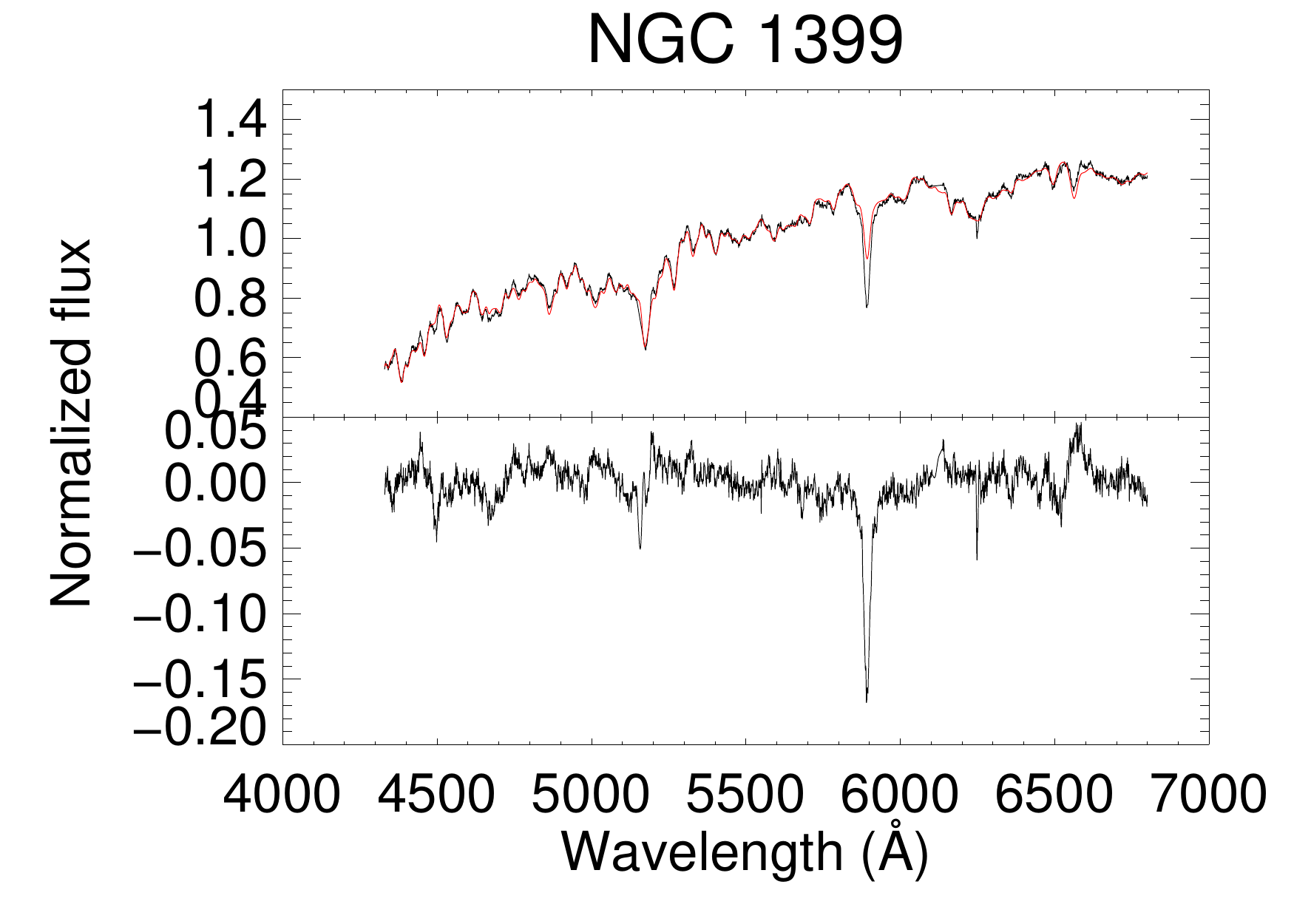}
\includegraphics[width=70mm,height=55mm]{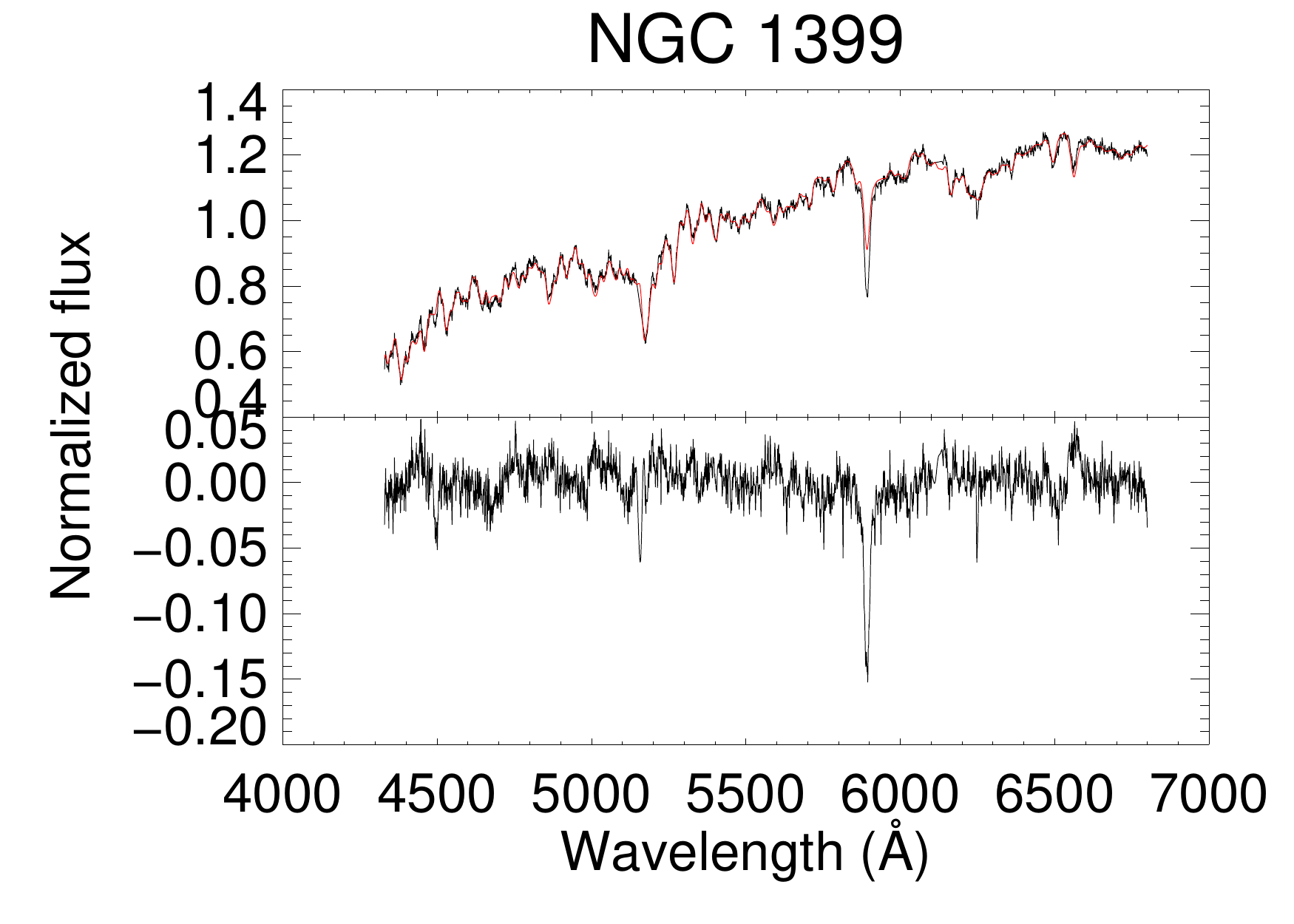}
\includegraphics[width=70mm,height=55mm]{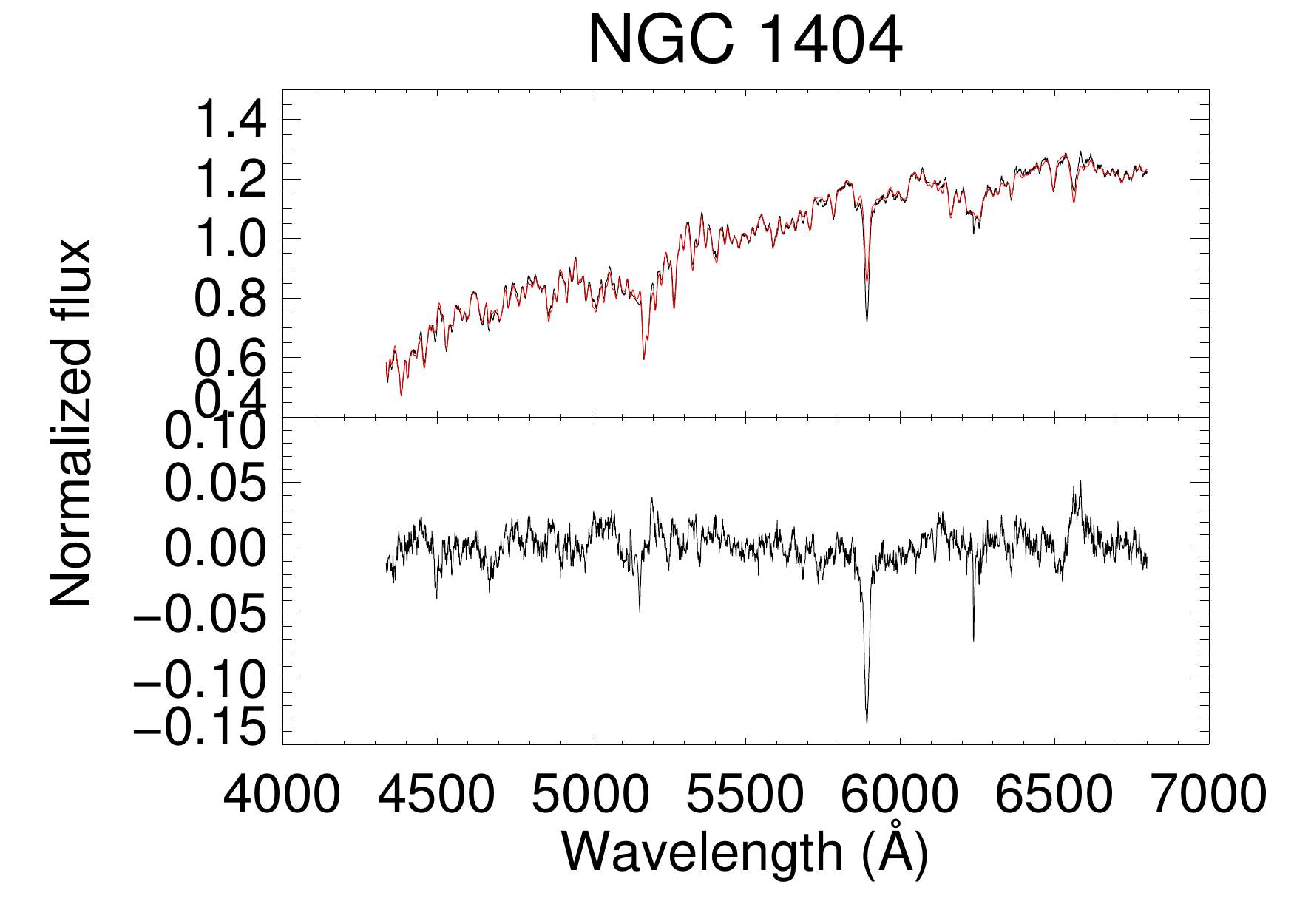}
\includegraphics[width=70mm,height=55mm]{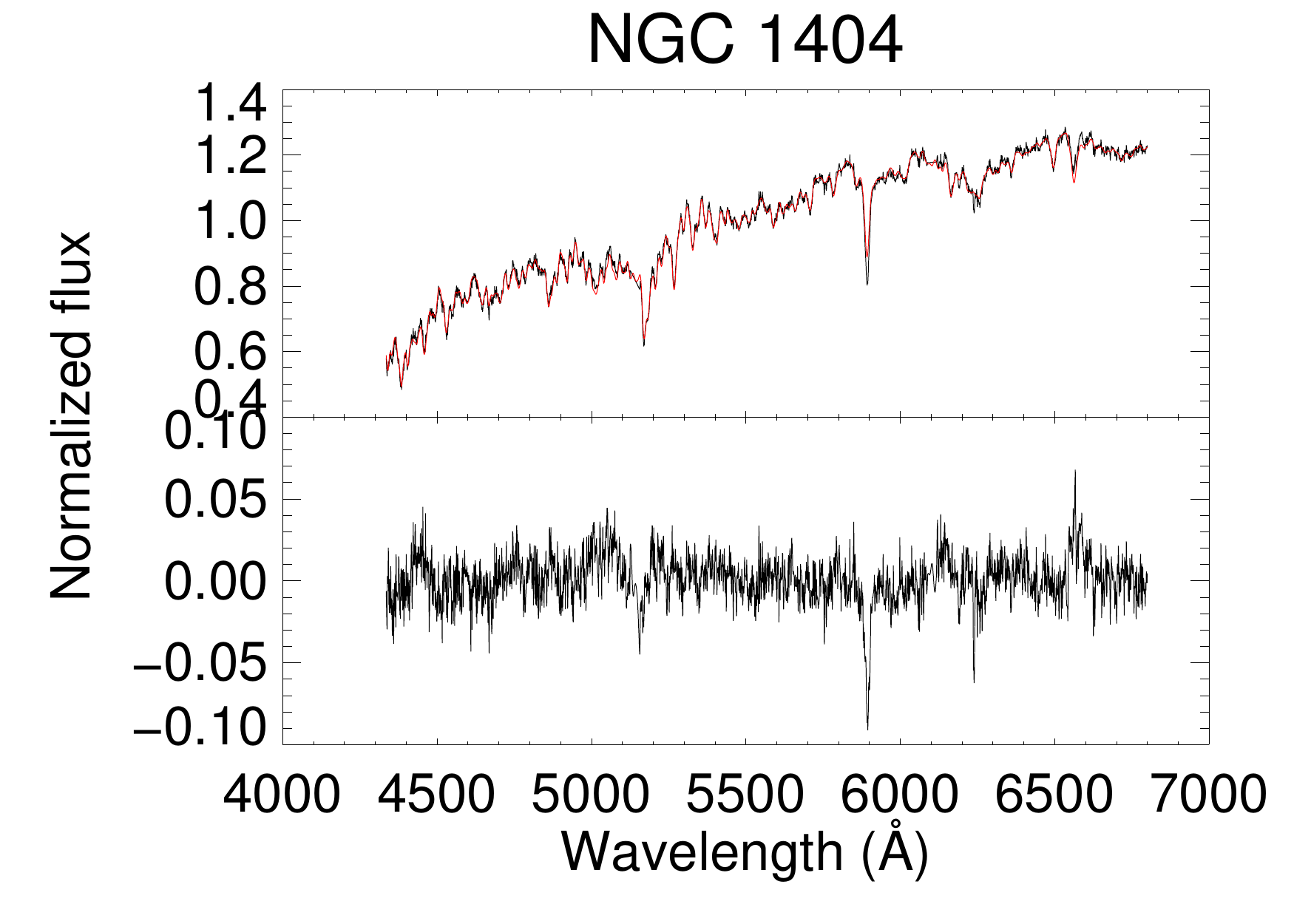}
\includegraphics[width=70mm,height=55mm]{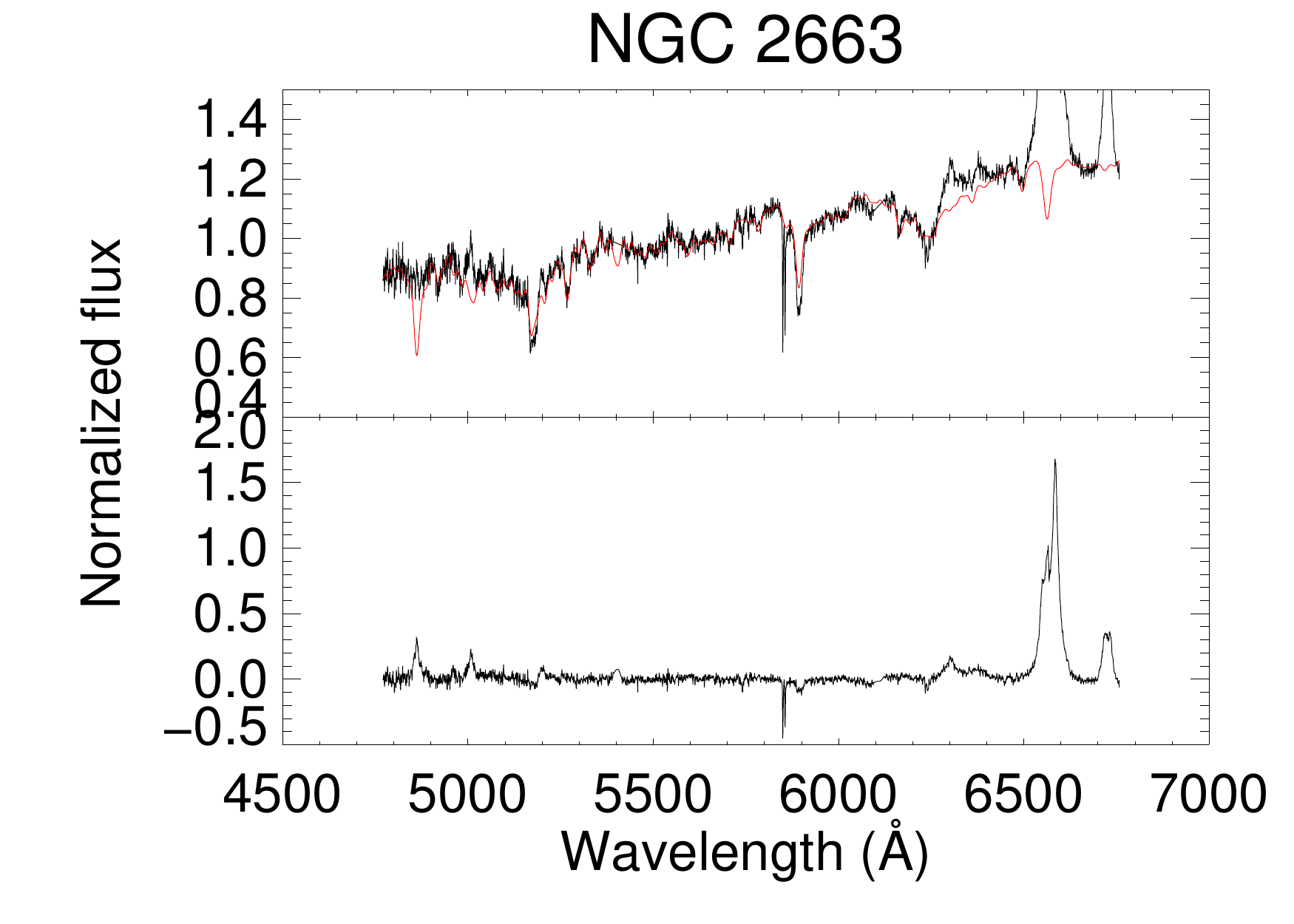}
\includegraphics[width=70mm,height=55mm]{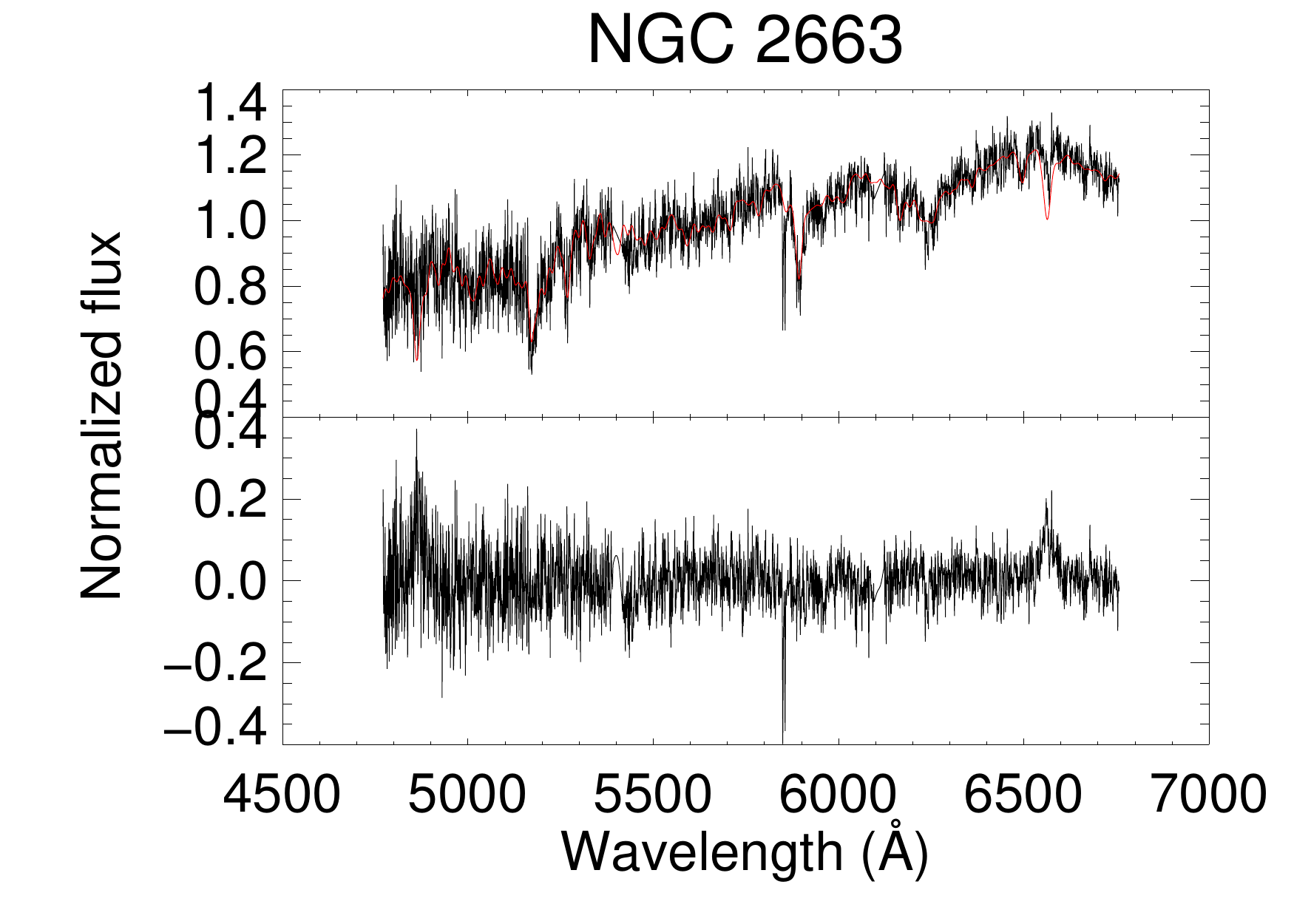}

\caption{continued \label{starlight_graf_2}
}
\end{figure*}

\addtocounter{figure}{-1}

\begin{figure*}
\includegraphics[width=70mm,height=55mm]{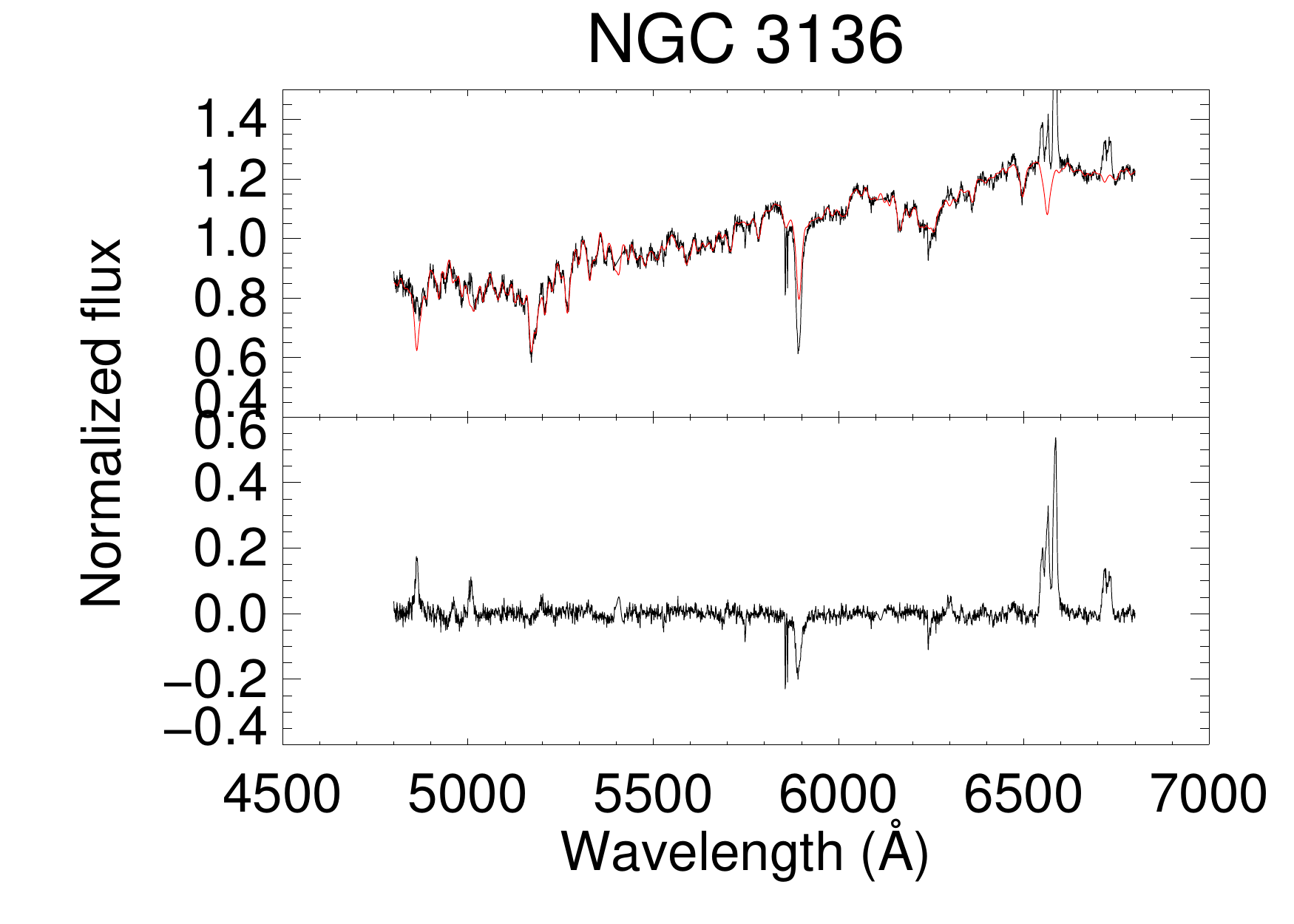}
\includegraphics[width=70mm,height=55mm]{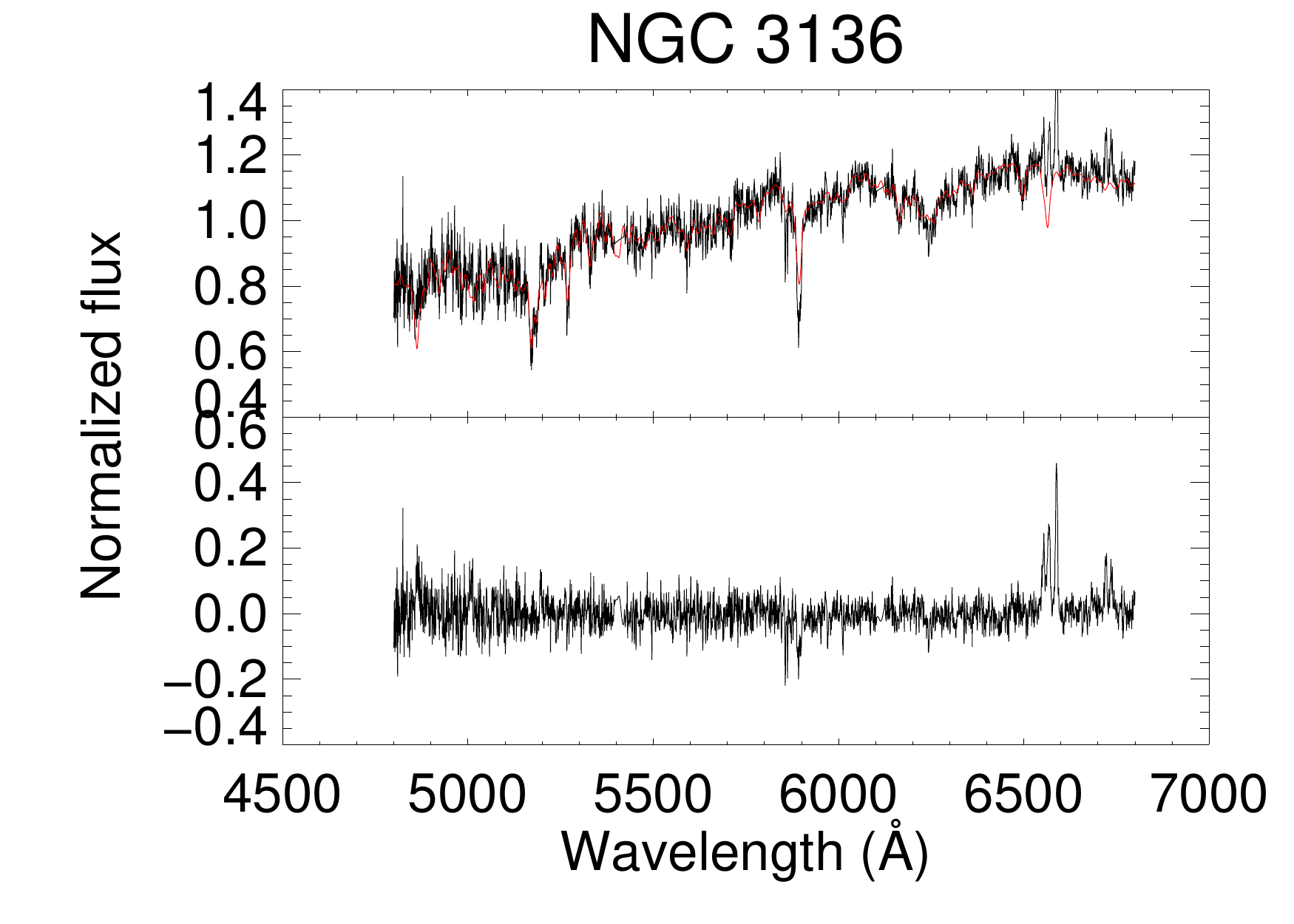}
\includegraphics[width=70mm,height=55mm]{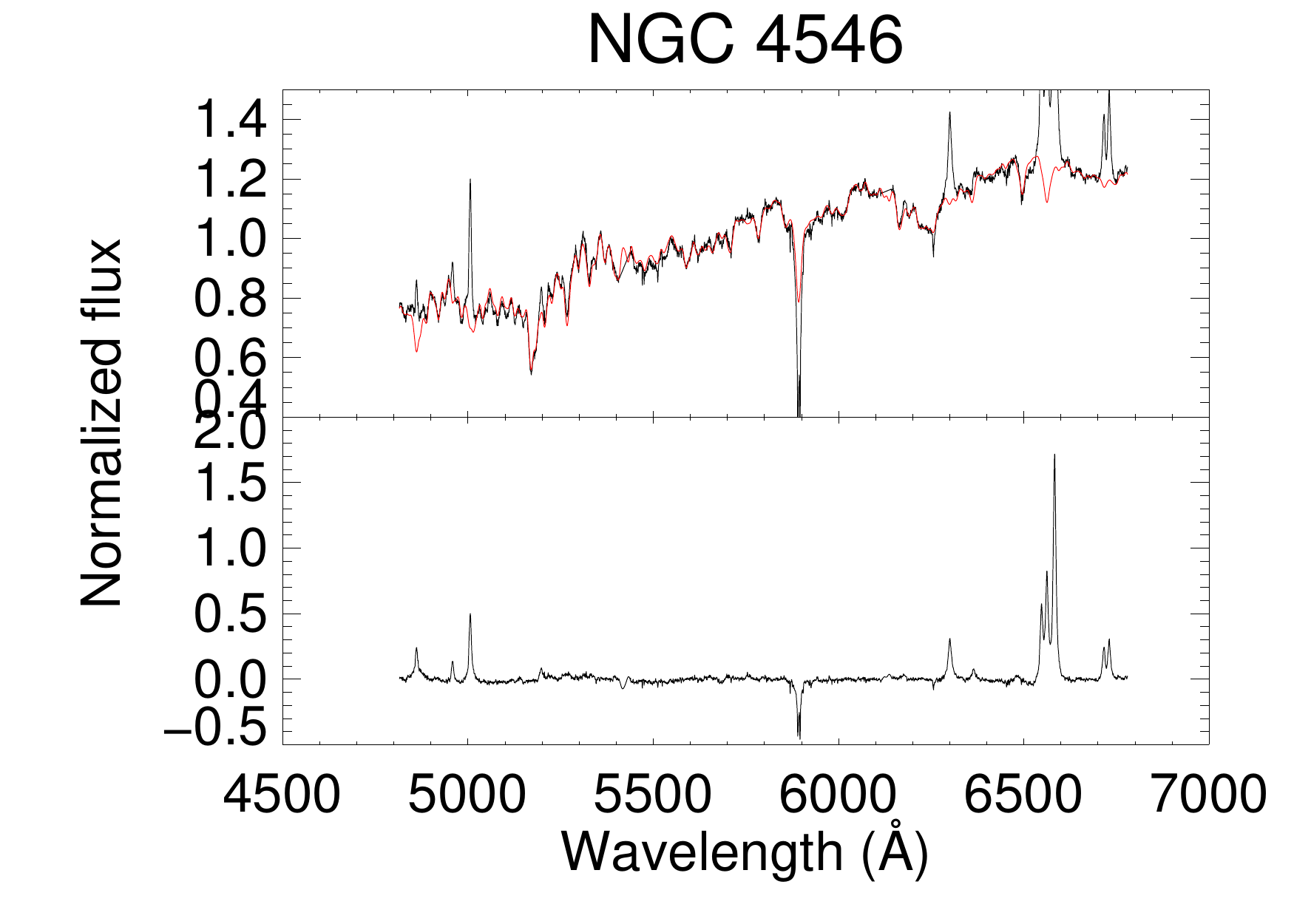}
\includegraphics[width=70mm,height=55mm]{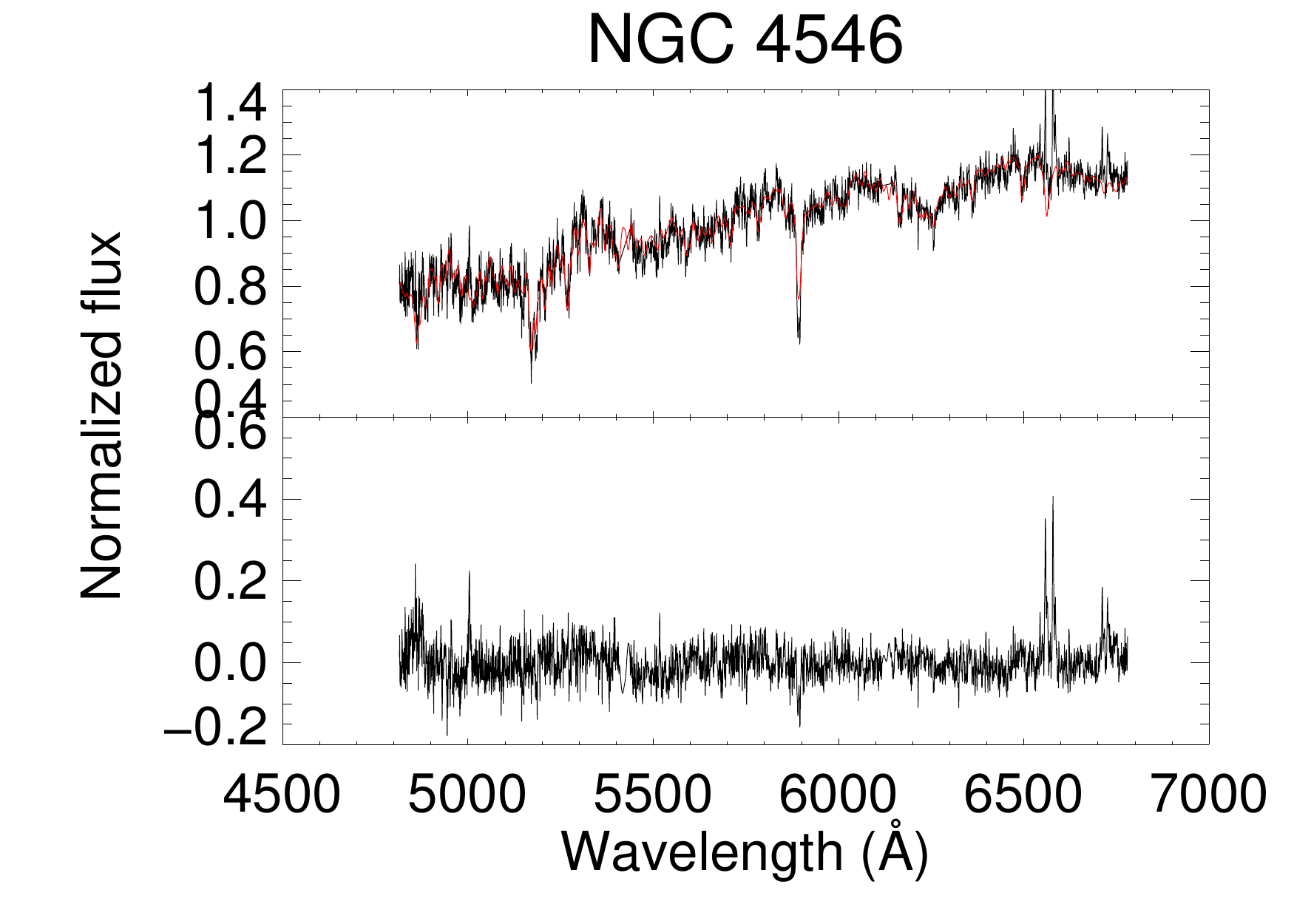}
\includegraphics[width=70mm,height=55mm]{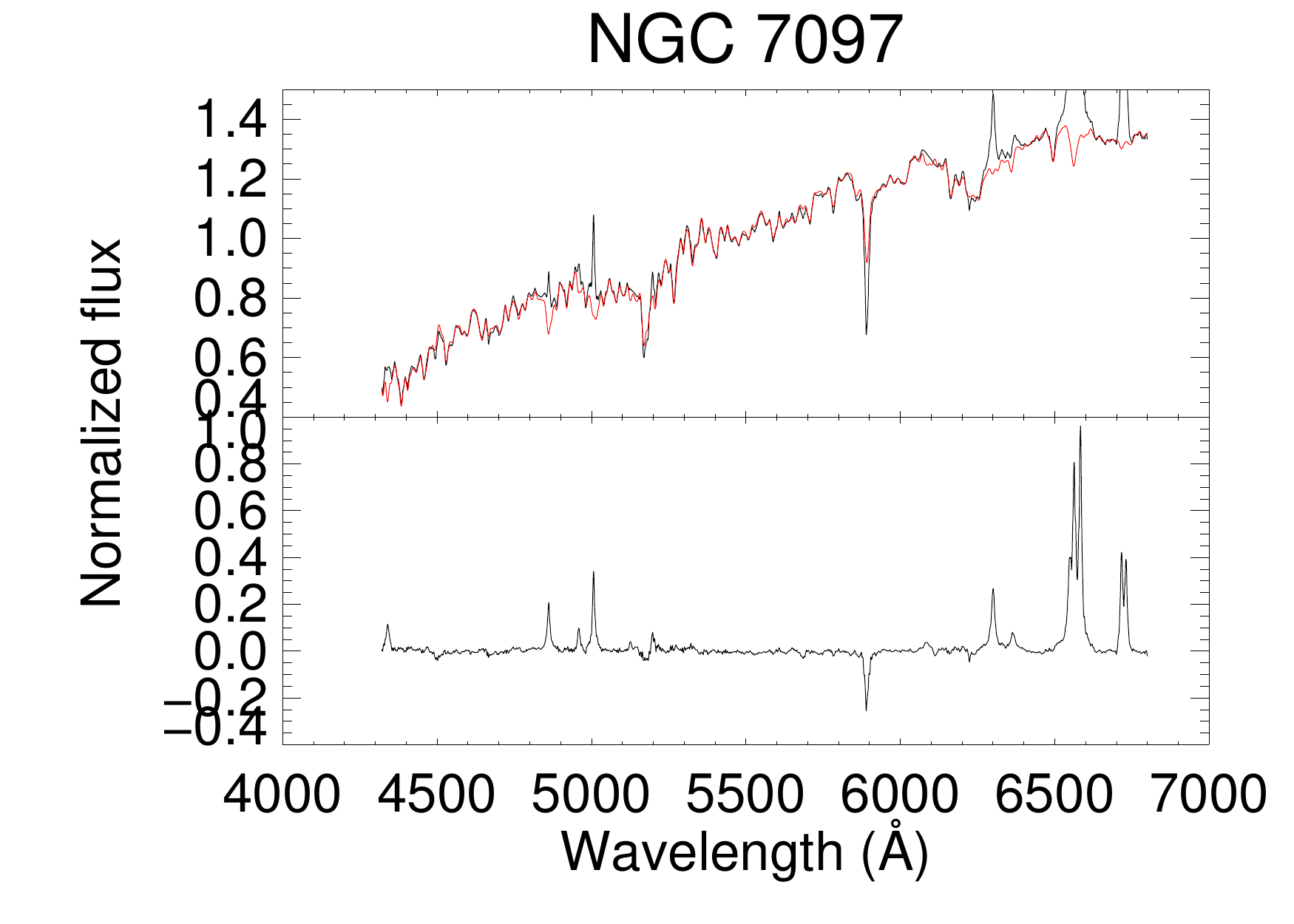}
\includegraphics[width=70mm,height=55mm]{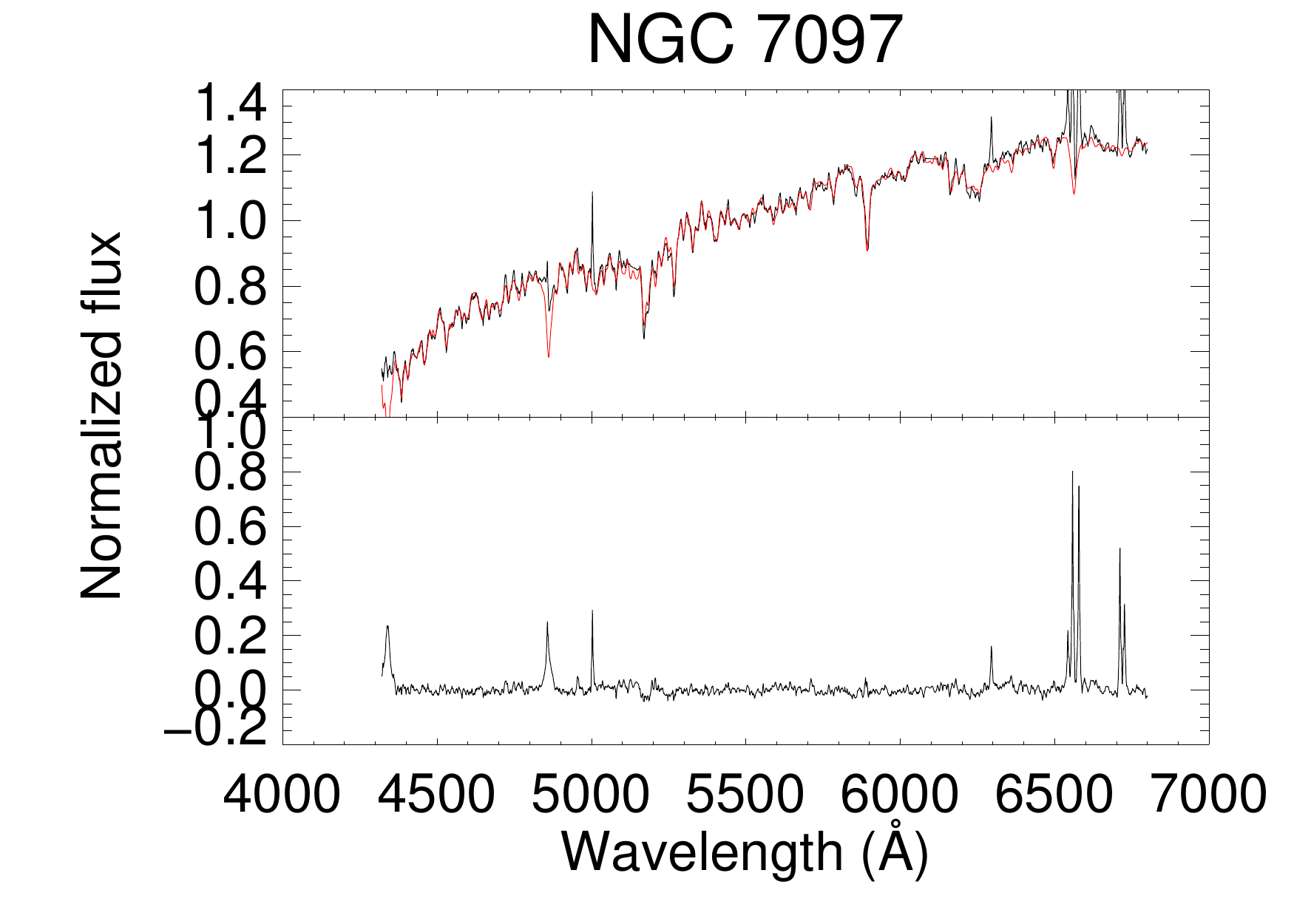}

\caption{continued  \label{starlight_graf_3}
}
\end{figure*}

\end{document}